\documentclass[iop,apj]{emulateapj}
\usepackage{amsmath,natbib,graphics,graphicx,apjfonts}

\usepackage{fnpct}

\input{buote_new.defs}
\newcommand\mlhband{{M_{\star}/L_H}}
\newcommand\src{\hbox{{Mrk~1216}}}

\begin{document}

\title{The Extremely High Dark Matter Halo Concentration of the \\ Relic Compact Elliptical Galaxy Mrk~1216}
\author{David A.\ Buote and Aaron J.\ Barth}
\affil{Department of Physics and Astronomy, University of California
  at Irvine, 4129 Frederick Reines Hall, Irvine, CA 92697-4575; \\ buote@uci.edu}


\slugcomment{Accepted for Publication in The Astrophysical Journal}

\begin{abstract}
  Spatially compact stellar profiles and old stellar populations have
  established compact elliptical galaxies (CEGs) as local analogs of
  the high-redshift ``red nuggets'' thought to represent the
  progenitors of today's early-type galaxies (ETGs). To address
  whether the structure of the dark matter (DM) halo in a CEG also
  reflects the extremely quiescent and isolated evolution of its
  stars, we use a new $\approx 122$~ks \chandra\ observation together
  with a shallow $\approx 13$~ks archival observation of the CEG \src\
  to perform a hydrostatic equilibrium analysis of the luminous and
  relaxed X-ray plasma emission extending out to a radius
  $0.85\rtwofiveh$. We examine several DM model profiles and in every
  case obtain a halo concentration $(c_{200})$ that is a large
  positive outlier in the theoretical \lcdm\ $c_{200}-M_{200}$
  relation; i.e., ranging from $3.4\,\sigma - 6.3\, \sigma$ above the
  median \lcdm\ relation in terms of the intrinsic scatter. The high
  value of $c_{200}$ we measure implies an unusually early formation
  time that firmly establishes the relic nature of the DM halo in
  \src. The highly concentrated DM halo leads to a higher DM fraction
  and smaller total mass slope at $1R_e$ compared to nearby normal
  ETGs. In addition, the highly concentrated total mass profile of
  \src\ cannot be described by MOND without adding DM, and it deviates
  substantially from the Radial Acceleration Relation.  Our analysis
  of the hot plasma indicates the halo of \src\ contains
  $\approx 80\%$ of the cosmic baryon fraction within $\rtwoh$.  The
  radial profile of the ratio of cooling time to free-fall time varies
  within a narrow range $(\tc/\tff\approx 14-19)$ over a large central
  region ($r\le 10$~kpc) suggesting ``precipitation-regulated
  AGN feedback'' for a multiphase plasma, though presently there is
  little evidence for cool gas in \src. Finally, other than its
  compact stellar size, the stellar, gas, and DM properties of \src\
  are remarkably similar to those of the nearby fossil group NGC~6482.

\end{abstract}

\section{Introduction}
\label{intro}

Massive early-type galaxies (ETGs) are widely believed to have formed
in a two-phase process~\citep[e.g.,][]{oser10a}. Phase~1 occurs at
early times $(z\ga 2)$ when dissipative gas infall leads to rapid star
formation and, along with some dark matter (DM) halo
contraction~\citep[e.g.,][]{dutt15a}, produces a very compact ``red nugget.''
Subsequent evolution in Phase~2 is primarily non-dissipative driven by
collisionless (``dry'') mergers, the effect of which is mostly
accretive (i.e., increasing the size of the stellar halo) with little
or no star formation.  This later slow accretive phase is revealed by
the stellar mass-size evolution of
ETGs~\citep[e.g.,][]{dadd05a,dokkum08a,damj11a,wel14a} and through
multi-component decompositions of nearby ETGs~\citep{huan13b}.  To
study the end of Phase~1 requires mapping the radial mass profiles of
galaxies at $z\sim 2$. Unfortunately, even with stellar dynamics
detailed mass mapping is not possible at present since only an average
velocity dispersion within approximately the stellar half-light radius
$(\reff)$ can be measured for $z\sim 2$
galaxies~\citep[e.g.,][]{toft12a,rhoa14a,long14a,sande14a}.

With detailed mass mapping of red nuggets extremely challenging, an
alternative approach is to study local analogs~\citep[e.g.,][]{remco12a,truj14a}. \citet{remco15a}
conducted a local survey of galaxies based on (among other criteria)
the estimated size of the gravitational radius of influence of the
central super-massive black hole (SMBH). From this survey they
identified a sample of compact elliptical galaxies (CEGs) that have
remarkable properties~\citep[][hereafter Y17, and references
therein]{yild17a}. (1) They have very old ($\ga 13$~Gyr) stellar
populations~(e.g., Y17; \citealt{ferr17a}). (2) They have compact
stellar surface brightness profiles that obey the stellar mass-size
relationship for $z\sim 2$ galaxies instead of $z=0$. (3) Some of the
CEGs have evidence for over-massive SMBHs with respect to the
$\msigma$
relation~\citep[e.g.,][see also \citealt{savo16a}]{ferr15a,yild15a,wals15a,wals17a}. Properties (1) and
(2) suggest that these CEGs are ancient relic galaxies that have
skipped the ``Phase 2'' of slow accretion of an extended stellar
envelope. In other words, they are likely passively evolved direct
descendants of the high-redshift red nugget population, and therefore
provide a new and more accessible avenue for studying the detailed
structure of red nuggets.

It is presently unknown whether the DM profiles corroborate the
interpretation of CEGs as relic galaxies. The scatter about the median
\lcdm\ $c_{200}-M_{200}$ relation reflects the halo formation time, history, and
environment~\citep[e.g.,][]{bull01a,neto07a,ludl16a,raga18a}. Consequently, if
the CEGs are truly red nugget analogs, their halo concentrations
should reflect the early formation epoch and isolated evolution and thus appear as large,
positive outliers in the local $c_{200}-M_{200}$ relation.

Motivated primarily by the desire to map the gravitating mass profiles
of CEGs, in \citet[hereafter Paper~1]{buot18a} we described the
results of the first systematic search for extended, luminous X-ray
emission in CEGs suitable for detailed hydrostatic equilibrium (HE)
analysis of their mass profiles.  Of the 16 CEGs studied by Y17, we
identified two objects -- \src\ and PGC~032873 -- that are extremely
promising for X-ray study and presented initial constraints on their
mass profiles~\citep[see also][]{wern18a}. Only for \src\ were the
existing \chandra\ Cycle 16 data of sufficient quality for a detailed
HE mass analysis from which we obtained the first tentative evidence
for an above average halo concentration for a CEG. We also placed a
tentative constraint on the SMBH mass consistent with the large
(``over-massive'') value obtained from stellar dynamics
by~\citet{wals17a}.

To confirm and strengthen these initial results, we submitted a
\chandra\ proposal for a deep 130~ks observation of \src\
which was approved and allocated time in Cycle 19. Here we report a
detailed analysis of the Cycle 19 image and spectra in conjunction
with an updated analysis of the shallow archival Cycle 16 data studied
in Paper~1. Some properties of \src\ are listed in
Table~\ref{tab.prop}.

The paper is organized as follows. We describe the \chandra\ X-ray
observations and the data preparation in \S\ref{obs}. In \S\ref{image}
we perform a detailed analysis of the image morphology to search for
features associated with AGN feedback. In \S\ref{spec} we describe the
spectral analysis.  We define the spectral model in \S\ref{specmod}
and present the results of the spectral fitting in
\S\ref{specresults}. We present the HE models in \S \ref{he}, the
fitting methodology in \S\ref{fitproc}, the results of the HE mass
analysis in \S\ref{results}, and the error budget in \S\ref{sys}.  We
discuss several topics in \S\ref{disc} and present our conclusions in
\S \ref{conc}.

\section{Observations and Data Preparation}
\label{obs}

\begin{table*}[t] \footnotesize
\begin{center}
\caption{Target Properties}
\label{tab.prop}
\begin{tabular}{lccccccccc}   \hline\hline\\[-7pt]
& & Distance & Scale & $N_{\rm H}$ & $L_{\rm H}$ & $R_e$ & $\sigma_e$ & $L_{\rm x}$ & $\ktemp$ \\
Name & Redshift & (Mpc) & (kpc/arcsec) & ($10^{20}$~cm$^{-2}$) & $(10^{11}\, L_{\odot})$  & (kpc) & (km/s) & ($10^{42}$~ergs~s$^{-1}$) & (keV)\\
\hline \\[-7pt]
\src\ &  0.021328 & 97.0 & 0.45 & 4.0 & 1.14 & 2.3 & 308 & $1.7 \pm 0.1$ & $0.73\pm 0.01$\\
\hline \\
\end{tabular}
\tablecomments{The redshift is taken from
  NED\footnote{http://ned.ipac.caltech.edu}. We compute the distance
  using the related redshift (also taken from NED) corrected to the
  reference frame defined by the 3~K cosmic background radiation
  assuming $\Omega_{\rm m,0}=0.3$, $\Omega_{\Lambda,0}=0.7$, and
  $H_0=70$~km~s$^{-1}$~Mpc s$^{-1}$.  We calculate the Galactic column
  density using the HEASARC {\sc w3nh} tool based on the data
  of~\citet{kalb05a}.  The total $H$-band luminosity, circularized
  effective radius $(R_e$), and stellar velocity dispersion are taken
  from Y17.  $L_{\rm x}$ and $\ktemp$ are, respectively, the projected,
  emission-weighted luminosity (0.5-7.0~keV) and temperature computed using the
  best-fitting hydrostatic model for the galaxy within a projected
  radius of 100~kpc (\S\ref{results}).}
\end{center}
\end{table*}

\begin{table}[t] \footnotesize
\begin{center}
\caption{Observations}
\label{tab.obs}
\begin{tabular}{cclccc}   \hline\hline\\[-7pt]
& & & & & Exposure\\
Cycle & Obs.\ ID & Obs.\ Date & Instrument & Active CCDs & (ks)\\
\hline \\[-7pt]
16 & 17061 & 2015  Jun.\ 12 & ACIS-S & S1,S2,S3,S4 & 12.9\\
19 & 20342 & 2018 Jan.\ 9 & ACIS-S & I2,I3,S2,S3 & 31.7\\
19 & 20924 & 2018 Jan.\ 9 & ACIS-S & I2,I3,S2,S3 & 29.7\\
19 & 20925 & 2018 Jan.\ 12 & ACIS-S & I2,I3,S2,S3 & 31.4\\
19 & 20926 & 2018 Jan.\ 14 & ACIS-S &  I2,I3,S2,S3 & 29.7\\
\hline \\
\end{tabular}
\tablecomments{The exposure times refer to those obtained after
  filtering the light curves (\S\ref{obs}), which resulted in a negligible amount of
  excluded time for each observation. The total clean exposure for
  the Cycle~19 observation is 122.4~ks.}
\end{center}
\end{table}

We list the details of the \chandra\ observations in
Table~\ref{tab.obs}.  In Cycle~19 \src\ was observed with the ACIS
CCDs during 2018 from January~9 to January~14 in four exposures for
$\sim 30$~ks each. The aim point of the telescope was located on the
S3 chip (i.e., ACIS-S configuration), although a non-standard chip set
was used (notably with the I2 and I3 chips both active) to allow for a
simultaneous measurement of the background. We prepared the data for
imaging and spectral analysis using the \ciao\
(v4.10)\footnote{http://cxc.harvard.edu/ciao/} and \heasoft\
(v6.24)\footnote{https://heasarc.gsfc.nasa.gov/docs/software/heasoft/}
software suites along with version 4.8.1 of the \chandra\ calibration
database\footnote{http://cxc.harvard.edu/caldb/calibration/}.

We begin by reprocessing each each Cycle 19 exposure with the latest
calibration information. To clean these exposures of periods of high
particle background, we created broad-band light curves extracted from
regions without obvious point sources and excluding most of the
emission from \src. We filtered the light curves with a $3\sigma$ clip
procedure (see \ciao\ {\sc deflare} and {\sc lc\_clean} tasks) which
resulted in almost no time removed for a combined total exposure of 122.4~ks. The
cleaned times for each exposure are listed in Table~\ref{tab.obs}.

To generate images for the entire Cycle 19 data set, we first combine
the individual events lists into a single file. We begin by correcting
the absolute astrometry for each exposure using the {\sc ciao} task
{\sc reproject\_aspect} along with initial point source lists obtained
from their 0.5-7.0~keV images using the {\sc ciao} task {\sc
  wavdetect}. We combined the aligned exposure into a single events
list from which an image and exposure map was created using the {\sc
  ciao} task {\sc merge\_obs}. In this way we create merged images of
the entire Cycle 19 observation of varying energy ranges and pixel
sizes.

Since our focus is on the diffuse emission, we generate a source list
using {\sc wavdetect} applied to the $0.5-7.0$~keV image. We verify
the detected point sources by visual inspection while
excluding the detection of the center of \src. We assign a radius for
each source to correspond to the 95\% encircled energy fraction for a
1-keV monochromatic point source appropriate for its off-axis location
in the ACIS field.

While most of our imaging and spectral analysis employs a local
background measured directly from the \chandra\ observations of \src, we
nevertheless, as described below, make some use of the background
derived instead from regions of nominally blank sky. For
each of the Cycle 19 observations we created such ``blank sky'' images using the
{\sc ciao} tasks {\sc blanksky} an {\sc blanksky\_image}. We co-add the
images of each exposure to obtain a total blank-sky background image
matching the energy band and spatial binning for the corresponding
source image. 
 
For our primary spectral analysis, we defined a series of concentric,
circular annuli positioned very near to the optical center (\S
\ref{image}) while masking out point sources, chip gaps, and other off-chip
regions. There is significant latitude in choosing the widths of
the annuli depending on the scientific objectives. We balanced the need
for source counts with the need to sample the radial profile within
$1R_e$ arriving at a criterion of $\approx 1000$ background-subtracted
counts in the $0.5-2.0$~keV image (using the softer band to emphasize
the $\ktemp=0.5-1$~keV hot plasma contribution). In addition, to better probe the
gravitational effect of a central SMBH, we required the central
aperture to have a radius of 2 pixels ($0.982\arcsec$ radius),
enclosing $\approx 90\%$ of the point spread function, and containing
a little below 600 source counts. The annulus definitions are listed in
Table~\ref{tab.gas}. Note that all the annuli listed in
Table~\ref{tab.gas} lie entirely on the S3 chip except annulus~10 for
which almost 30\% of the area lies on the S2 chip. Finally, to constrain
the local background, we also included a single large annulus
($R=4.1\arcmin-14.8\arcmin$, not listed in Table~\ref{tab.gas}) with
negligible source counts but containing most of the available area of the S2,
I2, and I3 chips.

We extracted a spectrum and created counts-weighted redistribution
matrix (RMF) and auxiliary response (ARF) files using the {\sc ciao}
task {\sc specextract} for each region and Cycle 19 exposure. Then
for each region we created a combined spectrum, RMF, and ARF files
using the {\sc ciao} task {\sc combine\_spectra}. Combining the RMFs
and ARFs in this way is a convenience and should be appropriate for
\src\ since constraints on the spectral models are dominated by
statistical rather than systematic errors in the response. Nevertheless, to verify
this expectation we have also analyzed the un-merged spectra (\S
\ref{specresults} and Appendix \ref{joint}). 

We also examined whether enhanced Solar Wind Charge Exchange (SWCX)
emission may have significantly affected the Cycle 19 observations. We
used the Level~2 data from
SWEPAM\footnote{http://www.srl.caltech.edu/ACE/ASC/level2/lvl2DATA\_SWEPAM.html}
to obtain the solar proton flux during each \chandra\ observation. All
4 Cycle 19 exposures have solar proton flux below
$\approx 2\times 10^8$~cm$^{-2}$~s$^{-1}$ indicating significant
proton flare contamination is not expected~\citep{fuji07a}.

Finally, we have updated and prepared the Cycle 16 observation as above,
but without merging it with the Cycle 19 data, and maintaining the same
annuli definitions used in Paper~1.

\section{Image Morphology: Search for Structural Evidence of AGN Feedback}
\label{image}

\begin{figure*}[t]
\begin{center}
\includegraphics[scale=0.55,angle=0]{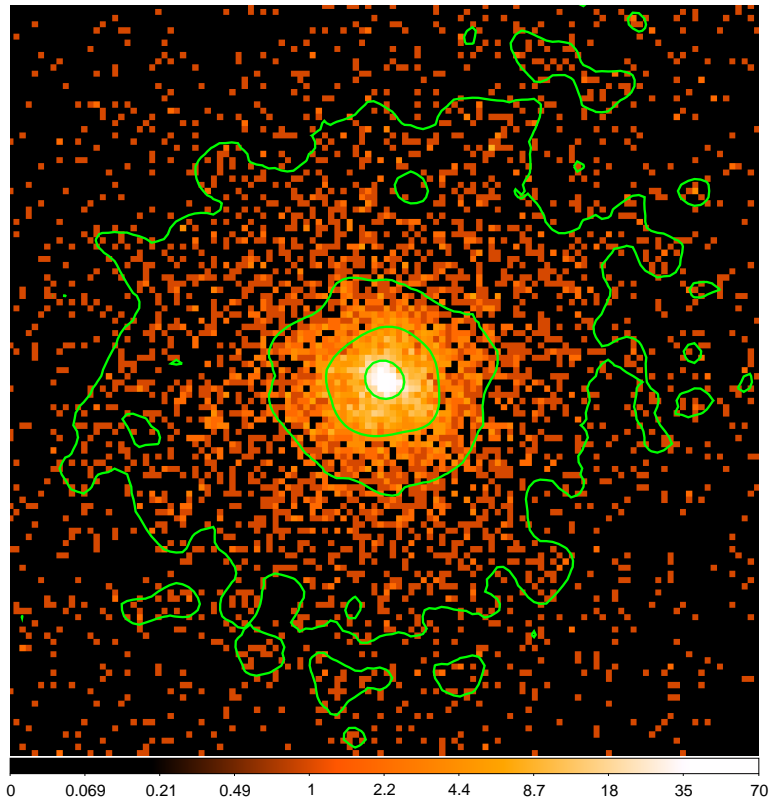}
\end{center}
\vskip 1.5cm
\caption{\footnotesize 0.5-2.0~keV raw image of the central $1\arcmin
  \times 1\arcmin$ region with smoothed, logarithmically spaced
  contours overlaid. The bottom color bar shows the counts per
  pixel, where 1 pixel is $0.492\arcsec\times 0.492\arcsec$.}
\label{fig.image}
\end{figure*}

\begin{table}[t] \footnotesize
\begin{center}
\caption{Ellipticity Profile of the Central 10~kpc}
\label{tab.ellip}
\begin{tabular}{cccc}   \hline\hline\\[-7pt]
$a$ & $a$ & $\epsilon$ & PA\\
(arcsec) & (kpc) & & (deg N-E)\\
\hline \\[-7pt]
      1.23 &      0.55 & $ 0.20 \pm   0.08$ & $   66 \pm    45$\\
     3.69 &      1.66 & $ 0.22 \pm   0.06$ & $   67 \pm    30$\\
     6.15 &      2.77 & $ 0.06 \pm   0.03$ & $   93 \pm    26$\\
     8.61 &      3.88 & $ 0.09 \pm   0.03$ & $   87 \pm    12$\\
    11.32 &      5.09 & $ 0.07 \pm   0.03$ & $   83 \pm    20$\\
    14.27 &      6.42 & $ 0.05 \pm   0.03$ & $   98 \pm    28$\\
    17.96 &      8.08 & $ 0.04 \pm   0.03$ & $  110 \pm    28$\\
    22.39 &     10.07 & $ 0.05 \pm   0.02$ & $  106 \pm    22$\\
\hline \\
\end{tabular}
\tablecomments{The ellipticity and position angle of the 0.5-2.0~keV
  surface brightness as a function of semi-major axis $a$ obtained
  using a moment analysis (\S\ref{image}).}
\end{center}
\end{table}

\begin{figure*}[t]
\begin{center}
\includegraphics[scale=0.42,angle=-90]{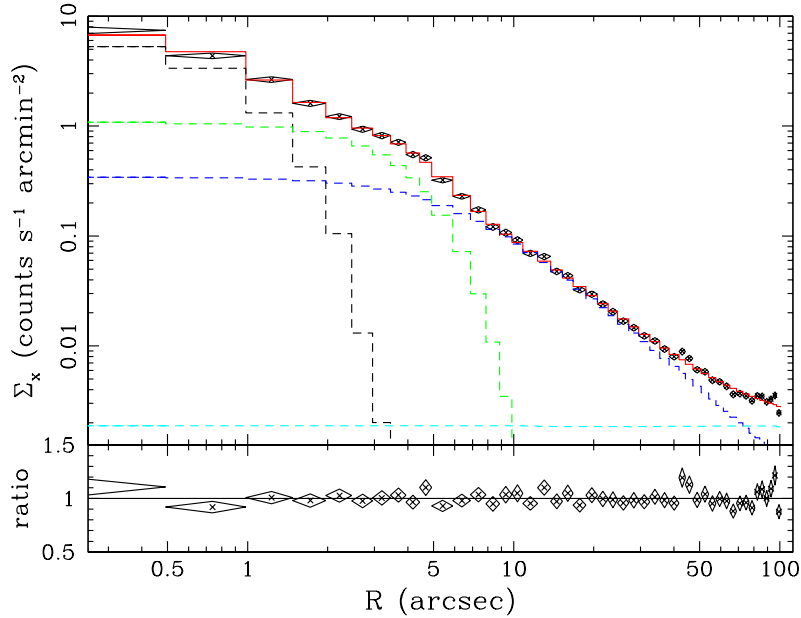}
\end{center}
\caption{\footnotesize 0.5-2.0~keV surface brightness profile and
  best-fitting model (Table~\ref{tab.surf}) of the central
  $\sim 100\arcsec$. The model is fully two-dimensional but has been
  binned radially for display purposes only ($\sim 200$ counts per
  bin). The total model is shown by the solid red bins. The individual
  model components are as follows: gauss~1~(dashed black),
  gauss~2~(dashed green), beta~(dashed blue), and constant
  background~(dashed cyan). The bottom panel shows the data/model
  ratio.}
\label{fig.surf}
\end{figure*}

\begin{table*}[t] \footnotesize
\begin{center}
\caption{Surface Brightness Model Parameters}
\label{tab.surf}
\begin{tabular}{ccccccc}   \hline\hline\\[-7pt]
& $A$ & FWHM & & PA & $r_c$\\
Model & (cts s$^{-1}$ arcmin$^{-2}$) & (arcsec) & $\epsilon$  & (deg N-E) & (arcsec) & $\beta$\\
\hline \\[-7pt]
gauss 1 & $6.01_{-0.41}^{+0.44}$ & $2.07_{-0.12}^{+0.13}$ & $0.29_{-0.05}^{+0.05}$ & $50_{-6}^{+6}$ & $\cdots$ & $\cdots$\\
gauss 2 & $1.09_{-0.12}^{+0.11}$ & $6.75_{-0.29}^{+0.34}$ & $0.09_{-0.04}^{+0.04}$ & $92_{-14}^{+14}$ & $\cdots$ & $\cdots$\\
beta & $0.34_{-0.10}^{+0.13}$ & $\cdots$  & $0$  & $\cdots$ & $6.2_{-1.3}^{+2.0}$ & $0.52_{-0.02}^{+0.03}$\\
const bkg & $0.0019_{-0.0002}^{+0.0002}$  & $\cdots$ & $\cdots$ & $\cdots$ & $\cdots$ & $\cdots$\\
\\[-7pt]
\hline \\
\end{tabular}
\tablecomments{Best-fitting parameters and $1\sigma$ errors of the two-dimensional,
  multi-component surface-brightness (0.5-2.0~keV) model (\S\ref{image}).}
\end{center}
\end{table*}

\begin{figure*}
\parbox{0.49\textwidth}{ 
\centerline{\includegraphics[scale=0.43,angle=0]{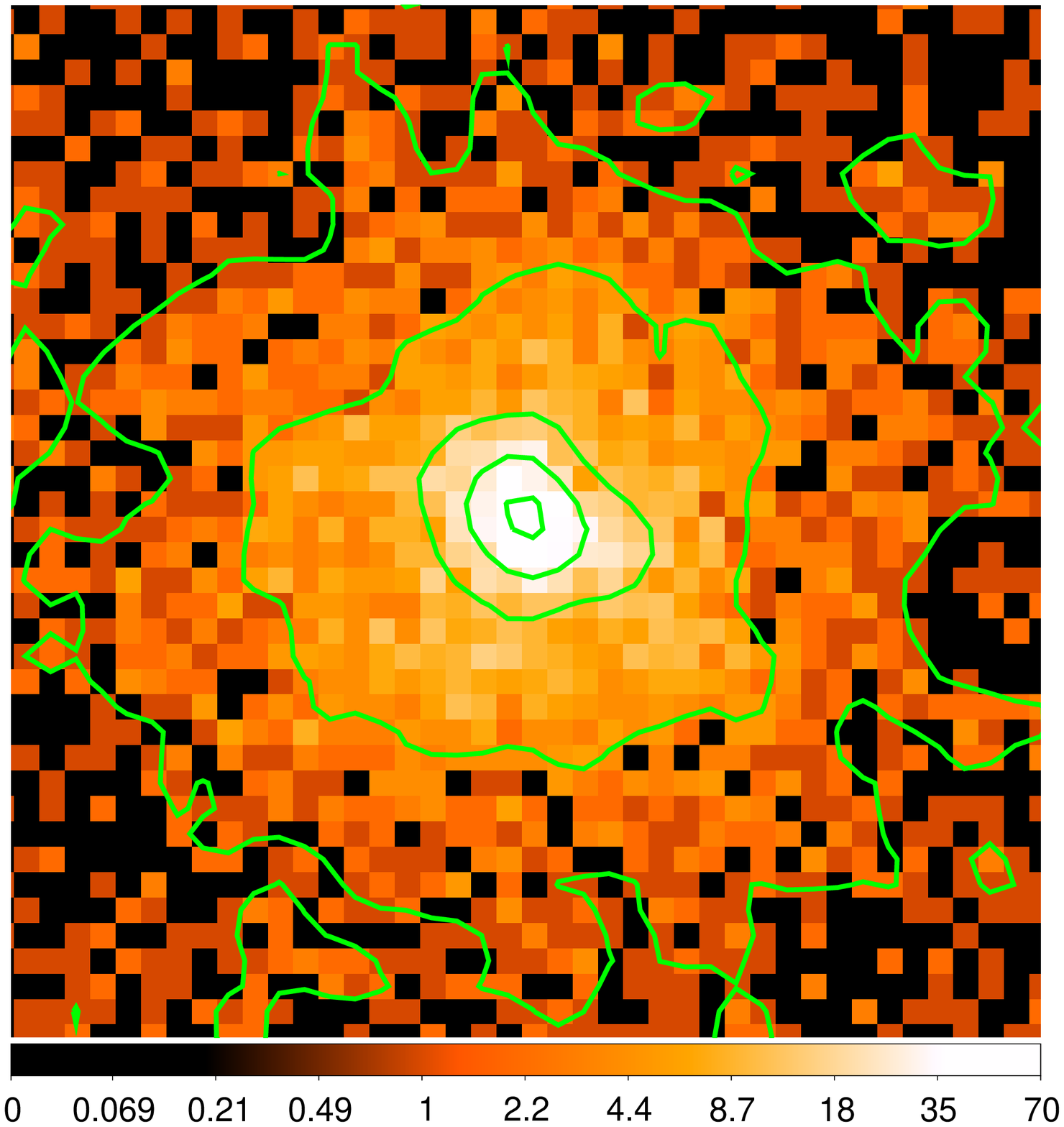}}}
\parbox{0.49\textwidth}{ 
\centerline{\includegraphics[scale=0.43,angle=0]{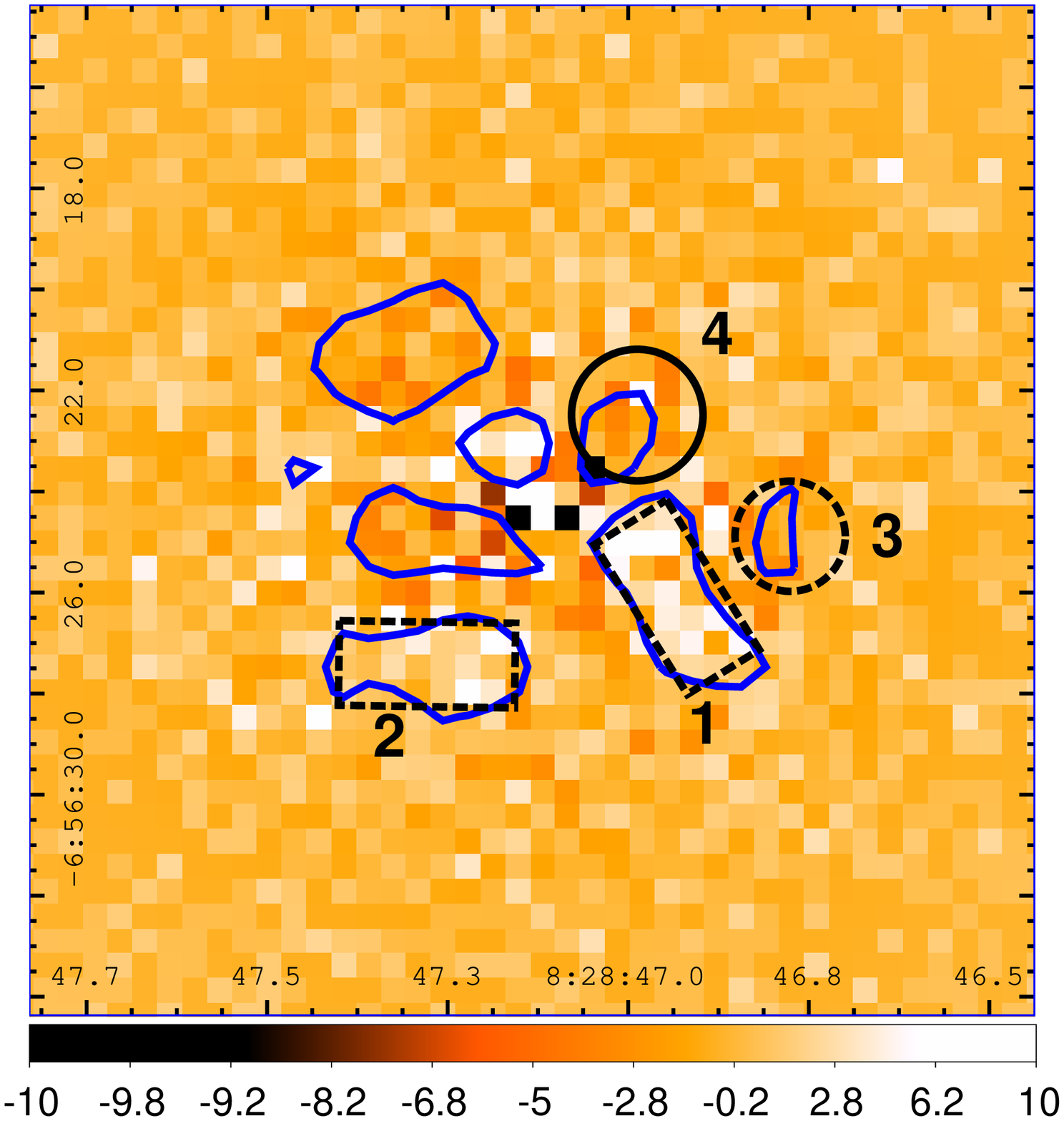}}}
\vskip 1.5cm
\caption{\label{fig.image.center} Residual image analysis.  ({\sl Left
      Panel}) 0.5-2.0~keV image of the central $20\arcsec\times
    20\arcsec$ region. The contours are different from those displayed
    in Fig.~\ref{fig.image} and use square-root spacing. In addition, this image has the point
sources filled in with local background (\S\ref{image}).  ({\sl Right
      Panel}) Residual image created by subtracting the smooth
    multi-component surface-brightness model (Table~\ref{tab.surf} and
    Fig.~\ref{fig.surf}). The smoothed contours are displayed with
    linear spacing. The black regions (dashed and solid) are those
    showing the most significant fluctuations with respect to their
    local environment (\S\ref{image}).}
\end{figure*}

\begin{table*}[t] \footnotesize
\begin{center}
\caption{Residual Map Region Properties}
\label{tab.resid}
\begin{tabular}{c|cc|rrr|lc}   \hline\hline\\[-7pt]
& \multicolumn{2}{c}{Center} & \multicolumn{3}{c}{Counts} & \multicolumn{1}{c}{Ratio} &  \multicolumn{1}{c}{sign(Ratio)$\sqrt{\rm |Ratio|}$}\\
Region & RA & Dec & \multicolumn{1}{c}{Image} & \multicolumn{1}{c}{Model} & \multicolumn{1}{c}{Residual} &  \multicolumn{1}{c}{(\%)} &  \multicolumn{1}{c}{(\%)}\\
\hline \\[-7pt]
    1 	& 8:28:46.950 & $-$6:56:26.092 &  239 & 175.5 & $+63.5$ & $+36.2\pm 8.8$ & $+16.7\pm 4.3$\\
    2 	& 8:28:47.282 & $-$6:56:27.420 &  143 & 108.1 & $+34.9$ & $+32.2\pm 11.1$ & $+15.0\pm 5.4$\\
    3 	& 8:28:46.799 & $-$6:56:24.885 &  36   & 57.1    & $-21.1$ & $-36.9_{-10.5}^{+12.3}$ & $-20.6_{-5.1}^{+6.0}$\\
    4 	& 8:28:47.002 & $-$6:56:22.486 &  147   & 169.3 & $-22.3$ & $-13.2\pm 7.1$ & $-6.8\pm 3.5$\\
\\[-7pt]
\hline \\
\end{tabular}
\tablecomments{Properties of notable regions of the $0.5-2.0$~keV
  residual map (\S\ref{image} and
  Figure~\ref{fig.image.center}). Regions~1-3 represent those with the
  largest residuals we studied; i.e., dashed regions in
  Figure~\ref{fig.image.center}.  Region~4 displays the most
  interesting spectral deviations (\S\ref{resid}); i.e., solid circle
  in Figure~\ref{fig.image.center}. The ``Image'' column gives the
  counts in the raw image, ``Model'' gives the counts predicted by the
  best-fitting model (Table~\ref{tab.surf} and Figure~\ref{fig.surf}),
  and ``Residual'' gives the counts resulting when the model is
  subtracted from the image. The ratios refer to the data divided by
  the model expressed as percentages above (positive) or below
  (negative) the model. The final column takes the square-root
  of the data / model ratio and converts it to a percentage while
  preserving the sign.  The
  result will indicate the hot gas density ratio if the differences
  between the plasma emissivities (especially the temperatures and
  iron abundances) of the data and model are negligible. The error bars
  on the ratios derive from Gaussian noise except for Region~3 where
  we used the Poisson error bars tabulated by \citet{gehr86a}. The
  definitions of the regions are as follows: Region~1 is a rectangle
  with sides of lengths $3.5\arcsec$ and $1.7\arcsec$ rotated by
  $121.8^{\circ}$ N-E; Region~2 is a rectangle also with sides of
  lengths $3.5\arcsec$ and $1.7\arcsec$ but rotated by only
  $0.9^{\circ}$ N-E; Region~3 is a circle with radius $1.1\arcsec$;
  Region~4 is a circle with radius $1.3\arcsec$.}
\end{center}
\end{table*}

Since the demise of the classical cooling flow paradigm brought about
by early observations with the \chandra\ and \xmm\
telescopes~\citep[e.g.,][]{pete06a}, it is now
generally accepted that in the central regions of cool-core clusters
and isolated massive galaxies episodic AGN feedback suppresses and
regulates gas cooling~\citep[e.g.,][]{mcna07a}. Although the details
of the feedback process are complex and are the subject of much
current research in the field, the fundamental mechanism by which the
AGN energizes the hot plasma is widely believed to be mechanical
feedback from AGN radio jets; i.e, the jet interacts with the hot
plasma and, e.g., inflates bubbles and cavities, generates weak shocks
and sound waves, which deliver energy to the hot plasma. Consequently,
in this section we have performed a detailed search for signs of AGN feedback in \src\
in the form of irregular features in the central part of the X-ray
image. (Spectral signatures are examined in \S\ref{quad} and \S\ref{spec.resid}).
Since radio observations of \src\ currently indicate only a
weak point source~\citep[limited to the single $9.2\pm 0.2$ mJy
detection in the 1.4~GHz NVSS,][]{nvss}, our present investigation of
signs of AGN feedback will consider mainly the X-ray image morphology.
We focus our analysis on the Cycle 19 data since the Cycle 16 data do
not provide strong constraints on the central image structure
(Paper~1).

We focus our analysis on the merged Cycle 19 image in the
$0.5-2.0$~keV band using a monochromatic 1-keV exposure map. In
Figure~\ref{fig.image} we show the raw image of the central
$1\arcmin \times 1\arcmin$ region at full resolution overlaid with
smooth contours. The image appears very regular with rather round
(though noisy) contours; i.e. the impact of AGN feedback on the image
of \src\ is not dramatic in the same way as observed for some
well-studied Virgo galaxies -- M84~\citep{fino08a} and
NGC~4636~\citep{bald09a}. It is possible, however, that features
similar to those seen in some Virgo galaxies are present in \src\ but
are merely less prominent owing to \src\ being 5-6 times more distant
than Virgo. Therefore, a quantitative assessment of image morphology
is required.

\subsection{Moment Analysis}
\label{moment}

To make a quantitative analysis of the X-ray image morphology, we
begin by computing the ellipticity $(\epsilon)$, position angle (PA),
and centroid evaluated within elliptical apertures as a function of
semi-major axis $a$. We apply an iterative scheme equivalent to
diagonalizing the moment of inertia tensor of the image region
(\citealt{cm}; see \citealt{buot94} for application to X-ray images of
elliptical galaxies). Before applying this technique, we replaced
detected point sources (\S\ref{obs}) with local background using the
\ciao\ {\sc dmfilth} tool.

In Table~\ref{tab.ellip} we list the ellipticity and position angle as
a function of $a$ within 10~kpc ($\approx 22\arcsec$). Not listed in
the table is the center position which is quite steady; e.g., the
center shifts by only $1.1\pm 0.4$ pixels ($0.5\arcsec\pm 0.2\arcsec$)
when comparing the centroids of the $a=1.23\arcsec,22.39\arcsec$
apertures. Within the relatively large statistical errors, the PA
within $\approx 10$~kpc is consistent with the $H$-band value of
$70.15^{\circ}$ reported by Y17, with some weak evidence it increases
near $a=10$~kpc (also see below). The ellipticity, however, displays a
clear radial variation. Within $a\approx 4\arcsec$ the image is
modestly flattened with $\epsilon\approx 0.20$. The ellipticity then
drops quickly for larger $a$ to a small value $\approx 0.05$ not
inconsistent with $\epsilon=0$. The X-ray morphology is thus broadly
similar to that observed for the relaxed, fossil-like elliptical
galaxy NGC~720~\citep[e.g.,][]{buot02b}; i.e., within $\approx 1R_e$
the X-ray image is moderately flattened (though rounder than the
stellar isophotes, $\epsilon=0.42$ -- Y17) and consistent with being
aligned with the stellar image before giving way to a much rounder
X-ray image at larger radius. Hence, the moment analysis of the
centroid, $\epsilon$, and PA within $a\approx 10$~kpc does not
indicate the presence of irregular surface brightness features.

\subsection{Two-Dimensional Model}

To search for more subtle features in the X-ray image we construct a
smooth two-dimensional model, subtract it from the image, and inspect
the residual image using the {\sc sherpa} fitting
package\footnote{http://cxc.cfa.harvard.edu/sherpa/} within {\sc
  ciao}. We initially defined a model consisting of an isothermal
$\beta$ model~\citep{beta} for the hot gas and a constant
background. Each model component was folded through the exposure map
and fitted (using the C-statistic) to the full-resolution image
within a radius of $\approx 100\arcsec$ from the center of the galaxy
(close to the edge of the S3). For our
fiducial analysis here and throughout the paper, we defined the center
of the X-ray image to be the centroid computed within a circular
aperture of radius $3\arcsec$ initially located at the emission
peak. This gives $\rm (R.A.,Dec)$ of (8:28:47.1410, $-$6:56:24.368)
which is very consistent with the stellar center determined by
\citet{springob14a}, though we note that there are $1\arcsec$-level
differences in the various position references collected by NED. (We
examine the effect of choosing a slightly different center in \S
\ref{center2}.)

We found that the initial model fit produced significant residuals
within central region. We noticed a substantial reduction in these
residuals upon adding two gaussian components with different widths;
i.e., a crude multi-gauss expansion. (Adding more gaussian components
produced comparatively minor changes.)  The best-fitting parameters
and $1\sigma$ errors are listed in Table~\ref{tab.surf}. In
Figure~\ref{fig.surf} we plot the best-fitting two-dimensional model
(and data/model ratio) binned as a radial profile where each bin
contains $\sim 200$ counts. Notice in particular the negligible
residuals within the central region. The gaussian components display
$\epsilon$ and PA values similar to the moment analysis
(Table~\ref{tab.ellip}); i.e., PA values broadly consistent with the
$H$-band value with a moderately flattened $\epsilon\approx 0.30$ that
drops to a much smaller value ($\epsilon\approx 0.10$) indicating
nearly round isophotes at larger radius.  (We emphasize that the
individual components of our surface brightness model -- $\beta$ model
and two gaussians -- should not be thought about as distinct
physically meaningful mass components.  We describe the physical
model(s) later; i.e., the fiducial HE model in Table~\ref{tab.fid}.)

When $\epsilon$ is allowed to vary for the $\beta$ model, we obtain
values $\epsilon\approx 0.13$ and $\rm PA\approx 135^{\circ}$, also
fully consistent with the moment analysis, indicating the PA begins to
deviate significantly from the stellar value near
$a\approx 30\arcsec$. Proper assessment of potential systematic errors
(e.g., from the treatment of embedded sources, accuracy of the
exposure map, etc.) on the values of $\epsilon$ and PA at these and
larger radii is beyond the scope of our paper and we defer such an
analysis to a future investigation. Consequently, we fixed
$\epsilon=0$ for the $\beta$ model for our present study, which we
found had negligible impact on the fit residuals within the central
$\approx 10$~kpc region which is our focus here.

\subsection{Residual Image}
\label{resid}

In Figure~\ref{fig.image.center} we show the raw image in the central
$20\arcsec\times 20\arcsec$ region and the corresponding residual
image constructed by subtracting the best-fitting two-dimensional
model from the image. As is readily apparent, the residual image is
noisy and lacks obvious bubbles or cavities or spiral features
indicative of ``sloshing.'' To guide the eye, we have over-plotted
smoothed contours (blue, linearly spaced) to trace subtle peaks and
valleys. These regions are located within a radius of $\sim 5\arcsec$
from the galaxy center without any obvious pattern in their locations.

To study further the properties of these regions, we approximated the
contour regions with simple boxes or circles, in some cases enlarging
the regions to obtain more counts. We also added a few regions
adjacent to the contours for comparison. Our region definitions
are meant simply to provide a reasonable sampling of the contour
regions and their surroundings and necessarily do not consider the
location(s) of any extended radio jet emission, for which there is
presently no evidence. Without having the radio jet emission as a
guide, the statistical significance we quote below for the regions are
over-estimates since we do not account for the ``look elsewhere
effect.'' Therefore, while the absolute values of the quoted
significances should be treated with caution, the {\it relative}
significance values of the regions should still be useful for guiding
future studies of the central image structure.

In all, we constructed 10 regions within a radius $\sim 8\arcsec$ from
the center. When compared to the model, 7/10 regions possess counts
within $2\sigma$ of the model. We denote the 3 most significant
deviations from the model by the dashed black regions in
Figure~\ref{fig.image.center} and list some of their properties in
Table~\ref{tab.resid}. Region~1, indicated by the rectangular region
to the SW of the center, possesses by far the most significant
$(4.1\sigma)$ difference from the model, and its effect is even
readily apparent in the raw image as a distortion in the third contour
from the center. Regions~2 and 3 have significances just below
$3\sigma$ and their manifestations are not obvious in the raw image.
Region~2 located to the SE is an excess of similar size ($\sim 30\%$
surface brightness deviation from the model) to Region~1 but less
significant.

Region~3 has a deficit of $\approx 37\%$ and a size well consistent with
those seen in well-studied cavity systems in
clusters~\citep[e.g.,][]{mcna07a}.  The fact that Region~1, an excess,
is adjacent to Region~3 is intriguing. The configuration might be a
rim bordering a cavity, though the relative placement
(cavity at larger radius than the rim) would not obviously favor this
interpretation.  Below in \S\ref{resid} we examine the spectra of these regions and
find gas parameters consistent with annular averages within the
sizable error bars due to the relatively few counts in these
regions. The most significant spectral difference we found is located
in the region denoted by the
solid black circle in Figure~\ref{fig.image.center} and
Region~4 in Table~\ref{tab.resid}. This region, however, corresponds
only to a $\approx 13\%$ deficit $(1.8\sigma)$, and we discuss it
further in \S\ref{resid}. 

We conclude that presently the \chandra\ X-ray image of \src\ does not
reveal obvious features of AGN feedback in the form of bubbles,
cavities, weak shocks or other irregularities in the central surface
brightness. Nevertheless, in this section we have identified regions
of the most prominent surface brightness deviations from a smooth
two-dimensional model as leading candidates for such feedback
signatures to be studied with future high sensitivity X-ray and radio
observations.

\section{Spectral Analysis}
\label{spec}

We used the \xspec\ v12.10.0e~\citep{xspec} software to fit the plasma
and background emission models to the \chandra\ spectra. The models
were fitted with a frequentist approach minimizing the
C-statistic~\citep{cstat} since it is largely unbiased compared to
$\chi^2$~\citep[e.g.][]{hump09a}. We also rebinned each spectrum so
that each PHA bin contained a minimum of 10 counts for each annulus
(\S\ref{proj.specresults} and \S\ref{deproj.specresults}), and 4
counts for each quadrant (\S\ref{quad}) and the residual region (\S
\ref{resid}). Although such rebinning is not required when using the
C-statistic, we find doing so typically reduces the time to achieve
the best fit.

Below in \S\ref{specmod} we summarize the model components and fitting
procedure we employ here and refer the reader to \S 3 of
\citet[hereafter B17]{buot17a} for a more detailed
description. Finally, we modified all the emission models (unless
otherwise stated) by foreground Galactic absorption with the {\sc
  phabs} model using the photoelectric absorption cross sections of
\citet{phabs} and a hydrogen column density,
$N_{\rm H} = 4.0\times 10^{20}$~cm$^{-2}$~\citep{kalb05a}.

\subsection{Spectral Models}
\label{specmod}

We model the interstellar plasma (``hot gas'') using the {\sc vapec}
optically thin coronal plasma model with version 3.0.9 of the atomic
database {\sc AtomDB}\footnote{http://www.atomdb.org} and the solar
abundance standard of~\citet{aspl}. In our implementation of the {\sc
  vapec} model, for elements heavier than He (which is kept fixed at
solar abundance) we fit the ratios of the metal abundances with
respect to iron; e.g., for Si we fit $Z_{\rm Si}/Z_{\rm Fe}$ rather
than $Z_{\rm Si}$ itself. The free parameters we considered for the
hot gas component in each spectrum are $\ktemp$, $Z_{\rm Fe}$,
$Z_{\rm Mg}/Z_{\rm Fe}$, $Z_{\rm Si}/Z_{\rm Fe}$, and the
normalization. All other elements heavier than He are fixed in their
solar ratios with iron. We do not fit the other elements individually
since they are too blended with iron (e.g., Ne), affected by
background (e.g., S), or simply too poorly constrained.

For several reasons, throughout most of this paper we fit the plasma
models directly to the observed spectra without performing any
onion-peeling--type deprojection.  While spectral deprojection to some
extent can help to mitigate possible biases associated with fitting a
single-temperature model to a multitemperature spectrum, this
advantage is outweighed by some key disadvantages.  Deprojection
generally, and onion-peeling in particular, amplifies noise especially
in the outer regions where the background dominates. Standard
deprojection procedures also do not easily self-consistently account
for the gas emission outside the bounding annulus, which can be a
sizable source of systematic
error~\citep[e.g.,][]{nuls95a,mcla99a,buot00c}. They also typically
assume the gas properties are constant within what are often wide
spherical shells (especially for systems like \src) introducing
additional systematic error for the radially varying gas properties.
(The assumption of constant properties per circular annulus on the sky
also applies to our default approach, but the errors associated with
this assumption do not propagate between annuli in the same way as the
spectral deprojection in which the model spectrum in any given shell
depends on all of those exterior to it.) Consequently, we relegate
deprojection analysis using the {\sc projct} mixing model in \xspec\
to a systematic error check (\S\ref{deproj.specresults}).

Although the emission from unresolved LMXBs and other stellar sources
is a small fraction of the X-ray emission of \src, we still included a
7.3~keV thermal bremsstrahlung
component~\citep[e.g.,][]{mats97a,irwi03a} to account for this
emission. We restricted the normalization of this component to lie
within a factor of 2 of the $L_x - L_K$ scaling relation for discrete
sources of \citet{hump08b} using the $K$-band luminosity
$(L_K=1.7\times 10^{11}\, L_{\odot})$ from the Two Micron All-Sky
Survey (2MASS) as listed in the Extended Source
Catalog~\citep{jarr00a}. Using the global $L_x$ from the scaling
relation, we assigned the expected range of $L_x$ for each annulus
according to the fraction of the total 2MASS $K$-band emission falling
into that annulus. Hence, for each annulus the flux of unresolved
discrete sources (with range restricted as noted) is a free
parameter. 

As described in \S 3.1.2 of B17 we model the Cosmic X-ray Background
(CXB) emission with multiple thermal plasma components for the
``soft'' CXB and a single power law for the ``hard'' CXB. By default
we fixed the soft CXB normalizations to those obtained from fitting
{\sl ROSAT} data using the HEASARC X-ray Background
Tool\footnote{https://heasarc.gsfc.nasa.gov/cgi-bin/Tools/xraybg/xraybg.pl}.
We examine the systematic error associated with this choice by
allowing the normalizations of the soft CXB components to vary within
a factor of 2 of the {\sl ROSAT} values (\S\ref{sys}). Hence, in our
default model the normalization of the power-law of the hard CXB
contribution for all annuli is the only free parameter of the CXB.

For the particle background we adopted a multicomponent model
consisting of a power-law with two break radii
along with three gaussians. Unlike the other models described above, we
do not fold the particle background model through the ARF; i.e., it is
``un-vignetted.''  However, since the particle backgrounds of the BI and FI
chips are not identical, we fitted separate versions of the model to
the data on the BI (S3) and FI chips.

The Cycle 16 and Cycle 19 data are fitted separately. For each data
set we fitted all annuli simultaneously, including large apertures at
the largest radii (not listed in Table~\ref{tab.gas}) dominated by
background.

\subsection{Results}
\label{specresults}

\begin{figure*}
\parbox{0.49\textwidth}{
\centerline{\includegraphics[scale=0.35,angle=-90]{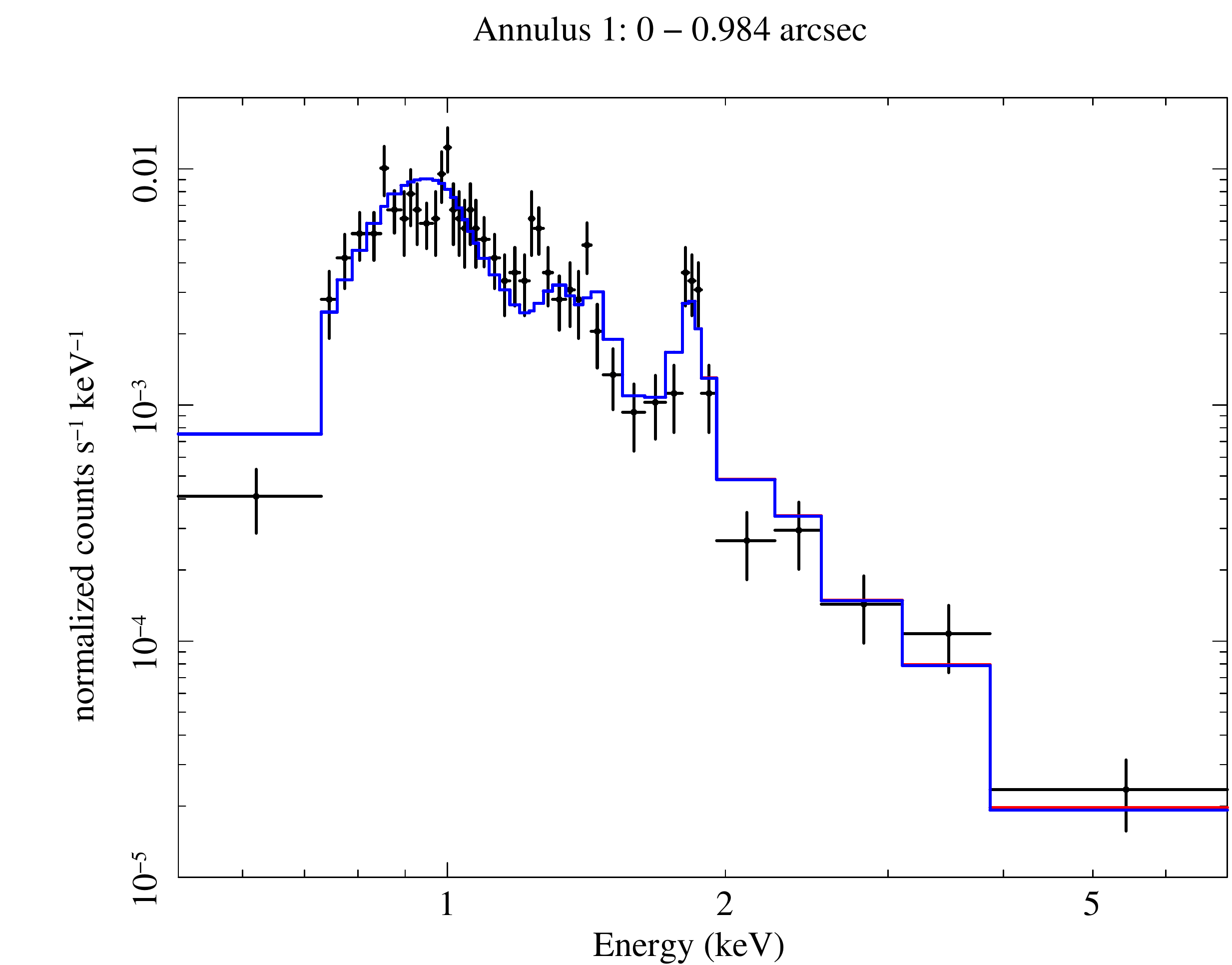}}}
\parbox{0.49\textwidth}{
\centerline{\includegraphics[scale=0.35,angle=-90]{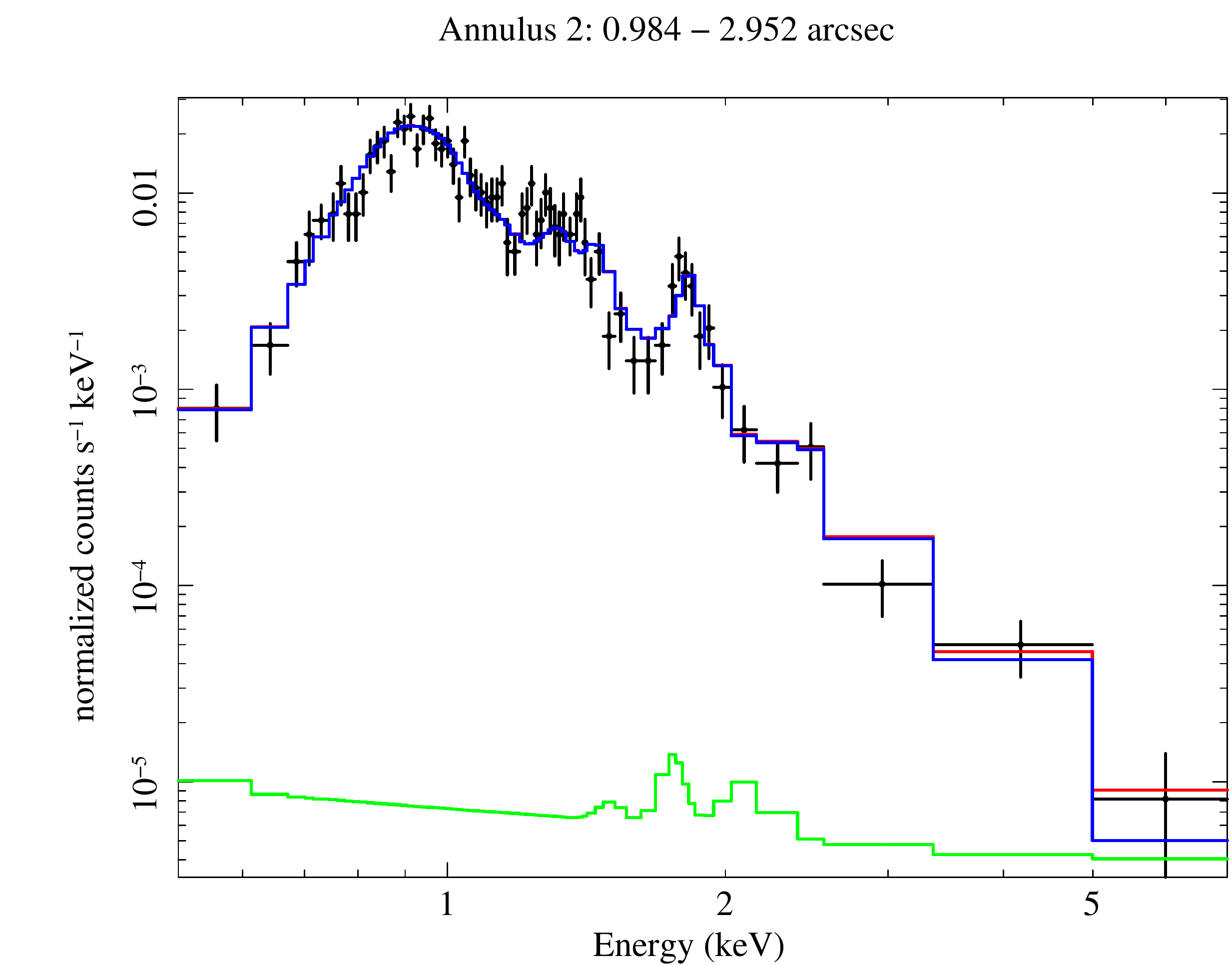}}}

\vskip 0.2cm

\parbox{0.49\textwidth}{
\centerline{\includegraphics[scale=0.35,angle=-90]{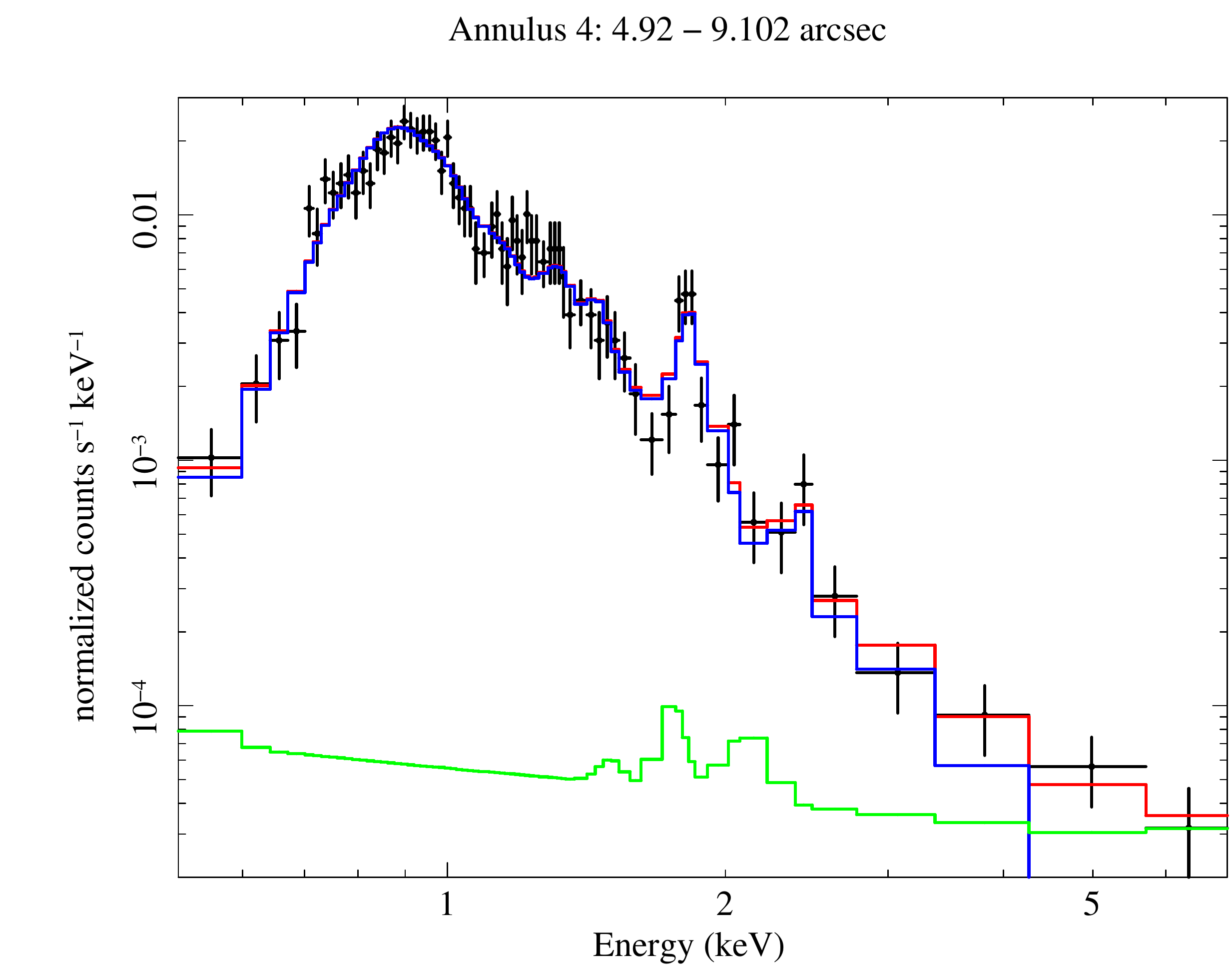}}}
\parbox{0.49\textwidth}{
\centerline{\includegraphics[scale=0.35,angle=-90]{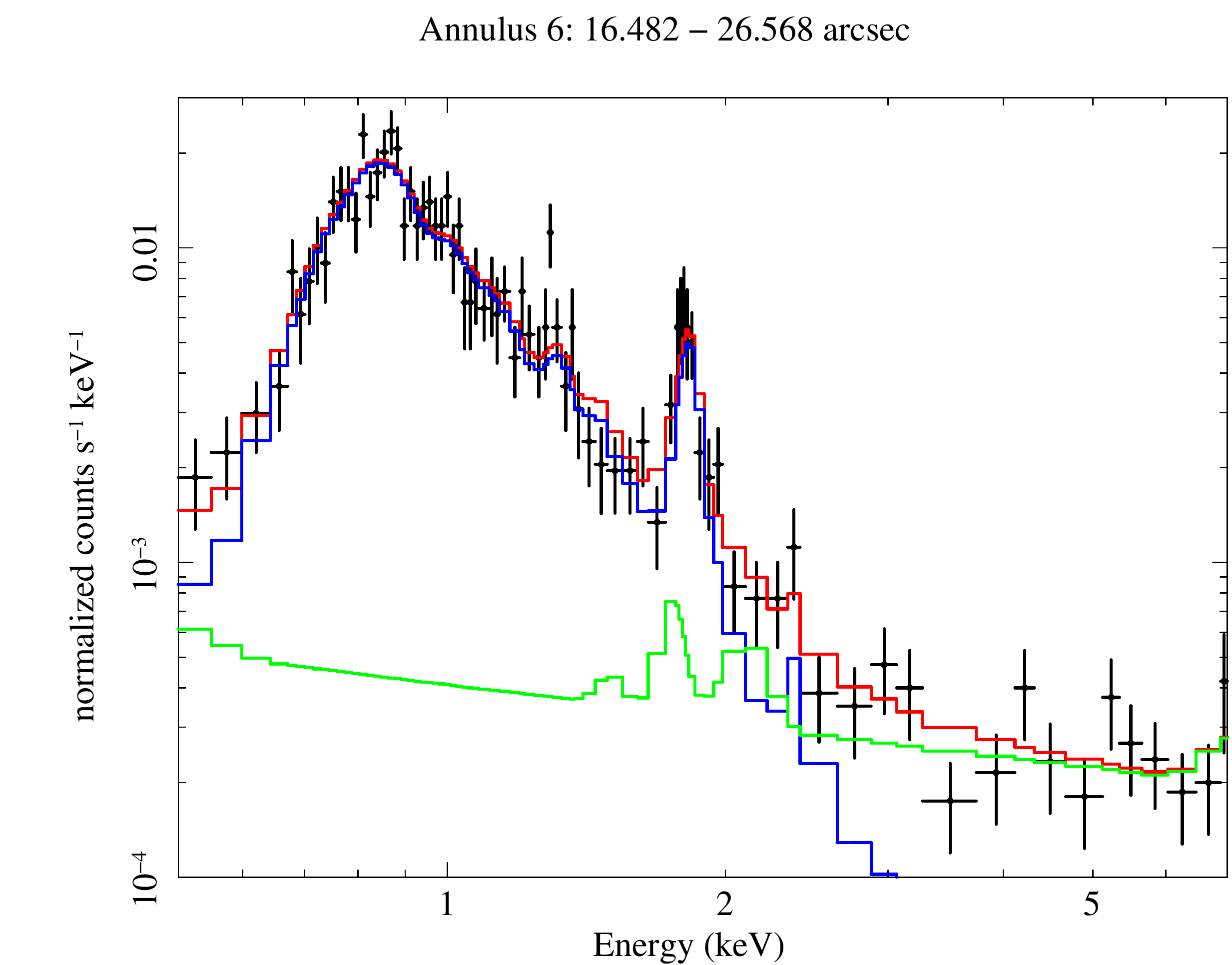}}}

\vskip 0.2cm

\parbox{0.49\textwidth}{
\centerline{\includegraphics[scale=0.35,angle=-90]{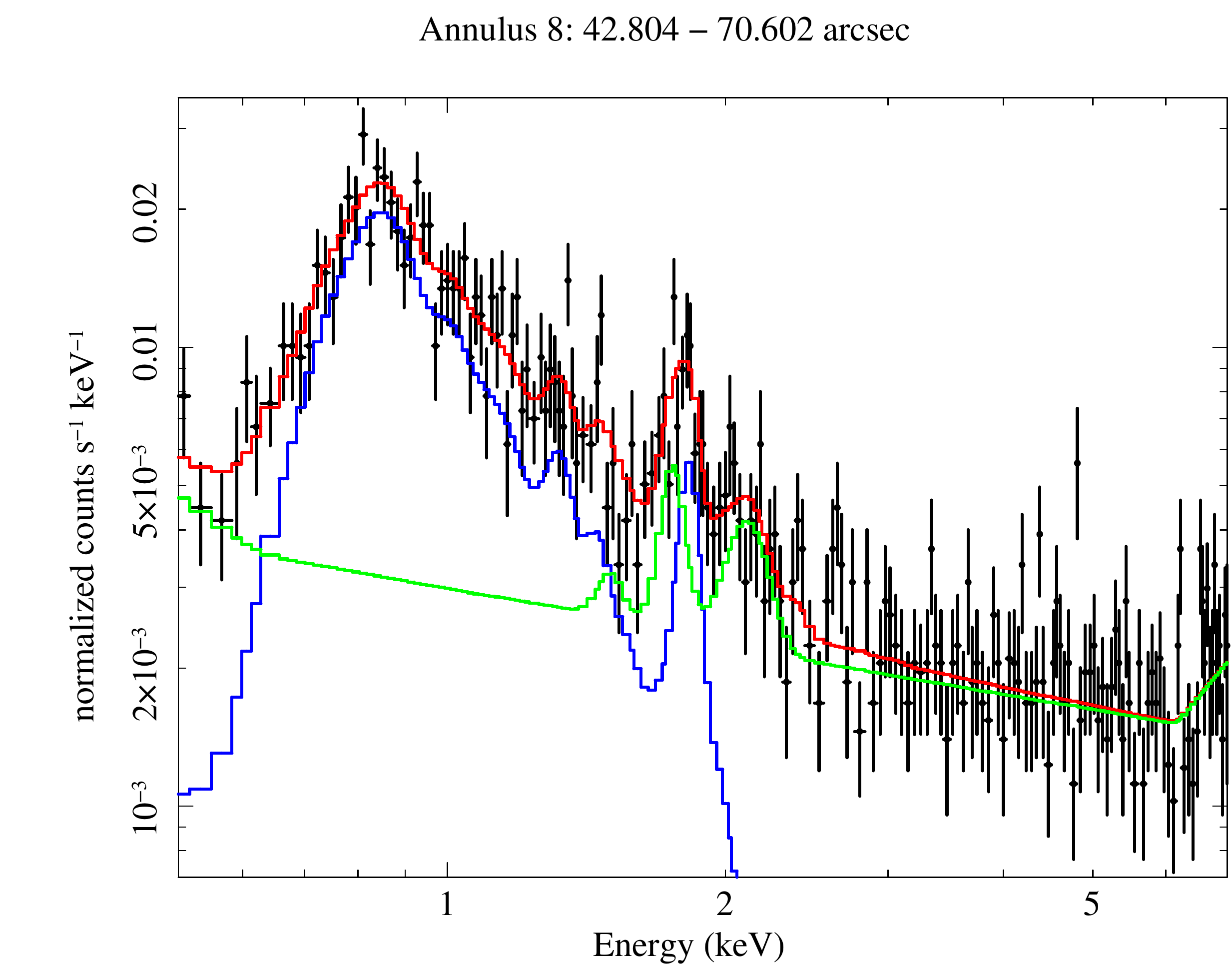}}}
\parbox{0.49\textwidth}{
\centerline{\includegraphics[scale=0.35,angle=-90]{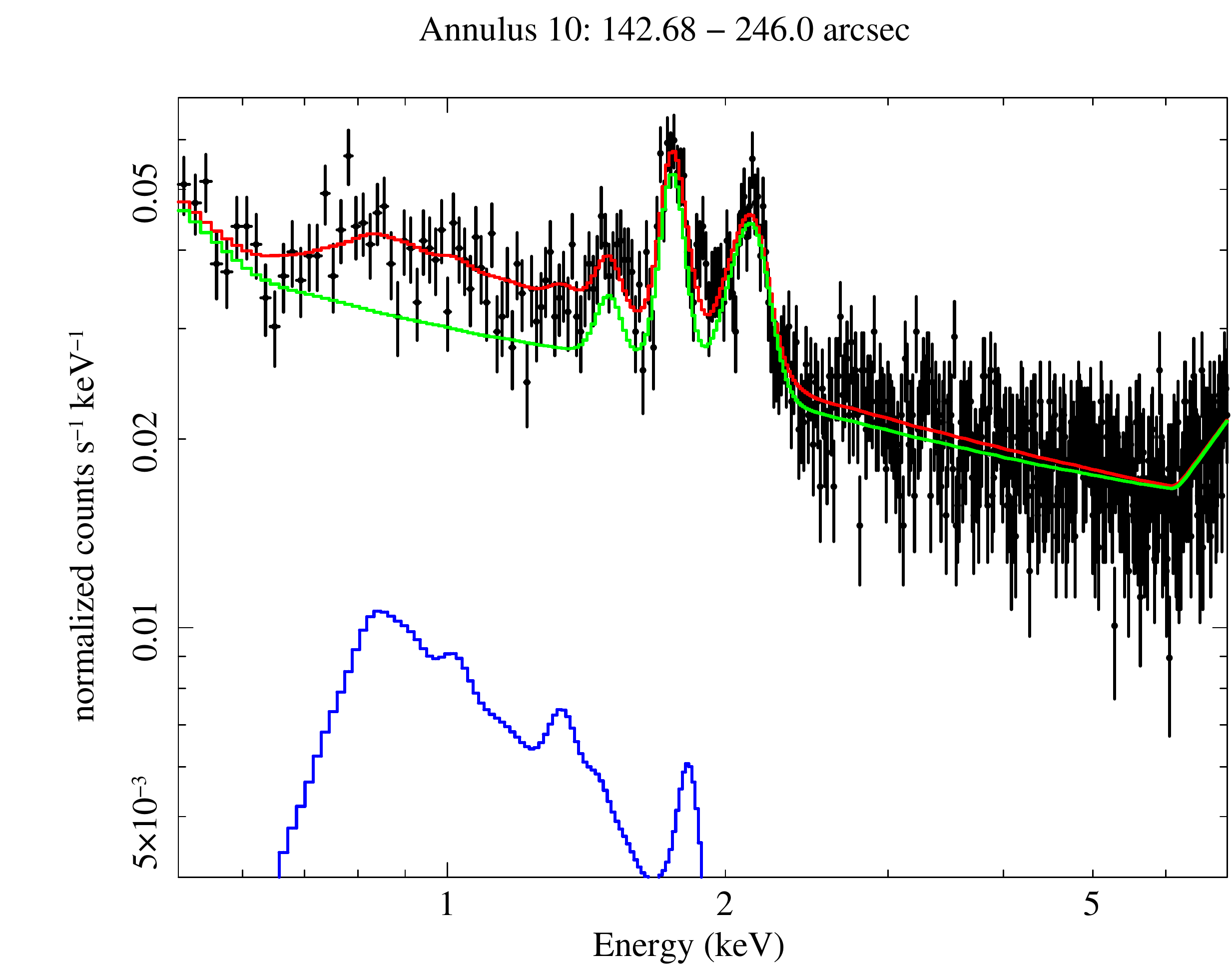}}}

\caption{\label{fig.spec} Representative combined \chandra\ Cycle~19 spectra in the
  0.5-7.0~keV band without any background subtraction. Also plotted
  are the best-fitting models (red) broken down into the separate contributions
  from the following:  (1) hot gas and unresolved LMXBs from \src\
  along with the CXB (blue), and (2) particle background
  (green). For the inner annuli, the broad peak near 1~keV is dominated by a great number of unresolved
  Fe~L~shell emission lines. The prominent feature near 1.8~keV is
  dominated by emission from He-like Si~K$\alpha$. The modest bump
  near 1.4~keV is mostly H-like Mg~K$\alpha$ with some contribution
  from other lines, most notably Fe~L, blended in. At large radii,
  background lines of Si and S near 2~keV become increasingly
  apparent. Note all of the displayed spectra except that of
  Annulus~10 derive entirely from the ACIS-S3 (BI) CCD, while the majority of
  the emission in Annulus~10 derives from several FI CCDs (see \S\ref{obs}).}
\end{figure*}

\subsubsection{Analysis of Projected Spectra}
\label{proj.specresults}

\begin{figure*}
\parbox{0.49\textwidth}{
\centerline{\includegraphics[scale=0.43,angle=0]{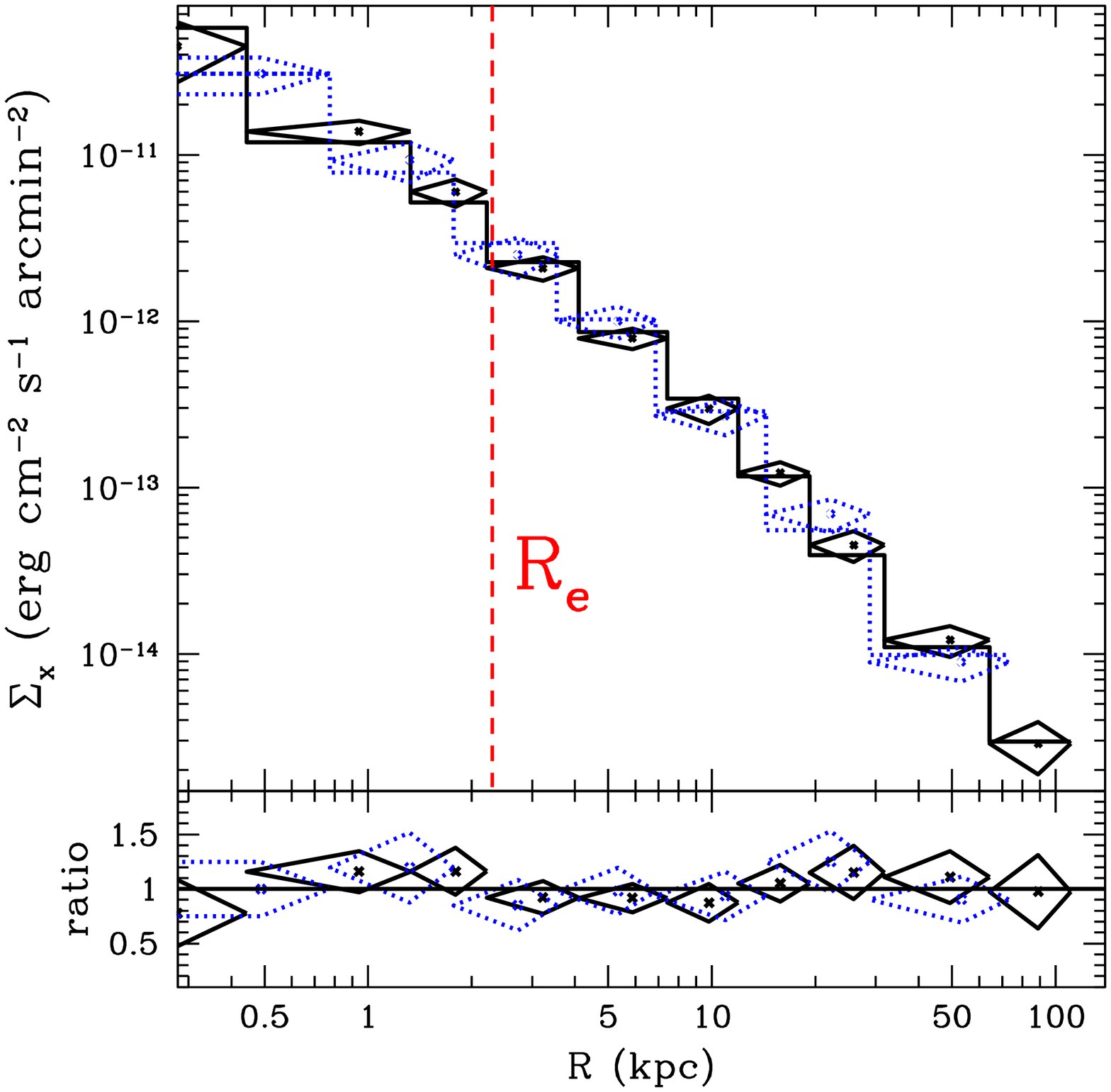}}}
\parbox{0.49\textwidth}{
\centerline{\includegraphics[scale=0.43,angle=0]{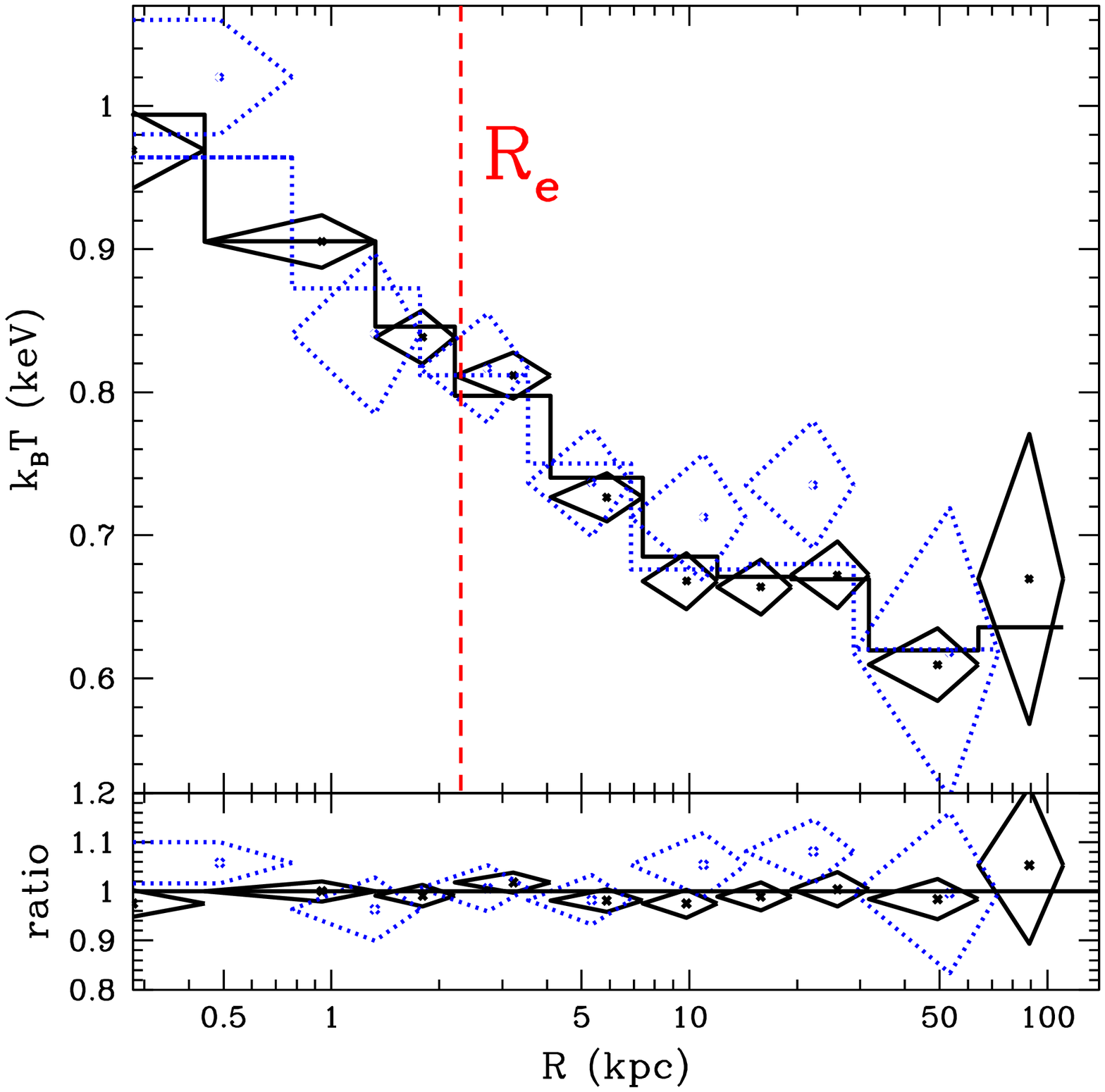}}}
\caption{\label{fig.best} The Cycle 19 \chandra\ data (solid black
  circles), $1\sigma$ errors (solid diamonds), and the best-fitting
  fiducial hydrostatic model (solid binned line) in each circular
  annulus on the sky for \src. The corresponding quantities for the
  Cycle 16 data are plotted in dotted blue. ({\sl Left Panel}) Surface
  brightness (0.5-7.0~keV). See the notes to Table~\ref{tab.gas}
  regarding the error bars on $\xsurf$.  ({\sl Right Panel}) Projected
  emission-weighted temperature ($k_BT$). Also shown is the location
  of the stellar half-light radius ($R_e$). The bottom panels plot the
  data/model ratios. Note the displayed best-fitting model corresponds
  to the ``Max Like'' parameters (see \S\ref{Bayes}) and is also
  indistinguishable from the best frequentist fit.}
\end{figure*}

\begin{table*}[t] \footnotesize
\begin{center}
\caption{Hot Gas Properties}
\label{tab.gas}
\begin{tabular}{lcccccccccc}   \hline\hline\\[-7pt]
& & $R_{\rm in}$ & $R_{\rm out}$ & $\Sigma_{\rm x}$ (0.5-7.0~keV) & $k_BT$ & $Z_{\rm Fe}$  & $Z_{\rm Mg}/Z_{\rm Fe}$  & $Z_{\rm Si}/Z_{\rm Fe}$ \\
Observation & Annulus & (kpc) & (kpc) & (ergs cm$^2$ s$^{-1}$ arcmin$^{-2}$) & (keV) & (solar) & (solar) & (solar)\\
\hline \\[-7pt]
\\ \chandra\ Cycle~19\\
&   1 & 0.00 & 0.44 & $\rm   4.50e-11 \pm   1.76e-11$ & $  0.969 \pm   0.027$ & $   1.04 \pm    0.14$ & $   1.66 \pm    0.43$ & $   1.64 \pm    0.32$\\
&   2 & 0.44 & 1.33 & $\rm   1.38e-11 \pm   2.24e-12$ & $  0.905 \pm   0.018$ & $   1.00 \pm    0.11$ & $   1.21 \pm    0.16$ & $   0.79 \pm    0.08$\\
&   3 & 1.33 & 2.21 & $\rm   5.99e-12 \pm   1.11e-12$ & $  0.838 \pm   0.019$ & $   1.00 \pm    0.14$ &  tied &  tied\\
&   4 & 2.21 & 4.10 & $\rm   2.08e-12 \pm   3.36e-13$ & $  0.812 \pm   0.016$ & $   0.91 \pm    0.08$ & $   0.81 \pm    0.11$ &  tied\\
&   5 & 4.10 & 7.42 & $\rm   7.90e-13 \pm   1.12e-13$ & $  0.727 \pm   0.017$ & $   0.74 \pm    0.07$ &  tied &  tied\\
&   6 & 7.42 & 11.96 & $\rm   2.98e-13 \pm   5.85e-14$ & $  0.668 \pm   0.020$ & $   0.75 \pm    0.12$ & $   0.51 \pm    0.10$ & $   1.30 \pm    0.13$\\
&   7 & 11.96 & 19.26 & $\rm   1.23e-13 \pm   1.97e-14$ & $  0.664 \pm   0.019$ & $   0.68 \pm    0.08$ &  tied &  tied\\
&   8 & 19.26 & 31.77 & $\rm   4.51e-14 \pm   9.55e-15$ & $  0.672 \pm   0.023$ & $   0.71 \pm    0.10$ & $   0.70 \pm    0.14$ &  tied\\
&   9 & 31.77 & 64.20 & $\rm   1.21e-14 \pm   2.58e-15$ & $  0.610 \pm   0.025$ &  tied &  tied &  tied\\
&  10 & 64.20 & 110.70 & $\rm   2.89e-15 \pm   9.96e-16$ & $  0.670 \pm   0.101$ & $0.36$ &  tied &  tied\\
\\ \chandra\ Cycle~16\\
&   1 & 0.00 & 0.78 & $\rm   3.07e-11 \pm   7.62e-12$ & $  1.020 \pm   0.040$ & $   0.98 \pm    0.16$ & $   0.48 \pm    0.17$ & $   0.63 \pm    0.20$\\
&   2 & 0.78 & 1.77 & $\rm   9.33e-12 \pm   2.51e-12$ & $  0.841 \pm   0.056$ &  tied &  tied &  tied\\
&   3 & 1.77 & 3.54 & $\rm   2.50e-12 \pm   6.80e-13$ & $  0.817 \pm   0.038$ &  tied &  tied &  tied\\
&   4 & 3.54 & 6.86 & $\rm   1.00e-12 \pm   2.19e-13$ & $  0.737 \pm   0.038$ & $   0.78 \pm    0.12$ &  tied &  tied\\
&   5 & 6.86 & 14.39 & $\rm   2.69e-13 \pm   6.39e-14$ & $  0.713 \pm   0.044$ &  tied &  tied &  tied\\
&   6 & 14.39 & 28.78 & $\rm   6.93e-14 \pm   1.57e-14$ & $  0.735 \pm   0.044$ & $   0.63 \pm    0.12$ &  tied &  tied\\
&   7 & 28.78 & 73.06 & $\rm   8.90e-15 \pm   2.08e-15$ & $  0.618 \pm   0.101$ &  tied &  tied &  tied\\
\hline \\
\end{tabular}
\tablecomments{1~kpc = 2.22\arcsec. The listed values of $\xsurf$
  correspond to the entire sky area of each annulus; i.e., for those
  annuli where portions of their sky area were masked out due to point
  sources and chip features (gaps, edges) we have rescaled the
  measured $\xsurf$ to account for the lost area. (Almost all annuli
  exclude no or $<1\%$ area. The largest excluded area for the
  Cycle~19 observation annuli listed is $\approx 11\%$ for Annulus~9
  due to the gap between the ACIS-S2 and ACIS-S3.) Annuli where an
  abundance is linked to the value in the previous annulus are
  indicated as ``tied.''  See \S\ref{proj.specresults} and
  \S\ref{abun} regarding the fiducial value $\zfe=0.36\,\zsun$ used
  for Annulus~10 of the Cycle~19 observation and the range of $\zfe$
  values explored as a systematic error. Note that the definition of
  $\xsurf$ is essentially the emission measure (i.e., \xspec\ {\sc
    norm} parameter in equation~\ref{eqn.norm}), which is the
  parameter actually fitted to the spectral data) multiplied by the
  plasma emissivity divided by $\pi\theta^2$, where $\theta$ is the
  aperture radius in arcminutes. Rather than quote the results for
  {\sc norm} itself, we have used the best-fitting plasma emissivity
  for each annulus (i.e., the plasma emissivity evaluated using the
  best-fitting $\ktemp$ and element abundances) to convert {\sc norm}
  into a surface brightness unit. Consequently, the error bars quoted
  for $\xsurf$ are directly proportional to the error bars for {\sc
    norm}.}
\end{center}
\end{table*}

The spectral model described above describes well the Cycle 19 data
with  a minimum C-statistic value of 2023.5 (in 2002 pha bins) with
1932 degrees of freedom (dof).  The success of the model is on display
in Figure~\ref{fig.spec} where we plot the spectra and best-fitting
models for 5 representative annuli. The most noticeable spectral features
are the broad bump near 1~keV dominated by a forest of emission lines from the Fe
L shell and the strong Si K$\alpha$ line near 1.85~keV. Less
noticeable, though still prominent in the inner annuli, is the Mg
K$\alpha$ line complex near 1.4~keV. (Note for the Cycle 16 data we
achieve a fit consistent with that obtained in Paper~1.)

The innermost and outermost annuli deserve special mention. Annulus~1
displays the most significant residuals from the best-fitting model
resulting in a fit there that is of formally marginal quality. In
Appendix~\ref{specresults.center} we study the central spectrum in
detail and examine several possibilities to improve the fit, though at
present we cannot confidently recommend a specific modification to the
fiducial model. Since the background dominates in the outermost
annulus (Annulus~10), the properties of the gas component there cannot
be robustly determined. We found it necessary in the spectral fitting
to restrict more strongly the gas parameter ranges there: i.e.,
$\ktemp=0.4-0.9$~keV. Guided by the average radial profiles of $Z_{\rm Fe}$ for groups and
clusters obtained by \citet{mern17a}, by default we fixed
$Z_{\rm Fe}=0.36$~$Z_{\odot}$ in Annulus~10. \citet{mern17a} quote a
scatter of $\sim \pm 0.09$ for $\zfe$ over the radial range
corresponding to Annulus~10, and we therefore use the range $Z_{\rm
  Fe}=0.27-0.45$~$Z_{\odot}$ as a systematic error in our HE models (\S\ref{abun}).

We list the gas parameters measured for each annulus in
Table~\ref{tab.gas} for the data sets. (We
express the emission measure of the gas as a surface brightness unit
-- see the notes to the table.) In Figure~\ref{fig.best} we plot the
radial profiles of the surface brightness ($\xsurf$) and temperature
($\ktemp$). As expected, $\xsurf$ and $\ktemp$ are consistent in their
overlap region (as is $\zfe$). The lack of a big temperature jump in Annulus~1 of the
Cycle 19 observation has implications for the mass of the SMBH
(\S\ref{smbh}). The Cycle 19 data confirm and strengthen the 
similarity of the temperature profile of \src\ to that of the fossil group
NGC~6482 (B17).

The profile of $\zfe$ decreases with radius but is nearly constant
over large stretches; i.e., $\zfe\approx 1\, \zsun$ for $R\la 2$~kpc
and $\zfe\approx 0.7\, \zsun$ for $R\approx 4-60$~kpc. This negative
gradient in $\zfe$ is very similar to those seen in several X-ray
bright, massive elliptical galaxies and small groups like NGC~6482
(B17), NGC~5044~\citep{buot03b} and
others~\citep[e.g.,][]{buot00c,hump06a,rasm09b,mern17a}.

Although the Mg and Si abundances are not as well constrained as Fe, we
find the Cycle 19 observation does place interesting constraints on
the radial variation of both $\zmgfe$ and $\zsife$. The $\zmgfe$ ratio
appears to peak in the central $R\approx 2$~kpc with a value at least
solar and is consistent with a constant ratio of $\approx 0.7$ solar
at larger radius. The $\zsife$ profile is broadly similar to $\zmgfe$ out to
$R\approx 10$~kpc after which it increases significantly to
$\zsife\approx 1.3$ solar. Since at large radius the Si abundance
measurement becomes especially more challenging due to the increasing
background level (both the continuum and the presence of an
instrumental line), we regard our measurement there as
provisional. Nevertheless, the $\zmgfe$ and $\zsife$ profiles are
broadly similar to mean profiles obtained from \xmm\ for groups by
\citet{mern17a}. 

For comparison, if we do not allow for a radial variation in either Mg
or Si we obtain $\zmgfe = 0.83 \pm 0.06$ solar and
$\zsife = 0.97 \pm 0.07$ solar for the Cycle 19 data (which also gives
$\zfe\approx 0.80\,\zsun$ in Annulus~1). We use these results to
perform a systematic error check on our fiducial models in \S
\ref{sys}. Notice also that these constant values of $\zmgfe$ and
$\zsife$ for the Cycle 19 data are consistent within the
$\approx 1.5\sigma$ errors with the values obtained for the Cycle 16
data (Table~\ref{tab.gas}), for which interesting constraints on the
radial variation were not obtained.

Finally, the Cycle 19 results we have described in this section for
the combined data are fully consistent with those obtained when
performing a joint fit of the individual exposures (see
Appendix~\ref{joint} and Table~\ref{tab.gas.joint}).

\subsubsection{Spectral Deprojection in Spherical Shells}
\label{deproj.specresults}

If instead we perform spectral deprojection of the hot plasma in
spherical shells with the {\sc projct} model in {\sc xspec} we find
the C-statistic is reduced with respect to the fiducial case just
described by only 0.6 with no obvious effect on the fractional
residuals; e.g., the fit residuals for Annulus~1 look the same as
obtained without deprojection (Figure~\ref{fig.spec} and
\ref{fig.center}).

We list the gas parameters for the deprojected case in
Table~\ref{tab.gas.deproj} in Appendix~\ref{deproj.appendix}. The
results for $\ktemp$ and the abundances are typically consistent
within $\sim 1\sigma$ with
those obtained without deprojection. As expected, the sizes of the
error bars on all the parameters are larger, often twice as large,
compared to those obtained with the fiducial projected model. We note
the relatively large best-fitting value of $\ktemp=1.23$~keV obtained
for Annulus~1 for the Cycle 16 data that is $\approx 2\sigma$ larger than the
projected case, but also very consistent with the result obtained with
{\sc projct} by \citet{wern18a}.

The deprojected results just described made no account of any
gas emission expected to exist exterior to Annulus~10 in the Cycle 19
data (or Annulus~7 of the Cycle 16 data). For comparison, we also
examined adding a fixed gas contribution to the background annulus (\S
\ref{obs}) using the gas emission predicted by our best-fitting fiducial hydrostatic
equilibrium model (Table~\ref{tab.fid}). In this case the fit quality
is unchanged, and the main differences are somewhat lower $\ktemp$ and
density values in Annulus~10.

Since these deprojected models do not improve the fit, lead to larger
parameter errors, and do not self-consistently address the emission
expected outside the bounding annuli, throughout the paper we focus on
the results obtained from the {\it projected} spectra.

\subsubsection{Central Region: Quadrants}
\label{quad}

\begin{table*}[t] \footnotesize
\begin{center}
\caption{Hot Gas Properties in Quadrants}
\label{tab.quad}
\begin{tabular}{cc|ccccc}   \hline\hline\\[-7pt]
  & & $k_{\rm B}T$ & {\sc norm} & $\tilde{n}_e$ & $\tilde{S}$ & $\tilde{P}$ \\
 Annulus & Quadrant  & (keV) & ($10^{-5}$~cm$^{-5}$) & (cm$^{-3}$) & (keV cm$^2$) & ($10^{-10}$~erg~cm$^{-3}$) \\
\hline \\[-7pt]
2 & 1 & $0.951 \pm 0.038$ & $1.07 \pm 0.07$ & $0.156 \pm 0.005$ & $3.29 \pm 0.15$ & $2.375 \pm 0.120$\\
2 & 2 & $0.842 \pm 0.040$ & $1.26 \pm 0.07$ & $0.169 \pm 0.005$ & $2.75 \pm 0.14$ & $2.279 \pm 0.127$\\
2 & 3 & $0.898 \pm 0.045$ & $1.08 \pm 0.07$ & $0.156 \pm 0.005$ & $3.10 \pm 0.17$ & $2.250 \pm 0.132$\\
2 & 4 & $0.892 \pm 0.037$ & $1.28 \pm 0.07$ & $0.171 \pm 0.005$ & $2.90 \pm 0.13$ & $2.440 \pm 0.122$\\
2 & Scatter & $0.043 \pm 0.022$ & $0.085 \pm 0.030$ & $0.043 \pm 0.015$ & $0.067 \pm 0.024$ & $0.032 \pm 0.027$\\
\hline \\
3 & 1 & $0.839 \pm 0.033$ & $1.04 \pm 0.07$ & $0.091 \pm 0.003$ & $4.13 \pm 0.19$ & $1.230 \pm 0.063$\\
3 & 2 & $0.828 \pm 0.038$ & $1.10 \pm 0.07$ & $0.094 \pm 0.003$ & $4.00 \pm 0.20$ & $1.249 \pm 0.069$\\
3 & 3 & $0.812 \pm 0.039$ & $0.97 \pm 0.06$ & $0.088 \pm 0.003$ & $4.10 \pm 0.22$ & $1.147 \pm 0.067$\\
3 & 4 & $0.913 \pm 0.048$ & $0.83 \pm 0.06$ & $0.081 \pm 0.003$ & $4.86 \pm 0.28$ & $1.191 \pm 0.076$\\
3 & Scatter & $0.045 \pm 0.026$ & $0.105 \pm 0.032$ & $0.054 \pm 0.017$ & $0.080 \pm 0.030$ & $0.033 \pm 0.028$\\
\hline \\
4 & 1 & $0.782 \pm 0.032$ & $1.58 \pm 0.09$ & $0.044 \pm 0.001$ & $6.27 \pm 0.28$ & $0.551 \pm 0.027$\\
4 & 2 & $0.827 \pm 0.035$ & $1.37 \pm 0.08$ & $0.041 \pm 0.001$ & $6.96 \pm 0.32$ & $0.543 \pm 0.028$\\
4 & 3 & $0.807 \pm 0.036$ & $1.19 \pm 0.08$ & $0.038 \pm 0.001$ & $7.10 \pm 0.35$ & $0.494 \pm 0.027$\\
4 & 4 & $0.808 \pm 0.036$ & $1.24 \pm 0.08$ & $0.039 \pm 0.001$ & $7.01 \pm 0.34$ & $0.506 \pm 0.028$\\
4 & Scatter & $0.020 \pm 0.020$ & $0.110 \pm 0.032$ & $0.054 \pm 0.015$ & $0.048 \pm 0.022$ & $0.046 \pm 0.026$\\
\hline \\
\end{tabular}
\tablecomments{See \S\ref{quad} for definitions of the \xspec\
  {\sc norm} parameter, pseudo electron number density $\tilde{n}_e$,
  pseudo entropy $\tilde{S}=\ktemp \tilde{n}_e^{-2/3}$, and pseudo
  pressure $\tilde{P}=\tilde{n}_e\ktemp$. The
  annuli numbers correspond to those in Table~\ref{tab.gas}. The
  quadrants are defined with respect to the $H$-band stellar position
  angle, $\theta_{\star} = 70.15\degr$ (N-E) from Y17: $(\theta_{\star}, \theta_{\star} + 90\degr)$ for
  quadrant~1, $(\theta_{\star}+90\degr, \theta_{\star} + 180\degr)$ for
  quadrant~2, $(\theta_{\star}+180\degr, \theta_{\star} + 270\degr)$ for
  quadrant~3, $(\theta_{\star}+270\degr, \theta_{\star} + 360\degr)$ for
  quadrant~4. The ``Scatter'' is the mean fractional scatter of the
  quadrants for a particular annulus defined by equation~\ref{eqn.scatter}.}
\end{center}
\end{table*}

Since the main objective of our paper is to infer the gravitating mass
distribution using a hydrostatic equilibrium analysis, ideally we
would want kinematic information for the hot plasma to inform and
correct our analysis, especially for the central region where AGN
feedback is expected to periodically inject energy into the hot gas. 
High spectral resolution observations of the Perseus cluster with {\sl
  Hitomi}~\citep{hit16a} and in massive elliptical galaxies / small
groups with the \xmm\ RGS~\citep[e.g.,][]{ogor17a} indicate low amounts of
turbulent pressure at the centers of these systems. 

Another manifestation of non-hydrostatic behavior is via spatial
fluctuations in the gas properties, in particular through the
azimuthal scatter of properties within subregions of circular
annuli~\citep[e.g.,][]{vazz11a}.  The high spatial resolution combined
with the moderate energy resolution of \chandra\ ACIS is well-suited
for studying such azimuthal fluctuations in the central, high S/N
regions of \src.  We were able to obtain useful constraints on the gas
properties when dividing up Annuli~2, 3, and 4 into four quadrants,
where for each annulus we fixed the metal abundances and background
levels to the best-fitting results obtained for the whole annulus in \S
\ref{proj.specresults}.

Since hydrostatic equilibrium is a balance between pressure and
gravity at any point, we use the inferred projected gas properties to
construct three-dimensional proxies for the gas density, entropy, and
pressure as follows. The normalization of the {\sc vapec} model in
{\sc xspec},
\begin{equation}
{\rm \sc norm} \equiv 10^{-14}\int n_en_{\rm H}{\rm dV} /
\left(4\pi\left[\rm {D_A}\left(1+z\right)^2\right]\right),
\label{eqn.norm}
\end{equation}
is proportional to the emission measure, $\int n_en_{\rm H}{\rm
  dV}$. We define a pseudo-electron number density ($\tilde{n}_e$) by
dividing this emission measure by
$V_{ij}^{\rm int}= (r_j^2-r_i^2)^{3/2}$, taking the square root, and
converting $n_{\rm H}$ to $n_e$. Here $r_i$ and $r_j$ are the inner
and outer radii of the annulus on the sky representing the volume of
the spherical shell intersected by the cylindrical annulus along the
line-of-sight of the same radii~\citep[e.g.,][]{kris83}. We then use
this pseudo-electron number density and the projected $\ktemp$ to
define corresponding a pseudo-entropy and pseudo-pressure.

In Table~\ref{tab.quad} we list the results for these gas properties
obtained for each quadrant for Annuli 2-4; see the caption to the
table for the definitions of the quadrants. (We obtain consistent
results for the scatter if we rotate the quadrants by 45 degrees.)  To
quantify the scatter between the quadrants of each sector, we follow
\citet{vazz11a} and define the quantity,
\begin{equation}
{\rm Scatter} \equiv \sqrt{\frac{1}{N}\sum_{i=1}^N\left(\frac{y_i - \bar{y}}{\bar{y}}\right)^2},
\label{eqn.scatter}
\end{equation}
where $y_i$ is the quantity in quadrant $i$, $\bar{y}$ is the average
value of the quantity over the whole annulus, and $N=4$. The errors
quoted for the scatter assume normal error propagation.

The largest scatter ($\approx 0.10$) occurs for {\sc norm} while the
other parameters have smaller scatters ($\la 0.05$) in most cases. The
hydrostatic equilibrium equation may be expressed in terms of the
entropy and pressure having a dependence $S^{3/5}$ and $P^{2/5}$
respectively~\citep[e.g.,][]{hump08a}. The scatter in the
pseudo-entropy and pseudo-pressure listed in  Table~\ref{tab.quad}
suggest $\la 5\%$ azimuthal fluctuations in the hydrostatic equilibrium
equation in the central region of \src. Errors of this magnitude are
less than the statistical errors on the inferred mass properties
(e.g., Table~\ref{tab.mass}).

\subsubsection{Central Region: Image Residuals}
\label{spec.resid}

The regions defined according to the residual image discussed in
\S\ref{resid} have too few counts to clearly
distinguish any differences in their spectra from the average
spectrum of the annulus (or annuli) where they are located. The most
significant differences we found occur for Region~4 defined in
Table~\ref{tab.resid}. This region has only 147 counts over
0.5-2.0~keV and straddles roughly equally the boundary between
Annuli~2 and 3.

Since the background level is low in this small region ($1.3\arcsec$
radius), we found it convenient to use the blank-sky spectrum
(\S\ref{obs}) to account for both the CXB and particle background.  If
we fix the metal abundances to the average values of Annuli~2-3, we
obtain a good fit with $\ktemp=0.79\pm 0.05$~keV, consistent the
average values within $\approx 1.5\sigma$.  The C-statistic is 24.4
for 31 pha bins and 28~dof.

If $\zfe$ is allowed to vary, the C-statistic falls by $\approx 5$ and
$\zfe$ fits to a much lower value, $\zfe=0.60\pm 0.15\,\zsun$, a little less than
$3\sigma$ below the average value of $\approx 1\,\zsun$. We strongly
suspect that this reflects the Fe Bias~\citep[e.g.,][]{buot00a}, since we obtain nearly the
same reduction in the C-statistic if instead we add a second
temperature component with the same abundances all fixed at 1
solar. The resulting 2T fit gives best-fitting $\ktemp$ values of
0.6~keV and 1.1~keV for the temperature components with approximately
the same emission measure for each. However, if we allow $\zfe$ to
vary in the 2T model, it still fits to $\approx 0.60\,\zsun$. These
solutions provide potentially interesting evidence for inhomogeneous
gas cooling and metal enrichment, though we believe the fact this
region is a deficit does not favor a cooling scenario. Higher quality
data are needed to clarify the gas properties and their implications.

\section{Hydrostatic Equilibrium Models}
\label{he}

\begin{table*}[t] \footnotesize
\begin{center}
\caption{Fiducial Bayesian Hydrostatic Equilibrium Model}
\label{tab.fid}
\begin{tabular}{ccc|cr|c|ccc}   \hline\hline\\[-7pt]
& & & \multicolumn{2}{c}{Prior}\\
Component & Model & Parameter & Type & Range & Units & Best Fit & Max Like & Std. Dev.\\
\hline \\[-7pt]
Boundary Condition & $P_{\rm gas}(r=10\, \rm kpc)$ & $P_{\rm ref}$ & Flat Log & $10^{-12}-10^{-9}$ & $10^{-11}$~erg~cm$^{-3}$               & 1.67 & 1.69 & 0.06 \\ \hline \\[-7pt]                     
  Entropy & Broken Power Law & $s_0$ & Flat & $0-10$ & keV~cm$^2$                                                                           & 1.62 & 1.25 & 0.32 \\                                     
& \& constant & $s_1$ & Flat Log & $0.001-1$ & keV~cm$^2$                                                                                   & 0.16 & 0.20 & 0.15 \\                                     
& & $\alpha_1$ & Flat & $1-5$  &                                                                                                            & 3.09 & 3.38 & 0.91 \\                                     
  & & $r_{\rm b,1}$ & Flat & $0.25-1$ & kpc                                                                                                 & 0.66 & 0.46 & 0.12 \\                                     
& & $\alpha_2$ & Flat & $0.1-3$   &                                                                                                         & 0.78 & 0.77 & 0.05 \\                                     
& & $r_{\rm b,2}$ & Flat & $5-18.5$ & kpc                                                                                                   &   16 &   18 &    2 \\                                     
& & $\alpha_3$ & Flat & $1-5$    &                                                                                                          & 2.25 & 3.54 & 0.74 \\                                     
& & $r_{\rm b,3}$ & Flat & $19-25$ & kpc                                                                                                    &   22 &   20 &    2 \\                                     
& & $\alpha_4$ & Flat Log & $0.05-5$ &                                                                                                      & 0.61 & 0.67 & 0.19 \\                                     
& & $r_{\rm b,4}$ & Flat & $26-80$ & kpc                                                                                                    &   58 &   50 &   14 \\                                     
& & $\alpha_5$ & Flat & $0.5-2$  &                                                                                                          & 1.08 & 0.99 & 0.31 \\ \hline \\[-7pt]                     
Black Hole & Point Mass & $M_{\rm BH}$ & Lognorm & $M_{\rm BH}-\sigma$  & $10^9\, M_{\odot}$                                                &  0.8 &  0.6 &  0.4 \\ \hline \\[-7pt]                     
Stellar Mass & MGE $H$-band & $M_{\rm stars}/L_{\rm H}$ & Flat &  $0.2-5$ & $M_{\odot}/L_{\rm H,\odot}$                                     & 1.19 & 1.22 & 0.11 \\ \hline \\[-7pt]                     
& (sphericalized)  \\ \hline \\[-7pt] 
Dark Matter & NFW  & norm & Flat Log & 0.5-100 & $10^{12}\, M_{\odot}$                                                                      &  1.9 &  1.9 &  0.4 \\                                     
& & $r_s$ & Flat & $2-60$ & kpc                                                                                                             & 12.1 & 12.0 &  2.2 \\                                                                                                               
  \hline \\
\end{tabular}
\tablecomments{See \S\ref{he} for definitions of the model
  components, \S\ref{Bayes} for details of the adopted Bayesian
  fitting procedure and priors, and \S\ref{entropy} for details
  specific to the entropy model.}
\end{center}
\end{table*}

We use a spherical, entropy-based solution of the hydrostatic
equilibrium (HE) equation to infer the mass distribution from the
radial gas properties measured from the \chandra\ observations
(\citealt{hump08a}; for a review of the relative virtues of this and
other HE methods see \citealt{buot12a}); we discuss the spherical
approximation in \S\ref{sph}. Our implementation of the entropy-based
method follows closely that described by B17, and we refer the reader
to that paper for details. Below we summarize the principal models
used for \src.

\begin{itemize}

\item{\it Entropy:} We represent the entropy proxy
  ($S \equiv k_{\rm B}Tn_e^{-2/3}$) in units of keV~cm$^{2}$ by,
  $S(r) = s_0 + s_1f(r_{0.25})$, where $s_0$ is a constant,
  $f(r_{0.25})$ is a dimensionless power-law with possibly one or more
  break radii, $r_{0.25}$ is the radius in units of 0.25~kpc, and
  $s_1=S(0.25\, {\rm kpc}) - s_0$; see equation (3) of B17.  Our
  fiducial model employs four break radii. In addition, we demand that
  at large radius the radial logarithmic entropy slope match the value $\approx 1.1$
  from cosmological simulations with only
  gravity~\citep[e.g.,][]{tozz01a,voit05a}. We adopt a fiducial radius
  of $r_{\rm b, baseline}=150$~kpc above which the 1.1 slope applies. 

\item{\it Pressure:} We express the free parameter associated with the
  boundary condition for the HE equation as a pressure located at a radius
  10~kpc which we denote as the ``reference pressure'' $P_{\rm ref}$. 

\item{\it Stellar Mass:}  We employ a spherically averaged version of
  the ellipsoidal multi-gauss expansion (MGE)
  model of the {\sl HST} $H$-band light reported by Y17. We convert the stellar
  light profile to stellar mass with the stellar
  mass-to-light ratio ($M_{\rm stars}/L_{\rm H}$) parameter that is
  free to vary.

\item{\it Dark Matter:} Our fiducial model represents the dark matter
  halo by an NFW profile~\citep{nfw} with free parameters a scale
  radius ($r_s$) and normalization that we convert to a halo
  concentration and mass computed for a radius of a specific
  overdensity. For comparison we also consider the
  Einasto~\citep{eina65} and CORELOG~\citep[e.g.,][]{buot12c}
  models. Finally, we also explore models that modify these DM
  profiles by ``adiabatic contraction'' (AC). Here we consider two
  variants of AC -- classic ``strong'' AC as originally proposed by
  \citet{blum86a} and a ``weak'' AC model proposed by \citet{dutt15a}
  which they call ``Forced Quenched.'' See B17 for details of our
  implementation of these AC models.

\item{\it SMBH:} We include a central point mass to represent a SMBH
  with mass $M_{\rm BH}$. Unlike Paper~1 where we fixed
  $M_{\rm BH}=4.9\times 10^9\, M_{\odot}$ to the stellar dynamical
  value obtained by \citet{wals17a} in our fiducial model, here we
  allow it to be a free parameter subject to different priors (see
  below in \S\ref{fitproc}).

\end{itemize}

We list the free parameters for the fiducial HE model in
Table~\ref{tab.fid}.  The 16 free parameters of the model are
constrained by 17 measurements each of $\ktemp$ and $\xsurf$; i.e., 34
total data points -- 14 and 20 respectively from the Cycle 16 and 19
observations.

\section{Model Fitting Methodology}
\label{fitproc}

\subsection{Bayesian Method}
\label{Bayes}

The primary method we adopt to fit the HE models to the \chandra\ data
employs a Bayesian nested sampling procedure based on the MultiNest
code v2.18~\citep{multinest} (see B17 for details). In
Table~\ref{tab.fid} we list the priors adopted for each free
parameter. We use flat priors in most cases and flat priors on the
decimal logarithm for a few parameters with ranges spanning multiple
orders of magnitude. For the black hole mass our fiducial
prior is based on the $M_{\rm BH}-\sigma$ relation of
\citet{remco16a}; i.e., a median value $\mbh=2.1\times 10^9\,\msun$
using the stellar velocity dispersion for \src\ listed in
Table~\ref{tab.prop}. Since \citet{remco16a} quote an intrinsic
scatter of 0.49 in $\log_{10}\mbh$, we adopt a lognormal prior with
mean $\log_{10}\mbh=9.32$ and standard deviation equal to the
intrinsic scatter.

For the Bayesian analysis we quote two ``best'' values for each free
parameter:  (1) the mean parameter value of the posterior which we
call the ``Best Fit'', and (2) the parameter value that maximizes the
likelihood, which we call ``Max Like.'' Unless otherwise stated, all errors quoted for the
parameters are the standard deviation ($1\sigma$) of
the posterior.

\subsection{Frequentist Method}
\label{freq}

We also perform frequentist $\chi^2$ fits of the HE models for
comparison to the results obtained with the Bayesian fits. Despite its
shortcomings~\citep[e.g.,][]{andr10a}, we also prefer the frequentist
  $\chi^2$ approach for model selection which, unlike Bayes factors,
  does not depend on the priors.  For the frequentist fits we use the
  {\sc minuit} fitting package that is part of the {\sc root}~v6.10
  software suite\footnote{https://root.cern}.

\section{Results}
\label{results}

\begin{table}[t] \footnotesize
\begin{center}
\caption{Quality of Frequentist $\chi^2$ Fits}
\label{tab.fitqual}
\begin{tabular}{rcc}   \hline\hline\\[-7pt]
Model & $\chi^2$ & dof\\
  \hline \\[-7pt]
            Fiducial &  11.9 &    18\\
Joint Fit of Cycle 19 Obs. &  13.7 &    18\\
            No Stars &  40.5 &    19\\
No DM Halo & 395.8 & 20\\
  Cycle 19 Only &   4.5 &     4\\
       0 Brk Entropy &  28.3 &    26\\
       1 Brk Entropy &  17.5 &    24\\
  Sersic Stars 2MASS &  11.0 &    18\\
          Einasto DM &  11.1 &    18\\
          Corelog DM &  10.9 &    18\\
           Strong AC &  14.8 &    18\\
             Weak AC &  11.2 &    18\\
     Weak AC Einasto &  10.7 &    18\\
Fixed Over-Massive BH &  14.2 &    19\\
 Fixed $M-\sigma$ BH &  12.6 &    19\\
\hline \\
\end{tabular}
\tablecomments{ The number of data points in these fits is 34
  ($\ktemp$ and $\xsurf$, see Table~\ref{tab.gas}). See \S\ref{he} and
  \S\ref{results} for definitions of these models. Note the frequentist
  Fiducial model is the same as the Bayesian version listed in
  Table~\ref{tab.fid} except for the priors. The parameters in the
  frequentist fits were allowed to vary over the same ranges as for
  the Bayesian priors. The one exception is $M_{\rm BH}$, for which we
  allow it to fit freely over the range $10^{8-10}\, M_{\odot}$. In
  all the cases listed in the table where $M_{\rm BH}$ is not held
  fixed, $M_{\rm BH}$ fits to a value consistent with the lower limit
  of this allowed range.}
\end{center}
\end{table}

\renewcommand{\arraystretch}{1.5}
%
%
\begin{table*}[t] \footnotesize
\begin{center}
\caption{Stellar and Total Mass}
\label{tab.mass}
\begin{tabular}{lc|c|cc|cc|cc}  \hline\hline\\[-7pt]
& $M_{\star}/L_H$ & $M_{\rm BH}$ 
& $c_{2500}$ & $M_{2500}$  & $c_{500}$ & $M_{500}$  & $c_{200}$ & $M_{200}$\\ 
& ($M_{\odot} L_{\odot}^{-1}$) & $(10^9\, M_{\odot})$&  & $(10^{12}\, M_{\odot})$ &  & $(10^{12}\, M_{\odot})$ &  & $(10^{12}\, M_{\odot})$\\
\hline \\[-7pt]
Best Fit & $1.19 \pm 0.11$  & $ 0.8 \pm  0.4$  & $11.1 \pm  1.6$ & $ 3.2 \pm  0.3$ & $21.1 \pm  3.0$ & $ 4.5 \pm  0.5$ & $30.4 \pm  4.3$ & $ 5.3 \pm  0.6$\\ 
(Max Like) & $(1.22)$ & $( 0.6)$ & $(10.9)$ & $( 3.2)$ & $(20.7)$ & $( 4.4)$ & $(29.9)$ & $( 5.3)$\\ 
\hline \\[-7pt]
1 Brk Entropy & $-0.04$ & $-0.1$ & $ 0.8$ & $-0.1$ & $ 1.5$ & $-0.2$ & $ 2.1$ & $-0.2$ \\ 
BH Flat Prior & $-0.03$ & $ 0.2$ & $ 0.4$ & $-0.1$ & $ 0.7$ & $-0.1$ & $ 1.0$ & $-0.1$ \\ 
BH Flat Logspace Prior & $0.03$ & $-0.5$ & $-0.1$ & $ 0.0$ & $-0.3$ & $ 0.0$ & $-0.4$ & $ 0.0$ \\ 
Fixed Over-Massive BH & $-0.46$ & $ 4.1$ & $ 3.7$ & $-0.4$ & $ 6.8$ & $-0.6$ & $ 9.8$ & $-0.8$ \\ 
Fixed $\msigma$ BH & $-0.13$ & $ 1.3$ & $ 1.0$ & $-0.1$ & $ 1.8$ & $-0.2$ & $ 2.6$ & $-0.3$ \\ 
Sersic Stars 2MASS & $0.04$ & $-0.0$ & $-1.5$ & $ 0.3$ & $-2.8$ & $ 0.5$ & $-4.0$ & $ 0.6$ \\ 
Einasto & $-0.07$ & $ 0.0$ & $-1.0$ & $ 0.4$ & $-2.1$ & $ 0.5$ & $-3.2$ & $ 0.5$ \\ 
Strong AC & $-0.40$ & $-0.0$ & $-3.3$ & $ 0.6$ & $-6.1$ & $ 1.1$ & $-8.8$ & $ 1.4$ \\ 
Weak AC & $-0.06$ & $-0.0$ & $-2.5$ & $ 0.3$ & $-4.6$ & $ 0.6$ & $-6.5$ & $ 0.8$ \\ 
Weak AC Einasto & $-0.13$ & $ 0.0$ & $-3.6$ & $ 0.9$ & $-6.6$ & $ 1.6$ & $-9.6$ & $ 1.9$ \\ 
Joint Fit of Cycle 19 Obs. & $-0.04$ & $-0.0$ & $ 1.6$ & $-0.2$ & $ 2.9$ & $-0.4$ & $ 4.2$ & $-0.5$ \\ 
Constant $Z_{\alpha}/\zfe$ & $0.13$ & $-0.0$ & $-0.6$ & $ 0.1$ & $-1.1$ & $ 0.2$ & $-1.6$ & $ 0.2$ \\ 
Annulus~10~$\Delta\zfe$ & $^{+0.02}_{-0.00}$ & $-0.0$ & $^{+ 0.4}_{-0.4}$ & $^{+ 0.2}_{-0.1}$ & $^{+ 0.7}_{-0.7}$ & $^{+ 0.2}_{-0.2}$ & $^{+ 1.0}_{-1.1}$ & $^{+ 0.3}_{-0.2}$ \\ 
Deproj & $-0.16$ & $ 0.0$ & $ 1.3$ & $-0.2$ & $ 2.5$ & $-0.3$ & $ 3.9$ & $-0.2$ \\ 
Distance & $-0.16$ & $ 0.0$ & $-1.1$ & $ 0.2$ & $-2.1$ & $ 0.4$ & $-2.9$ & $ 0.6$ \\   
\\ 
\hline \\
\end{tabular}
\tablecomments{Best-fit values and $1\sigma$ error estimates 
  for the free parameters of the mass components of the fiducial Bayesian
  hydrostatic equilibrium model; i.e., stellar mass-to-light ratio
  $(M_{\star}/L_H)$, black-hole mass $(M_{\rm BH})$, concentration, and enclosed total mass
  (BH+stars+gas+DM). We show the concentration and mass results obtained
  within radii $r_{\Delta}$ for overdensities $\Delta=200, 500, 2500.$
  In addition, we provide a budget of systematic errors where only the
  most significant / interesting systematics are shown (\S
  \ref{sys}). For each column we quote values with the same
  precision. In several cases, an error has a value smaller than the
  quoted precision, and thus it is listed as a zero; e.g., ``0.0'' or
  ``-0.0'', where the sign indicates the direction of the
  shift. Briefly, the various systematic errors are as follows, 
  \\
(``1 Brk Entropy''): Entropy profile has only one break radius.\\
(``BH Flat Prior''): Flat prior on $M_{\rm BH}$ ranging from $10^{8-10}\, M_{\odot}$.\\
(``BH Flat Logspace Prior''): Flat prior on $\log_{10}M_{\rm BH}$ ranging from $8-10$\\
(``Fixed Over-Massive BH''): $M_{\rm BH}$ fixed to $4.9\times 10^9\, M_{\odot}$~\citep{wals17a}.\\
(``Fixed $\msigma$ BH''): $M_{\rm BH}$ fixed to $2.1\times 10^9\, M_{\odot}$, the median of the $M-\sigma$ relation~\citep{remco16a}.\\
(``Sersic Stars 2MASS''): The stellar mass is represented by a Sersic model with $R_e$ from 2MASS (see \S\ref{stars}).\\
(``Einasto''): Einasto DM profile.\\
(``Strong AC''): Adiabatically contracted DM profile following~\citet{blum86a}.\\
(``Weak AC''): Weak adiabatically contracted DM profile following~\citet{dutt15a}; i.e., their ``Forced Quenched'' model implemented as described in B17.\\
(``Weak AC Einasto''): Same as ``Weak AC'' except here the Einasto DM profile is used.\\
(``Joint Fit of Cycle~19 Obs.''): Spectral fitting of \chandra\ Cycle~19 observation performed simultaneously on each observation segment (\S\ref{obs})\\
(``Constant $Z_{\alpha}$/$Z_{\rm Fe}$''): $Z_{\rm Mg}$/$Z_{\rm Fe}$ and $Z_{\rm Si}$/$Z_{\rm Fe}$ not allowed to vary with radius in the spectral analysis (\S\ref{spec})\\
(``Annulus~10~$\Delta\zfe$''): Spectral fitting of \chandra\ Cycle~19 observation performed for different choices of $\zfe$ in Annulus~10 (\S\ref{abun})\\
(``Deproj''): Spectral analysis performed with deprojection assuming constant spectral properties within each annulus (\S\ref{deproj.specresults} and \S\ref{sys.deproj})\\
(``Distance''): Use luminosity distance of 113~Mpc~(NED; \citealt{springob14a}).}
\end{center}
\end{table*}
\renewcommand{\arraystretch}{1}

The best-fitting Bayesian fiducial model is displayed along with the
$\ktemp$ and $\xsurf$ data points in Figure~\ref{fig.best}. The
corresponding best-fitting parameter values and $1\sigma$ errors are
listed in Table~\ref{tab.fid}. The fit is excellent for both the
Cycle 16 and Cycle 19 observations. The frequentist fit (not shown)
yields a nearly identical result with $\chi^2=11.9$ for 18 dof
(Table~\ref{tab.fitqual}).  The formal probability of obtaining a
smaller value of $\chi^2$ is $\approx 15\%$, indicating the fit is
marginally ``too good'' to happen by chance. However, we do not
believe the error bars have been overestimated. When fitting the
Cycle 19 data alone we obtain $\chi^2=4.49$ for 4 (dof, see
Table~\ref{tab.fitqual}) with a very reasonable
probability of $66\%$ for obtaining a smaller $\chi^2$ value (and
a corresponding significance of only $34\%$ to reject the model). The
Cycle 16 data also give a very reasonable $\chi^2$ value when fitted
on their own -- see Paper~1.

\begin{figure*}
\parbox{0.49\textwidth}{
\centerline{\includegraphics[scale=0.43,angle=0]{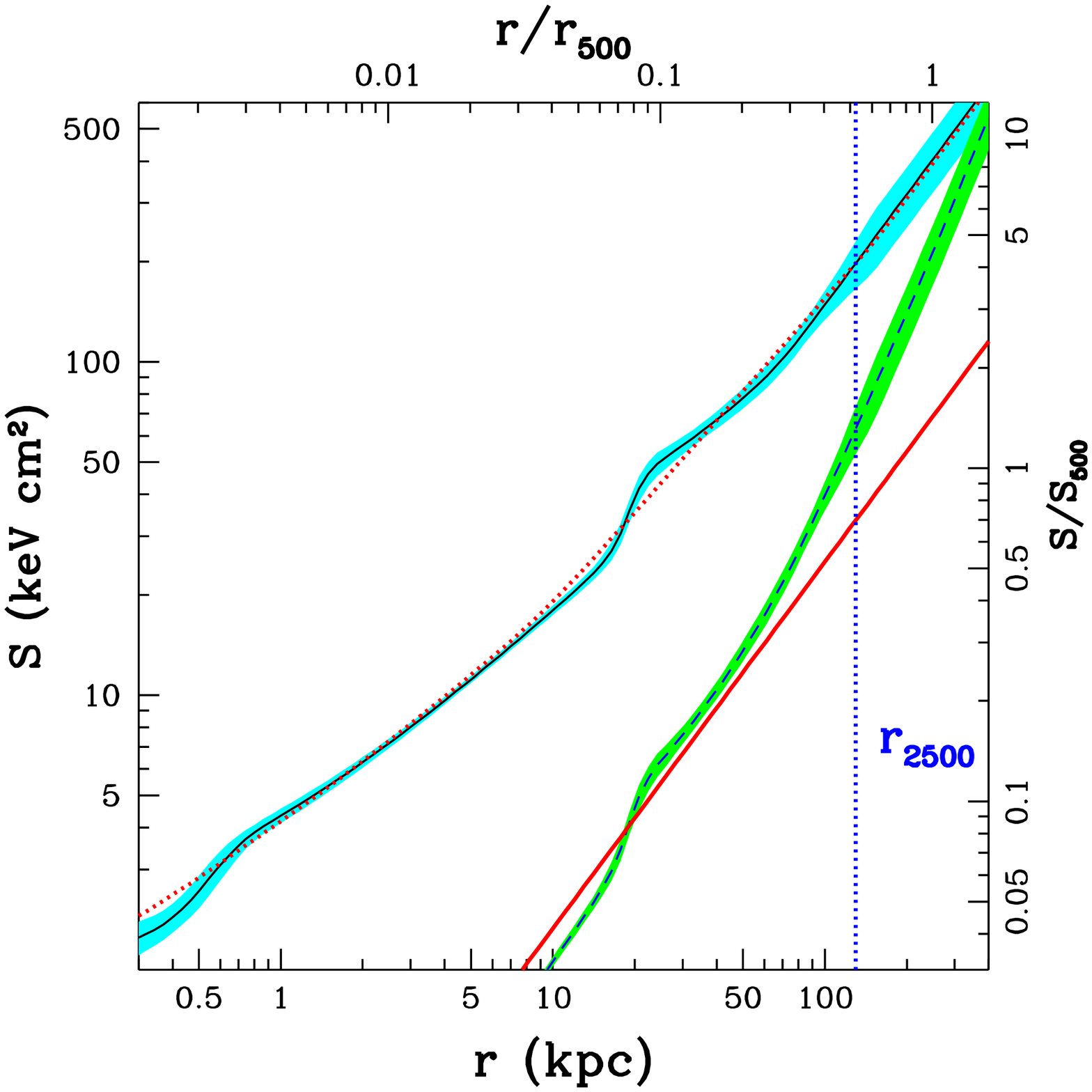}}}
\parbox{0.49\textwidth}{
\centerline{\includegraphics[scale=0.43,angle=0]{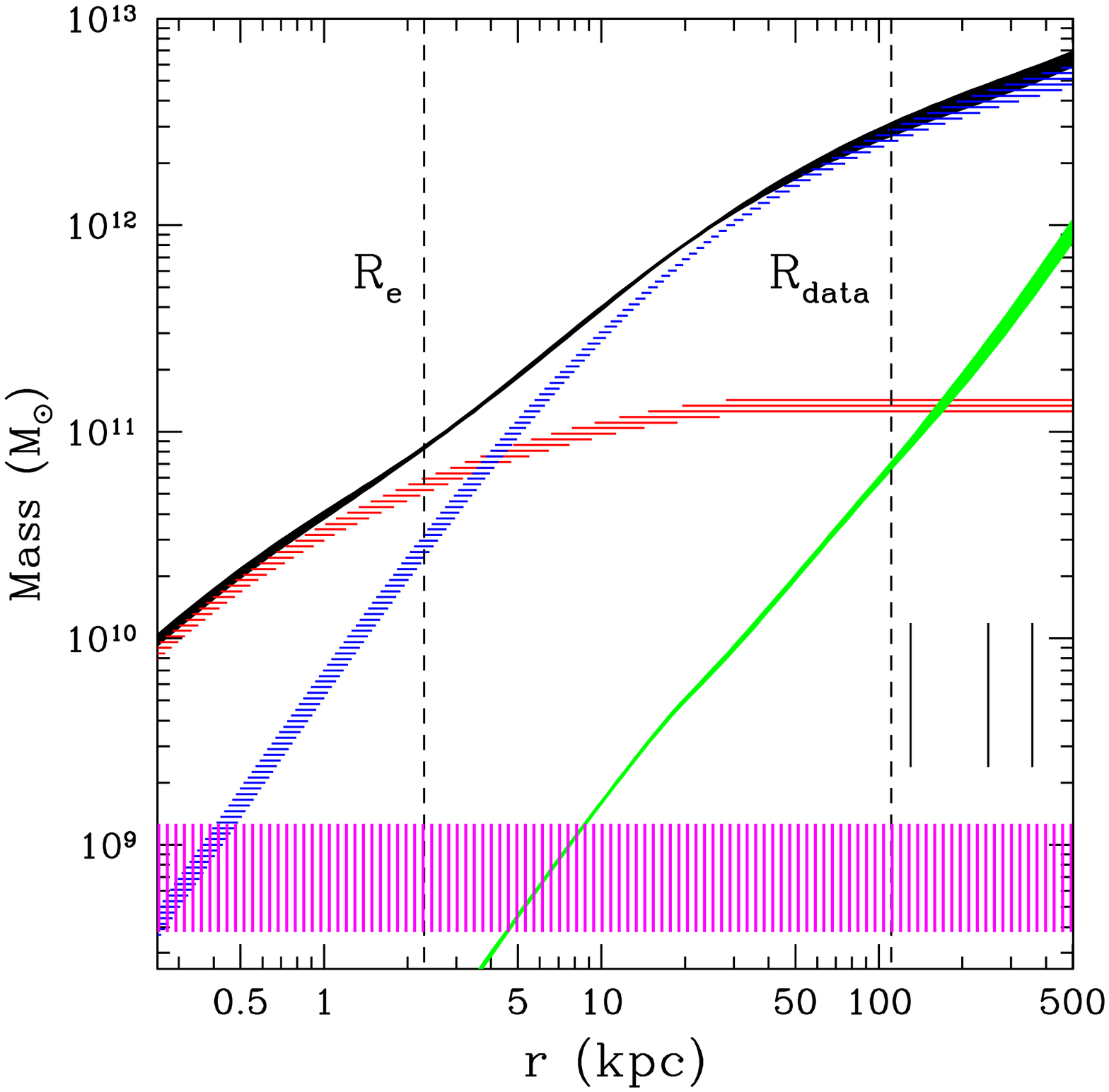}}}
\caption{\label{fig.he} Results for the Bayesian HE modeling of \src.
  The curved lines and associated shaded regions in both plots show
  the mean (i.e., ``best fit'') and standard deviation of the
  posterior as a function of radius for the quantity of interest;
  i.e., entropy or mass.  ({\sl Left Panel}) Radial profile of the
  entropy (black) and $1\sigma$ error region (cyan) for the fiducial
  hydrostatic model.  For comparison with a dotted red line we show
  the best-fitting entropy profile with only one break radius which
  follows the fiducial 4-break model very closely.  The upper
  horizontal axes gives the radius in units of $\rfiveh$ while the
  vertical axis on the right shows the entropy rescaled by
  $S_{500}=48.0$~keV~cm$^2$ (see eqn.\ 3 of \citealt{prat10a}). The
  baseline $r^{1.1}$ profile obtained by cosmological
  simulations~\citep{voit05a} with only gravity is shown as a red
  line. The result of rescaling the entropy profile by
  $\propto f_{\rm gas}^{2/3}$ \citep{prat10a} is shown by the black
  dashed line (and green $1\sigma$ region). ({\sl Right Panel}) Radial
  profiles of the total mass (black) and individual mass components of
  the fiducial hydrostatic model: total NFW DM (blue), stars (red),
  hot gas (green), and SMBH (magenta).  The black vertical lines in
  the bottom right corner indicate the best-fit virial radii; i.e.,
  from left to right: $r_{2500}$, $r_{500}$, and $r_{200}$. The
  vertical dashed lines indicate the location of the stellar
  half-light radius $(R_e)$ and the outer extent of the \chandra\ data
  analyzed $(R_{\rm data})$.}
\end{figure*}

In Figure~\ref{fig.he} we show the total mass profile of the Bayesian
fiducial model broken down into its constituents. Within the central
$0.44$~kpc corresponding to the radius of Annulus~1 of the Cycle 19
observation, the DM and SMBH contribute nearly equally to the total
mass, while the stellar mass dominates both. The DM and stars
contribute equally at $r\approx 3.9$~kpc ($\sim 1.7R_e$) after which
the DM dominates all components. The gas mass does not equal the
stellar mass until $r\sim 170$~kpc.

We will refer to the best-fitting ``virial radii'' of the fiducial
model evaluated at a few reference overdensities (see \S 5.3 of B17
for details on their computation): $\rtwofiveh=130\pm 5$~kpc,
$\rfiveh=248\pm 10$~kpc, and $\rtwoh=358\pm 14$~kpc.  The Cycle 19
gas measurements extend to $\approx0.85\rtwofiveh$. Since, however, the outer bin
width is rather large, more relevant for discussing the properties of the HE
models is the average bin radius (eqn.\ 10 of \citealt{mcla99a})
$\approx 90$~kpc or $\approx0.7\rtwofiveh$.

\subsection{Entropy}
\label{entropy}

In Paper~1 we found the Cycle 16 data were described well by an
entropy profile consisting of only a constant and a power-law with no
breaks in radius. Fitting such a model jointly to the Cycle 16 and
Cycle 19 still results in a formally acceptable fit: $\chi^2=28.3$ for
26 dof (``0 Brk Entropy'' in Table~\ref{tab.fitqual}). However, when
adding a break radius, the fit improves significantly ($\chi^2=17.5$
for 24 dof, ``1 Brk Entropy'' in Table~\ref{tab.fitqual}). The
1-break entropy profile has a poorly constrained break radius
$r_{\rm b,1}=7\pm 5$~kpc, and its best-constrained parameter is the
power-law exponent exterior to the break radius
($\alpha_2=0.93\pm 0.05$). Adding more breaks reduces $\chi^2$ but
does not lead to a statistically significant improvement in the fit.

Nevertheless, we add more break radii to provide greater flexibility
in the entropy model for the following reason.  When comparing models
with different assumptions in the mass components (e.g., fixed
over-massive SMBH, Einasto DM halo, AC, etc.)  we want potential
differences in the fits to be determined by differences in the form of
the mass components not the precise form of the assumed entropy
profile. HE and convective stability only requires that the entropy
profile increase monotonically with radius. Consequently, we added
break radii until the value of $\chi^2$ changed by less than 1.

We followed this procedure and arrived at an entropy profile with four
breaks having break radii $\approx 0.7$, 16, 22, and 58~kpc. (We
obtained this final result after initially using larger prior ranges
than indicated in Table~\ref{tab.fid}.) The inner break adds
flexibility when testing different SMBH priors while the others allow
for flexibility primarily for the DM halo models. The two break radii
in the middle are rather close together indicating a fairly abrupt
jump in entropy near 20~kpc. The slope parameter at large radius is
not well-constrained and is consistent with
the $r^{1.1}$ profile from gravity-only
simulations~\citep{voit05a}. Although this 4-break entropy profile is the
fiducial model we employ (\S\ref{he}, Table~\ref{tab.fid}), in
\S\ref{sys} we compare to results obtained with the 1-break entropy
profile as a systematics error check. (The best-fitting 1-break model is plotted in
Figure~\ref{fig.he} as a red dotted line.)

In Figure~\ref{fig.he} we plot the entropy profile of the Bayesian fiducial
model along with the $\sim r^{1.1}$ theoretical entropy profile produced by
gravity-only cosmological simulations~\citep{voit05a}. The entropy
profile of \src\ is remarkably similar to those we have obtained
previously for other massive, fossil-like elliptical galaxies,
NGC~720~\citep{hump11a}, NGC~1521~\citep{hump12b}, and NGC~6482
(B17). That is, within a radius $\sim\rtwofiveh$, the entropy profile
greatly exceeds the theoretical gravity-only profile. In addition, when
the observed entropy profile is rescaled by $\sim f_{\rm
  gas}^{2/3}$~\citep{prat10a}, the result broadly matches the
gravity-only profile. The success of this rescaling indicates that
for $r\la\rtwofiveh$ the feedback energy responsible for raising
the entropy in that region is consistent with having spatially
rearranged the gas rather than raising its temperature, as also is
observed for galaxy clusters~\citep{prat10a}.

We see, however, that as the entropy approaches $\rtwofiveh$ this
rescaling becomes increasingly less successful. Very similar behavior
is observed for NGC~1521~\citep{hump12b}, and NGC~6482 (B17). Perhaps
for \src\ and these galaxies at larger radius the feedback energy
heated the gas by raising its temperature indicating a different
feedback mechanism prevails for $r\ga\rtwofiveh$. It is also
noteworthy that interior to $\approx 0.6\rtwofiveh$ (77~kpc) the
cooling time is less than the age of the Universe (see
Figure~\ref{fig.tc}), and thus the details of gas cooling may be
responsible for the apparent transition in the rescaling behavior as
the radius approaches $\rtwofiveh$.

\subsection{Pressure}
\label{pressure}

The pressure profile of the fiducial Bayesian HE model (not shown) is
extremely similar to that of NGC~6482 (see \S 6.3 and Figure~5 of
B17). Between radii $\approx 0.1-0.7\,\rfiveh$ the pressure profile
broadly matches the ``universal'' pressure profile determined for
galaxy clusters by \citet{arna10a}, while at smaller and larger radii
the observed profile is clearly a different shape. While the results
for $r\ga 0.7\rfiveh$ are provisional owing to the limited data range,
the different shape at smaller radius is robust implying a breakdown
in the mass scaling of the universal pressure profile and, presumably,
the greater importance of non-gravitational energy in shaping the
thermodynamical properties of the gas in galaxy/group-scale halos.

\subsection{SMBH}
\label{smbh}

In Paper~1 we analyzed the Cycle 16 observation and obtained a
Bayesian constraint on the black hole mass,
$M_{\rm BH} = (5\pm 4)\times 10^{9}\, M_{\odot}$, for a flat prior
very consistent with the stellar dynamical measurement,
$\mbh=(4.9\pm 1.7)\times 10^{9}\, M_{\odot}$, of \citet{wals17a}.
When instead using a flat prior on $\log_{10}\mbh$ (``Flat Logspace Prior''), we obtained a
smaller value, $M_{\rm BH} = (1.4\pm 1.7)\times 10^{9}\, M_{\odot}$,
still consistent with the stellar dynamical measurement within
the large errors.  Our best-fitting model in Paper~1 for the flat
prior on $\mbh$ predicted a projected
emission-weighted temperature, $\ktemp\sim 1.18$~keV, within the
central $R=1\arcsec$; i.e,. the region corresponding to Annulus~1 of
the new Cycle 19 observation. 

As is readily apparent from the temperature profile displayed in
Figure~\ref{fig.best} and the $\ktemp$ values listed in
Table~\ref{tab.gas} we measure a much smaller temperature in Annulus~1
with the Cycle 19 observation: $\ktemp=0.969 \pm 0.027$~keV implying a
smaller value of $\mbh$ than indicated in Paper~1. (Note only for
Annulus~1 was the predicted $\ktemp$ from Paper~1 inaccurate; e.g.,
for Annulus~2 the predicted value was $0.910$~keV compared to the
value measured of $\ktemp=0.905 \pm 0.018$~keV.) For a flat prior, our
joint fit of the Cycle 16 and Cycle 19 data gives,
$\mbh=(1.0\pm 0.6)\times 10^9\,\msun$, with a 99.9\% upper limit,
$\mbh\le 3.9\times 10^9\,\msun$. The flat logspace prior fits to even
lower values approaching the lower limit of the adopted prior range,
$\mbh=(0.3\pm 0.2)\times 10^9\,\msun$ with a 99.9\% upper limit,
$\mbh\le 1.1\times 10^9\,\msun$ .

As mentioned in the notes to Table~\ref{tab.fitqual}, the frequentist
fit also gives $\mbh$ consistent with the adopted lower fit
boundary. However, when fixing $\mbh$ to the stellar dynamical value,
$\chi^2$ only increases by 2.3 (``Fixed Over-Massive BH'' in
Table~\ref{tab.fitqual}). According to the F-Test, the fiducial model
is $\approx 92\%$ more probable; i.e., the preference for a small
value of $\mbh$ is less than a $2\sigma$ effect for the frequentist
fit.

Fixing $\mbh$ to the stellar dynamical value leads to a smaller
inferred value for $\mlh$ (``Fixed Over-Massive BH'' in
Table~\ref{tab.mass}). The Bayesian fit gives, $\mlh=0.73\pm 0.13$
solar with a 99.9\% upper limit, $\mlh\le 0.99$ solar, significantly
less than $\mlh=1.2$~solar expected for a Kroupa IMF
(\S\ref{stars}). The frequentist fit gives $\mlh=0.96\pm 0.13$ solar
which is $\sim 2\sigma$ discrepant.

Therefore, the Bayesian fits clearly disfavor the over-massive SMBH,
as indicated by the values inferred for $\mbh$ when it is
fitted freely and for $\mlh$ when $\mbh$ is
fixed to the stellar dynamical value. (The significantly larger
$c_{200}$ value provides more evidence against the over-massive SMBH model
-- see \S\ref{cm} and \S\ref{highc}.) The frequentist analysis also
disfavors the over-massive SMBH but at a much lower 
significance level ($\sim 2\sigma$). 

Given the dependence of the constraints on the priors, by
default we adopt a prior on $\mbh$ consistent with the $\msigma$
relation of \citet{remco16a} -- see \S\ref{Bayes}. In \S\ref{bigbh} we
discuss the implications of the smaller values inferred for $\mbh$
compared to the stellar dynamics measurement.

\subsection{Stellar Mass and IMF}
\label{stars}

The \chandra\ data clearly require the stellar mass component. If the
stellar mass is omitted, the frequentist fit gives $\chi^2=40.5$ for
19 dof (Table~\ref{tab.fitqual}), which is an increase of 28.6 over the
fiducial model with one extra dof. The strong need for the stellar
component translates to a good constraint on stellar mass-to-light
ratio; i.e., $\mlh=1.19 \pm 0.11$ for the fiducial Bayesian model
(Table~\ref{tab.mass}). This value agrees very well with that expected from
single-burst stellar population synthesis (SPS) models with a Kroupa
IMF ($\mlhband = 1.2$), but is significantly lower than predicted for
a Salpeter IMF ($\mlhband = 1.7$ solar); see \citet{wals17a} for more
information about these SPS estimates.

Our fiducial measurement also agrees very well with the value $1.3\pm
0.3$~solar (statistical error) obtained by the stellar
dynamical study of \citet{wals17a}, with a factor of $\approx 2.5$
higher precision. Y17's stellar dynamical measurement gives a value
about $50\%$ larger, $\mlh=1.9^{+0.5}_{-0.4}$~solar, at $\approx
2\sigma$ significance, which agrees better with the SPS value for a Salpeter
IMF. In \S\ref{keyfactor} we discuss these stellar dynamical
measurements and their sensitivity to the assumed DM halo
concentration. 

Most of the systematic errors we considered do not change $\mlh$ by
more than the $1\sigma$ statistical error (Table~\ref{tab.mass}).  Of
key importance is that if we use a different model to represent the
stellar light profile (\S\ref{sys.stars}), we obtain $\mlhband = 1.23$
solar, very consistent with the MGE result. Much smaller values of
$\mlh$ are given by a few models, ``Fixed Over-Massive BH''
(\S\ref{smbh}), ``Strong AC'' (\S\ref{dm}), and to a lesser extent
``Deproj'' and ``Distance''', which
we believe is notable evidence disfavoring these models. The
``Constant $Z_{\alpha}/\zfe$'' test give the largest positive shift in
$\mlh$. 

The very good agreement we find with the value of $\mlh$ we measure
and SPS models with a Kroupa IMF is also very consistent
with what we find for other fossil-like massive elliptical galaxies
from X-ray HE analysis. This evidence for a Kroupa IMF is in stark
contrast to the support typically found for a Salpeter IMF from
stellar dynamics and lensing studies of massive elliptical galaxies
(see discussion in \S~8.3 of B17).

\subsection{Dark Matter: NFW and Alternatives}
\label{dm}

Of the SMBH, stars, and DM, by far the most important component needed
to describe the X-ray emission is the DM (Table~\ref{tab.fitqual}). If
the DM halo is omitted from the fiducial model, the frequentist fit
gives $\chi^2=395.8$ for 20 dof; i.e., the F-test gives a tiny
probability ($2\times 10^{-14}$) that the ``No DM Halo'' model is
preferred over the fiducial model with DM. While the DM component is
clearly required, the \chandra\ data do not distinguish between the
different DM models we investigated. The Einasto and CORELOG models
give minimum $\chi^2$ values that are $<1$ from that obtained for the
fiducial model and with the same number of dof
(Table~\ref{tab.fitqual}). The weak AC model applied to the NFW and
Einasto models changes the fit quality very little.  The strong AC NFW
model has a $\chi^2$ value that is a little larger (i.e., by 2.9) but is
still a good fit.

As noted above in \S\ref{stars}, the strong AC model gives a much
smaller $\mlh$, which we believe disfavors that model. The CORELOG
model (not shown in Table~\ref{tab.mass}) is peculiar in that the
``Best Fit'' and ``Max Like'' values for $\mlh$ differ significantly;
i.e. $\mlh\sim 0.7$~solar for ``Best Fit'' and
$\mlh\sim 1.2$~solar for ``Max Like'' with $1\sigma$ error $\pm
0.2$~solar, where the larger ``Max Like'' values are favored to match
the SPS models (\S\ref{stars}).

These results are very consistent with those obtained for NGC~6482
(B17).

\subsection{Halo Concentration and Mass}
\label{cm}

In Paper~1 we obtained tentative evidence for an ``over-concentrated''
NFW DM halo compared to the general halo population. Our Bayesian
fiducial model fitted to the Cycle 16 \chandra\ data yielded a Best
Fit  virial mass, $M_{200} = (9.6\pm 3.7)\times 10^{12}\, M_{\odot}$
and concentration, $c_{200}=17.5\pm 6.7$ considerably larger than the
median \lcdm\ value of $\approx
7$~\citep[e.g.,][]{dutt14a}. Intriguingly, we found the Max Like
best-fitting value suggested an even more extreme outlier from the \lcdm\ relation:
$c_{200}=25.9$ and $M_{200} = 5.1\times 10^{12}\, M_{\odot}$. 

Adding the much deeper Cycle 19 \chandra\ observations to our analysis confirms
the higher concentration solution from Paper~1. In
Table~\ref{tab.mass} we list the concentration and mass for
overdensities $\Delta=200,$ 500, and 2500.  Now the Best Fit and
Max Like  values agree extremely well. The values we measure,
$c_{200}=30.4\pm 4.3$ and
$M_{200} = (5.3\pm 0.6)\times 10^{12}\, M_{\odot}$, indicate an
extreme outlier to the median \lcdm\ relation (see \S \ref{highc}).

The AC models give statistically significant lower values of
$c_{200}$, with Strong AC giving a larger reduction $(\approx 9)$ than
Weak AC $(\approx 7)$. As noted above in \S\ref{stars}, the Strong AC
model gives an uncomfortably low $\mlh$ whereas Weak AC gives $\mlh$
fully compatible with the fiducial model. The Fixed Over-Massive BH
model gives the largest increase in $c_{200}(+9.8)$, which disfavors
the model as an even more extreme outlier from the median \lcdm\
relation (\S\ref{highc}). Note, however, that the frequentist fit for
this model increases $c_{200}$ by only $\approx 4$; i.e., the evidence
is mostly directed against the Bayesian version of the Fixed
Over-Massive BH model (also see above in \S\ref{smbh}).

The Einasto model and Weak AC Einasto models behave analogously to the
NFW versions except with $c_{200}$ shifted lower by $\sim 3$. We
reiterate that these various DM models cannot be distinguished in
terms of their fit quality (\S\ref{dm}).

All of the results described in this section, in particular the
evidence that \src\ is an extreme outlier in the \lcdm\
$c_{200}-M_{200}$ relation, are very consistent with the fossil
galaxy/group NGC~6482 (B17). In \S\ref{highc} we discuss further
implications of the high value measured for $c_{200}$.

\subsection{Density Slope and DM Fraction}
\label{slope}

\begin{table}[t] \footnotesize
\begin{center}
\caption{Mass-Weighted Total Density Slope and DM Fraction}
\label{tab.slope}
\begin{tabular}{rrcc|cc}   \hline\hline\\[-7pt]
Radius & Radius & \multicolumn{2}{c}{Fiducial} & \multicolumn{2}{c}{Weak AC + Einasto}\\
(kpc) & ($R_e$) & $\langle\gamma\rangle$ & $f_{\rm DM}$  & $\langle\gamma\rangle$ & $f_{\rm DM}$ \\
  \hline \\[-7pt]
  1.1 &   0.5 & $ 2.12 \pm  0.04$ & $ 0.18 \pm  0.04$       & $ 2.05 \pm  0.04$ & $ 0.27 \pm  0.05$    \\
  2.3 &   1.0 & $ 2.04 \pm  0.05$ & $ 0.34 \pm  0.05$       & $ 2.02 \pm  0.04$ & $ 0.43 \pm  0.06$    \\
  4.6 &   2.0 & $ 1.93 \pm  0.03$ & $ 0.54 \pm  0.05$       & $ 1.99 \pm  0.02$ & $ 0.60 \pm  0.05$    \\
  9.2 &   4.0 & $ 1.91 \pm  0.02$ & $ 0.71 \pm  0.03$       & $ 1.97 \pm  0.02$ & $ 0.74 \pm  0.03$    \\
 11.5 &   5.0 & $ 1.94 \pm  0.03$ & $ 0.76 \pm  0.03$      & $ 1.98 \pm  0.02$ & $ 0.78 \pm  0.03$    \\
 23.0 &  10.0 & $ 2.08 \pm  0.05$ & $ 0.85 \pm  0.01$     & $ 2.02 \pm  0.04$ & $ 0.87 \pm  0.02$   \\
\hline \\
\end{tabular}
\tablecomments{The mass-weighted slope $(\mslope)$ is evaluated for the Bayesian
  HE models using equation (2) of~\citet{dutt14b}. The DM fraction is
  defined at each radius $r$ as, $f_{\rm DM} = M_{\rm DM}(<r)/M_{\rm
    total}(<r)$. The Weak AC Einasto model is defined in the notes to Table~\ref{tab.mass}.}
\end{center}
\end{table}

In Table~\ref{tab.slope} we list the mass-weighted slope $(\mslope)$
and DM fraction $(\dmfrac)$ for the Bayesian fiducial model evaluated
at radii for several multiples of $\reff$. For $r=R_e$ we obtain
$\mslope =  2.04 \pm  0.05$ and $\dmfrac=0.34 \pm  0.05$ which differ
significantly from the average values obtained by Y17 for their sample of 16
CEGs ($\langle\gamma\rangle = 2.3$, $f_{\rm DM} = 0.11$). We discuss
the principal reason for the discrepancy between our results and those
of Y17 in \S\ref{keyfactor}. We note that the AC models give even
larger DM fractions for \src; e.g., the Weak AC+Einasto model result
shown in Table~\ref{tab.slope}.

The mass-weighted slope we measure for \src\ within $r=R_e$ is less
than the average slopes of local massive ETGs ($2.15 \pm 0.03$,
intrinsic scatter 0.10) determined by stellar
dynamics~\citep{capp15a}. The local ETGs also have a smaller DM
fraction (0.19, accounting for the higher mass range of the CEGs --see
Y17). These differences between the local ``normal'' ETGs and \src\ presumably
reflect the unusual and very high $c_{200}$ we measure for \src; i.e.,
higher DM concentration means more DM near the center (higher
$\dmfrac$) which translates to a smaller slope due to the higher
weighting of the NFW (or Einasto) profile with a flatter slope than
the stellar profile.

Studies of ETGs that combine strong lensing with stellar dynamics have
found a large range of DM fractions within $\reff$, including several
with values consistent with (or larger than) what we have found for
\src~\cite[e.g.,][]{barn11a,sonn15a}.  The large DM fractions in the
lensing/SD studies probably do not reflect unusually high $c_{200}$
like \src; i.e., those galaxies were not selected to be early forming
objects like \src\ (or NGC~6482), and thus they should not often possess
$c_{200}$ values that deviate extremely from the \lcdm\
$c_{200}-M_{200}$ relation.

Recently, \citet{wang18a} have used the IllustrisTNG simulation to
show that massive ETGs in \lcdm\ have $\mslope=2.003$ with a scatter
of $0.175$ over the radial range $0.4-4\,\reff$ with the slope
changing little for redshifts below 2. The slope we measure for \src\
is consistent with these results (Table~\ref{tab.slope}). Using the
same simulation \citet[][see their Figure~12]{love18a} obtain
$\dmfrac$ within $5R_e$ consistent with our measurements but within
$1R_e$ they find large values ($\dmfrac\sim 0.6$). Using a different
simulation \citet{remu17a} find a large range of DM fractions within
$1R_e$ consistent with what we find for \src. Interestingly, both
\citet{remu17a} and \citet{love18a} obtain smaller DM fractions
within $1R_e$ at $z=2$; e.g., $\dmfrac\approx
10\%$~\citep{remu17a}. If \src\ is truly a nearby analog of a
$z\approx 2$ galaxy, its higher DM fraction is in conflict with these
simulations. 

In \citet{hump10a} we showed that the total mass slope  approximated by
a power-law between $0.2-10\, \reff$ decreases with
halo mass and $\reff$ for halos ranging from massive galaxies ($\sim
10^{12}\,\msun$) up to massive clusters  ($\sim
10^{15}\,\msun$) with the mean relation,
$$\gamma=2.31 - 0.54\log (R_e/\rm kpc).$$
This slope-$R_e$ relation predicts $\gamma=2.11$ for \src.  The values
we obtain within $10\reff$ for the fiducial and Weak AC+Einasto models
listed in Table~\ref{tab.slope} are consistent with the relation
within the observed scatter~\citep[0.14 dex,][]{auge10a}.

\subsection{MOND and RAR}
\label{mond}

\begin{figure*}[t]
\parbox{0.49\textwidth}{
\centerline{\includegraphics[scale=0.42,angle=0]{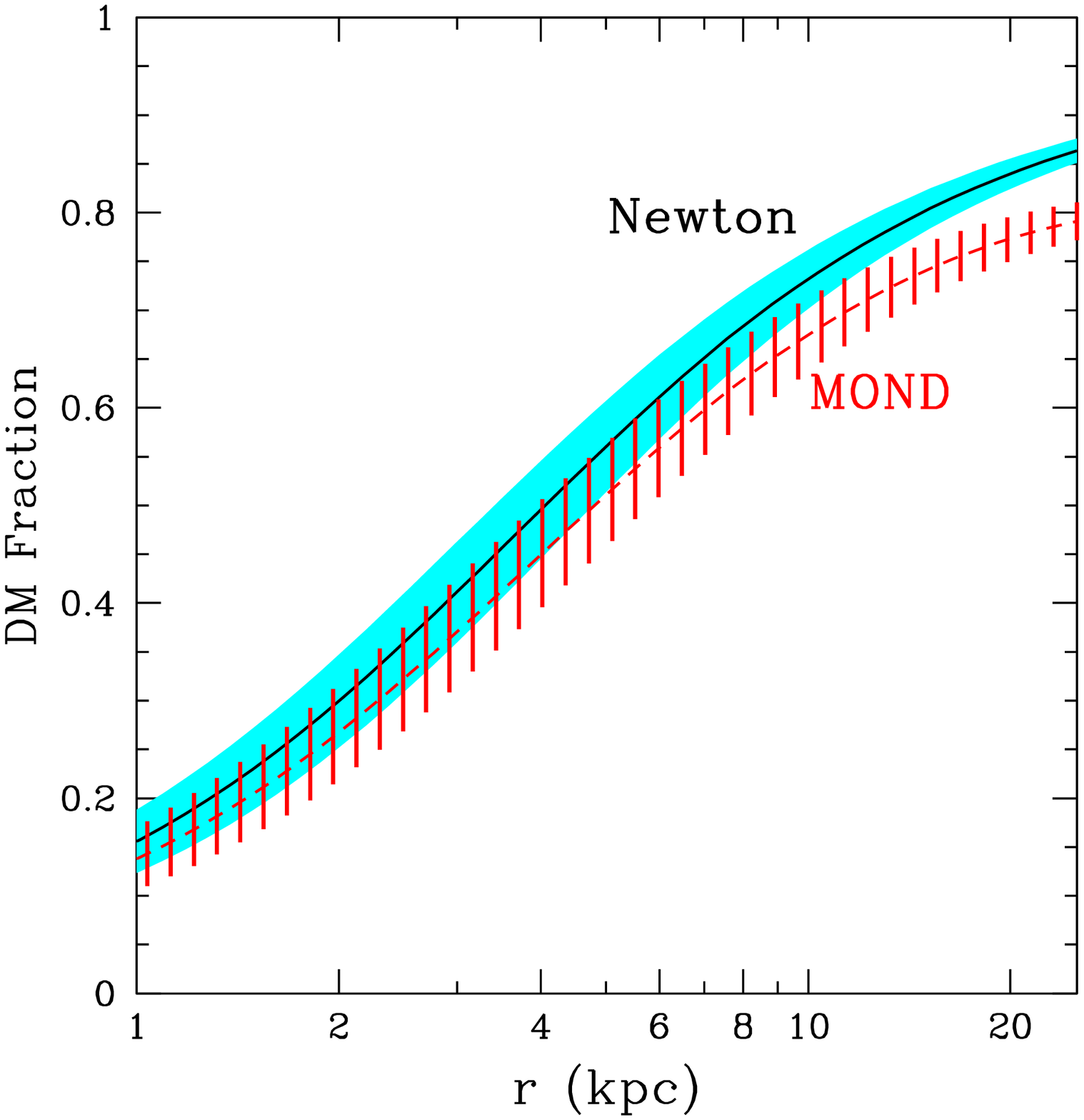}}}
\parbox{0.49\textwidth}{
\centerline{\includegraphics[scale=0.42,angle=0]{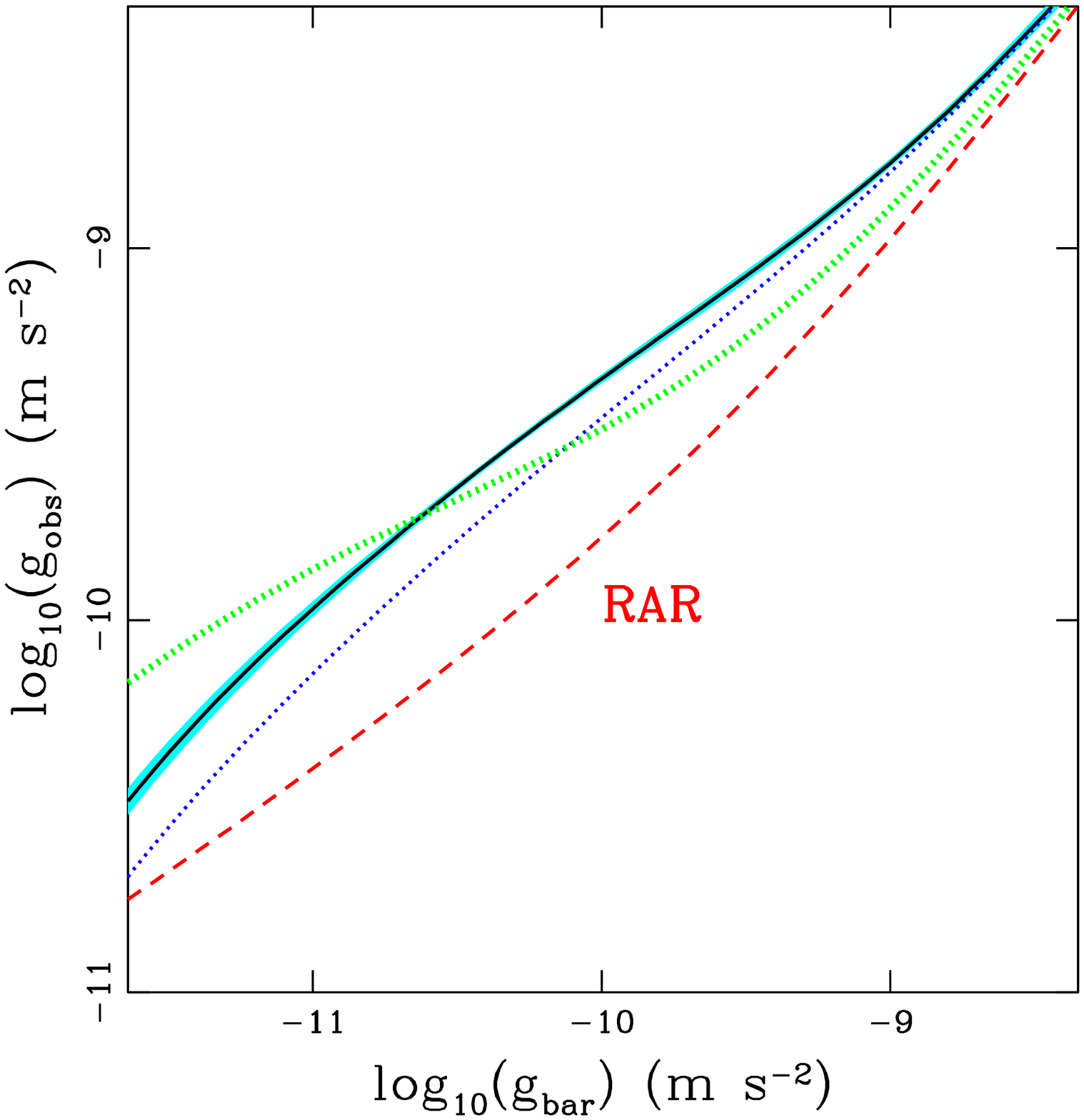}}}
\caption{\footnotesize  ({\sl Left Panel}) Radial profiles of the DM fraction within the
  central $\sim 25$~kpc for the fiducial HE model for the Newtonian
  (black and cyan) and MOND (red) cases. The shaded and hashed regions
  represent $1\sigma$ errors. ({\sl Right Panel})  The solid black line enclosed by the cyan region
 shows the mean (i.e., ``Best Fit'') and standard deviation of the
 posterior  
 of the fiducial Bayesian   HE model of \src\ for the radial  gravitational acceleration $(g_{\rm obs})$
plotted versus the acceleration arising from only the
  baryonic mass components (i.e., stars+SMBH+gas).  (Only the errors in $g_{\rm obs}$ are shown. The errors for
$g_{\rm bar}$ are much smaller.) The blue dotted line is the ``Best
Fit'' result for NGC~6482 computed using the corresponding fiducial HE
model from B17. The dashed red line is the ``universal'' RAR of
\citet{lell17a}. The range plotted for $g_{\rm bar}$ corresponds to
$1\le R\le 100.7$~kpc. The dotted green line is the ``Best Fit''
Bayesian HE model for \src\ with high-baryon mass and low central DM
concentration (see \S\ref{mond} and \S\ref{keyfactor}).}
\label{fig.mond}
\end{figure*}

We have also interpreted our HE analysis in terms of
MOND~\citep{milg83a} to compare to our traditional Newtonian
analysis. Our method follows \citet{sand99a} and \citet{angu08a} and
is most recently described in detail in \S 6.7 of B17. In
Figure~\ref{fig.mond} we display the DM fraction profile within a
radius of 25~kpc for our fiducial Bayesian HE model computed from the
Newtonian and MOND perspectives. As is clear in the figure, MOND
requires almost as much DM as in the Newtonian approach, very similar
to what we found for NGC~6482 (B17). 

The MOND acceleration scale $a_0\approx 1.2\times 10^{-8}$~cm~s$^{-2}$
is reached at a radius $\approx 11$~kpc (considering only the mass in
baryons -- stars+SMBH+gas, $\sim 42$~kpc otherwise). Even considering
a much smaller central DM contribution (i.e., setting $c_{200}=10$ in
the Newtonian analysis, too small to be consistent with the data --
see \S\ref{keyfactor}) with correspondingly much larger stellar mass
contribution ($\mlh\sim 1.6$~solar) only changes this radius by $\sim
1$~kpc; i.e., the need for DM in MOND occurs well within the
Newtonian regime of \src. Hence, \src\ and NGC~6482 provide strong
evidence that MOND requires DM on the massive galaxy scale, extending to
lower masses the results obtained from HE studies on the group
\citep[e.g.,][]{angu08a} and cluster \citep[e.g.,][]{poin05a} scales.

Given the close similarity between MOND and the Radial Acceleration
Relation~\citep[RAR,][]{lell17a}, it is also to be expected that these
galaxies deviate significantly from the RAR. Indeed, as shown in
Figure~\ref{fig.mond} we find that the gravitational acceleration we
derive from our HE analysis $(g_{\rm obs})$ for \src\ (and NGC~6482)
significantly exceeds the acceleration from only the baryons
$g_{\rm bar}$. For example, at the MOND acceleration scale,
$g_{\rm bar}=1.2\times 10^{-10}$~m~s$^{-2}$, the RAR predicts
$g_{\rm obs}=1.9\times 10^{-10}$~m~s$^{-2}$ whereas we measure for
\src\ $g_{\rm obs}=(4.99\pm 0.15)\times 10^{-10}$~m~s$^{-2}$.
\citet{lell17a} quote a scatter in the RAR of $\la 0.13$ dex
indicating at this one data point the discrepancy is $>3\sigma$. In
fact, as seen in Figure~\ref{fig.mond} this level of discrepancy
applies over a wide range in $g_{\rm bar}$, and thus the significance
of the discrepancy with the entire RAR is in fact much larger. 

Finally, we mention that in Figure~\ref{fig.mond} we also show
$g_{\rm obs}$ for the $c_{200}=10$ model mentioned above which has a
lower central DM and a higher baryon contribution for \src. Although
the shape of the $g_{\rm obs}$ profile is different from the fiducial
model, the level of discrepancy with the RAR is broadly similar. The
fractional errors (not shown) for $g_{\rm obs}$ of this model are
similar to the fiducial model.

\subsection{Gas and Baryon Fraction}
\label{baryfrac}

\renewcommand{\arraystretch}{1.5}
%
%
\begin{table*}[t] \footnotesize
\begin{center}
\caption{Gas and Baryon Fraction}
\label{tab.fb}
\begin{tabular}{lcc|cc|cc}   \hline\hline\\[-7pt]
& $f_{\rm gas, 2500}$ & $f_{\rm b, 2500}$ & $f_{\rm gas, 500}$ & $f_{\rm b, 500}$ & $f_{\rm gas, 200}$ & $f_{\rm b, 200}$\\
\hline \\[-7pt]
Best Fit 
 & $0.028 \pm 0.003$ & $0.071 \pm 0.006$ & $0.058 \pm 0.008$ & $0.089 \pm 0.010$ & $0.094 \pm 0.015$ & $0.120 \pm 0.016$\\ 
(Max Like) 
 & $(0.027)$ & $(0.071)$ & $(0.058)$ & $(0.089)$ & $(0.094)$ & $(0.120)$\\ 
\hline \\[-7pt]
1 Brk Entropy & $0.001$ & $0.001$ & $0.004$ & $0.004$ & $0.007$ & $0.008$ \\ 
BH Flat Prior & $0.000$ & $0.000$ & $0.001$ & $0.001$ & $0.002$ & $0.002$ \\ 
BH Flat Logspace Prior & $-0.000$ & $0.001$ & $-0.000$ & $0.000$ & $-0.000$ & $0.000$ \\ 
Fixed Over-Massive BH & $0.002$ & $-0.010$ & $0.007$ & $-0.001$ & $0.013$ & $0.007$ \\ 
Fixed $\msigma$ BH & $0.001$ & $-0.002$ & $0.002$ & $0.000$ & $0.004$ & $0.002$ \\ 
Sersic Stars 2MASS & $-0.001$ & $-0.003$ & $-0.004$ & $-0.006$ & $-0.008$ & $-0.010$ \\ 
Einasto & $-0.002$ & $-0.009$ & $-0.005$ & $-0.010$ & $-0.009$ & $-0.012$ \\ 
Strong AC & $-0.003$ & $-0.022$ & $-0.011$ & $-0.025$ & $-0.020$ & $-0.033$ \\ 
Weak AC & $-0.002$ & $-0.008$ & $-0.006$ & $-0.011$ & $-0.011$ & $-0.015$ \\ 
Weak AC Einasto & $-0.004$ & $-0.016$ & $-0.012$ & $-0.022$ & $-0.022$ & $-0.031$ \\ 
Joint Fit of Cycle 19 Obs. & $-0.000$ & $0.001$ & $0.001$ & $0.002$ & $0.003$ & $0.004$ \\ 
Constant $Z_{\alpha}/\zfe$ & $-0.001$ & $0.002$ & $-0.004$ & $-0.002$ & $-0.007$ & $-0.005$ \\ 
Annulus~10~$\Delta\zfe$ & $^{+0.002}_{-0.002}$ & $^{+0.003}_{-0.004}$ & $^{+0.005}_{-0.006}$ & $^{+0.006}_{-0.007}$ & $^{+0.009}_{-0.010}$ & $^{+0.010}_{-0.011}$ \\ 
Deproj & $0.010$ & $0.007$ & $0.023$ & $0.022$ & $0.040$ & $0.038$ \\ 
  Distance & $0.003$ & $0.006$ & $0.005$ & $0.007$ & $0.007$ & $0.008$ \\
  \\ 
\hline \\
\end{tabular}
\tablecomments{Best-fit values and $1\sigma$ error estimates 
  for the gas and baryon fractions of the fiducial Bayesian
  hydrostatic equilibrium model  quoted for several over-densities. See the notes to
  Table~\ref{tab.mass} regarding the other systematic error tests}
\end{center}
\end{table*}
\renewcommand{\arraystretch}{1}

\begin{figure}[t]
\begin{center}
\includegraphics[scale=0.42]{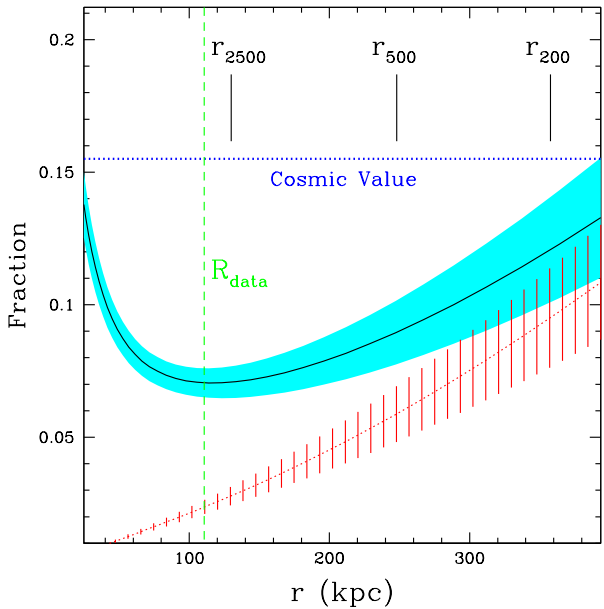}
\end{center}
\caption{\footnotesize Baryon fraction (solid black line, shaded cyan
  $1\sigma$ error region) and gas fraction (dotted red line, shaded
  red $1\sigma$ error region) of the fiducial Bayesian HE model.  The
  green vertical dashed line indicates the outer extent of the
  \chandra\ data analyzed $(R_{\rm data})$.}
\label{fig.baryfrac}
\end{figure}

We display the radial profiles of the gas $(\gasfrac)$ and baryon
fractions $(\baryfrac)$ in Figure~\ref{fig.baryfrac} for the fiducial
Bayesian HE model and list the results for these quantities within
radii $\rtwofiveh$, $\rfiveh$, and $\rtwoh$ in Table \ref{tab.fb}
along with the systematic error budget. Within $\rtwofiveh$, which
essentially represents the extent of the \chandra\ data, we measure
$\fbtwofiveh=0.071\pm 0.006$; i.e., $\approx 45\%$ of the cosmic mean
value $f_{\rm b,U}=0.155$ determined by \planck\ \citep{plan14a},
where the hot gas contributes $\approx 40\%$ of the measured
baryons. The gas and baryon fractions continue to increase with radius
until at $\rtwoh$ we have $\fbtwoh=0.120\pm 0.016$ comprising nearly
$80\%$ of the comic mean. Here the hot gas contributes $\approx 80\%$
of the measured baryons. 

Thus far we have only considered the stellar baryons associated with
the MGE decomposition of the HST $H$-band light from
Y17. \citet{yild15a} note there are only two galaxies known within a
1~Mpc radius of \src. Using NED we identify these galaxies as
2MASX~J08284832-0704316 (PGC152584) and 2MASX~J08291551-0647454
(PGC152635), which on the sky are located, respectively,
$\approx 220$~kpc ($\approx 0.9\rfiveh$) and $\approx 300$~kpc
($\approx 0.9\rtwoh$), from the center of \src. There is sparse
information on these galaxies. However, based on their 2MASS $K$-band
magnitudes they are each about 2 magnitudes fainter than \src,
suggestive of a fossil group. If we assume a similar contribution of
non-central baryons as for the fossil group NGC~6482, that would add
$\approx 0.04$ to the baryon fraction at $\rtwoh$ to give
$\fbtwoh\approx 0.16$, consistent with the cosmic mean value. 

Most of the systematic errors we have considered
(Table~\ref{tab.fb}) shift the gas and baryon fractions by less
than the statistical error. The spectral deprojection (``Deproj''
test) shifts are not significant within the larger $1\sigma$ erros in
the deprojection model; e.g., $1\sigma$ error is $\pm 0.041$ on
$\fbtwoh$.  The AC models have the largest effect
resulting in lower values. The Weak AC Einasto model, which is our
favored model (see \S\ref{highc}), yields $\fbtwoh$ lower by 0.036
which effectively cancels the expected increase from non-central
baryons. 

We conclude that the Bayesian HE models predict that $\fbtwoh$ is
close to the cosmic mean value for \src. In \S\ref{disc.baryfrac} we
discuss this measurement in relation to previous X-ray studies of
other fossil elliptical galaxies and consider the implications for
the ``Missing Baryons'' problem.

\subsection{Cooling Time and Free-Fall Time}
\label{tcool}

\begin{figure}[t]
\begin{center}
\includegraphics[scale=0.42]{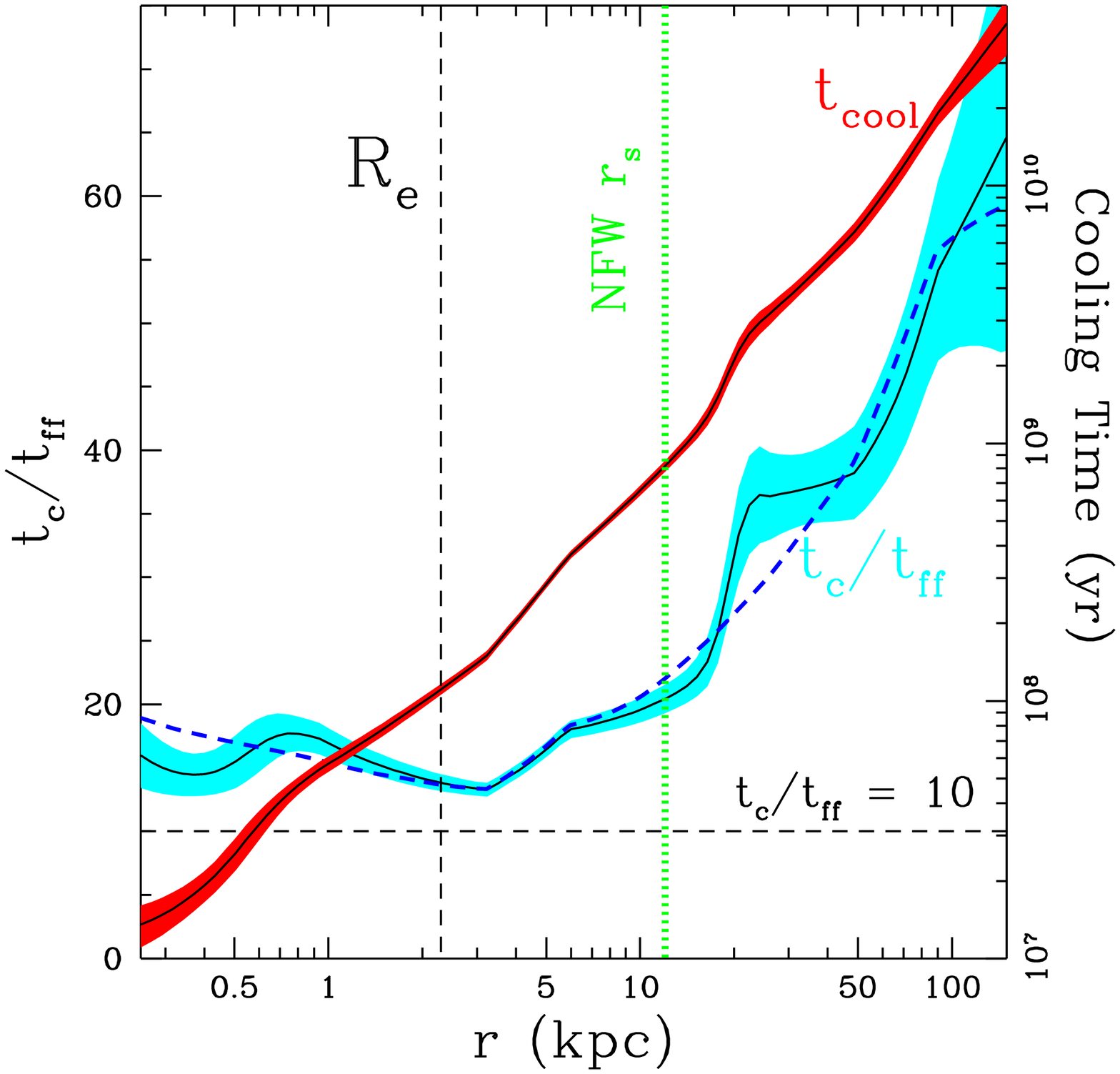}
\end{center}
\caption{\footnotesize Ratio of the cooling time $(t_c)$ to the
  free-fall time $(t_{\rm ff})$ versus radius for the fiducial
  Bayesian HE model. The solid black line is the mean (i.e., ``Best
  Fit'') and the associated cyan region is the standard deviation of
  the posterior. For comparison, we show with the dashed blue line the
  ``Best-Fit'' result for the model having only one break radius in
  the entropy profile (\S\ref{entropy}). The vertical black dashed
  line indicates the $H$-band stellar half light while the vertical
  green dotted line shows the ``Best Fit'' NFW scale radius $r_s$.
  The horizontal dashed black line shows the value
  $t_c/t_{\rm ff} = 10$ proposed as an approximate lower limit before
  the onset of multiphase
  cooling~\citep[e.g.,][]{shar12a,gasp12b,meec15a}. The corresponding
  cooling time is shown by the solid black line and associated red
  region with values plotted on the right vertical axes.}
\label{fig.tc}
\end{figure}

In Figure~\ref{fig.tc} we plot $\tc/\tff$, the ratio of cooling time
to free-fall time, as well as $\tc$ itself, as a function of radius
for the fiducial Bayesian HE model. \src\ displays a large central
region of approximately constant $\tc/\tff$; i.e., for
$r\le 10$~kpc, $\tc/\tff\approx 14-19$. Outside of this region,
$\tc/\tff$ increases with radius. The $\tc/\tff$ profile of \src\
resembles a scaled down version of the massive
cluster Hydra-A~\citep{hoga17a}, which has pronounced X-ray cavities
associated with AGN radio jets. Similarly, the observed $\tc/\tff$
profile more closely resembles massive elliptical galaxies and groups
with evidence for multiphase gas~\citep[see Figure~2
of][]{voit15a}. The $\tc/\tff$ profile suggests
``precipitation-regulated AGN feedback''~\citep[e.g.,][]{voit17a} for
\src, and we discuss further the implications of \src\ for feedback
models in \S\ref{feedback}.

We also remark that the approximately constant region of
$\tc/\tff\approx 17$ ends near the NFW scale radius
$r_s\approx 12$~kpc (and for $\tc\approx 10^8$~yr) for the fiducial
model. We obtain extremely similar $\tc/\tff$ profiles for the
Einasto DM model ($r_s\approx 14$~kpc) and other models (e.g.,
AC); i.e., in \src\ the DM scale radius approximately equals the
feedback radius.

\section{Error Budget}
\label{sys}

As in our previous studies we have examined the sensitivity of our
measurements for the mass profile to various choices we have made in
the spectral fitting and HE analysis. Many of these are listed in
Tables~\ref{tab.gas} and \ref{tab.fb} and have been discussed already
in the text (e.g., SMBH priors, various DM and AC models), and we will
not say anything further about them here.
We mention that the numbers
quoted for the systematic errors in Tables~\ref{tab.gas} and
\ref{tab.fb} are intended to provide the interested reader with some
idea of how sensitive are the fiducial model parameters to arbitrary
(but well motivated) changes to the fiducial model and/or analysis. As
such, these numbers should not be added in quadrature to produce a
single systematic error bar for a given parameter.

Below we provide more details for several tests in \S\ref{center2},
\S\ref{sys.deproj}, \S\ref{abun}, and \S\ref{sys.stars}. First we very
briefly list several notable tests that did not affect the measured HE
model parameters significantly.

\bigskip

\noindent{\it Entropy Profile: } We examined an entropy profile with
only a single break radius (see \S\ref{entropy}) with results listed
in Tables~\ref{tab.mass} and \ref{tab.fb} and the best-fitting model
shown as the red dotted line in Figure~\ref{fig.he}. We also studied
larger radii where the baseline gravity-only slope (see \S\ref{he})
sets in, $r_{\rm b, baseline}=250,500$~kpc.

\medskip 

\noindent{\it Joint Spectral Fitting of Individual Cycle 19 Observations: } 
We found no significant differences in the derived HE models when the
gas properties of the Cycle 19 observation are obtained without
summing the spectra of the individual exposures
(\S\ref{proj.specresults}, Appendix~\ref{joint}, Tables~\ref{tab.mass}
and \ref{tab.fb}).

\medskip 

\noindent{\it Radial Extent of Models: } By default we filled the
gravitational potential of our HE models with hot gas out to a radius
of $r_{\rm max}=1$~Mpc. We also examined $r_{\rm max}=0.5,1.5$~Mpc. 

\medskip 

\noindent{\it Soft CXB: } We examined rescaling the nominal fluxes of
the soft CXB components  (\S\ref{specmod}) by factors of 0.5 and 2.

\subsection{Choice of Center}
\label{center2}

As noted in \S\ref{moment}, the centroid of the X-ray brightness
changes by very little within the central $\sim 20\arcsec$. For our
analysis we adopted the center position obtained by computing the
centroid within a circle of radius $3\arcsec$ placed initially on the
nominal stellar galaxy position. The annuli used for spectral
extraction (Table~\ref{tab.gas}) were defined about this 
centroid position (8:28:47.141,$-$6:56:24.367).

To gauge the sensitivity of our results to this choice, we also
examined using the center position (8:28:47.131,-6:56:24.047) located
at the emission peak $\approx 0.35\arcsec$ to the NW. Using this
center has no measurable effect on the results.

\subsection{Deprojection}
\label{sys.deproj}

In \S\ref{deproj.specresults} we discussed the results obtained for
the spectral fitting when using a deprojection analysis. One issue
that requires clarification is the error bars reported in
Table~\ref{tab.gas.deproj}. Because of the correlations between annuli
introduced by the deprojection procedure, we estimate errors on the
gas parameters (temperature, normalization, and abundances) via
Monte Carlo simulations in \xspec\ as we have done in previous studies of
deprojected spectra~\citep[e.g.,][]{buot00c,buot03a}; i.e., we
performed 100 simulations of each set of spectra, computed the gas
properties (e.g., temperature) for each simulation, and then compute
the standard deviations on the parameters which we quote as the
$1\sigma$ errors in Table~\ref{tab.gas.deproj}.

For our analysis of the projected spectra we have also computed
parameter errors by simply finding the values of a given parameter
that change the C-statistic by 1 from its minimum value; i.e., using
the {\sc error} command in \xspec. (We quote these results by default;
e.g., Table~\ref{tab.gas}). As expected, for the projected spectral
analysis we find that both approaches -- $\Delta C=1$ and Monte Carlo --
give very consistent results. 

As mentioned in  \S\ref{deproj.specresults}, we performed deprojected spectral
fits for two cases: (1) no gas emission was accounted for outside of
the bounding annuli; (2) gas emission predicted by the Best Fit
fiducial HE model from the projected spectral analysis
(Table~\ref{tab.fid}) was assigned to the large background apertures
exterior to the bounding annuli. The results for case (2) are
presented in the ``Deproj'' entries  in Tables~\ref{tab.mass} and \ref{tab.fb}.

For most parameters the ``Deproj'' systematic test shifts the values
by an amount comparable to the $1\sigma$ statistical error. As noted
in \S\ref{baryfrac}, the listed positive shifts and the gas and baryon
fractions are not statistically significant when considering the
larger statistical uncertainties of the deprojection models. Case (1)
produces larger shifts that are modestly significant; e.g., $\fbtwoh$
is increased by $0.063\pm 0.038$. However, we prefer not to emphasize
case (1) given that it does not account for emission projected into
the bounding annuli.  Finally, we also mention that the minimum
$\chi^2$ achieved for the frequentist HE analysis for both deprojected
cases is $\approx 8$ larger than the projected fit; i.e., the fits
are formally acceptable for the deprojected cases, but the default
projected analysis gives a better fit.

We conclude that the constraints on the mass profile obtained
when using the spectral deprojection approach are overall very
consistent with the default projection analysis.

\subsection{Metal Abundances}
\label{abun}

When we do not allow the abundance ratios $\zmgfe$ and $\zsife$
to vary with radius in the spectral fits (\S\ref{proj.specresults}),
the effects on the HE models are listed in the ``Constant
$Z_{\alpha}/\zfe$'' entries in Tables~\ref{tab.mass} and
\ref{tab.fb}. This test shifts most of the parameters by an amount
comparable to the $1\sigma$ errors and, in the case of $\mlh$, by
$\approx 1.5\sigma$. Interestingly, the results for this test differ
noticeably for the Best Fit and Max Like values.
The shifts of marginal significance listed in the tables reflect only
the Best Fit values. If we compare Max Like values instead, then
all the shifts are $<1\sigma$. 

When we allow $\zfe$ to take values $0.27\,\zsun$ and $0.45\,\zsun$
for Annulus~10 of the Cycle 19 observation (see
\S\ref{proj.specresults}) representing the intrinsic scatter of the
average group/cluster profile of \citet[][]{mern17a}, we list the
parameter shifts for the HE models in the ``Annulus~10~$\Delta\zfe$''
entries in Tables~\ref{tab.mass} and \ref{tab.fb}. All the quoted parameter shifts are
less than the statistical errors.

\subsection{Radial Range}
\label{range}

Throughout the paper we have quoted global parameters (e.g.,
concentration, mass, etc.) for a series of ``virial'' radii
representing three common overdensities; i.e., $\rtwofiveh$,
$\rfiveh$, and $\rtwoh$ (see \S\ref{results}), where $\rtwofiveh$
essentially matches the radial extent of the Cycle 19 observation, while
$\rfiveh\approx 1.9\rtwofiveh$ and $\rtwoh\approx 2.8\rtwofiveh$ fall
outside the observed data range. Our analysis of the projected data
technically constrains the projected emission for radii outside the
data range in projection (i.e., $r>\rtwofiveh$). In reality, the
emission from radii much larger than the data extent projected onto
the observed sky annuli is dominated by the emission from within the
three-dimensional spherical shells corresponding to the radii of the
observed sky annuli. Hence, the global parameter values quoted at
radiii $\rfiveh$ and, particularly, $\rtwoh$ are to an increasing
amount extrapolations of our model outside the data range.

We do not expect significant systematic error in the extrapolated
parameter values (e.g., $c_{200}$, and $M_{200}$) provided (1) the
true DM profile is accurately described by the NFW/Einasto models and
(2) we have accurately measured the DM scale radius $r_s$. Previously
in \citet{gast07b} we emphasized that accurate and precise constraints
on $r_s$, and thus the NFW DM profile, from HE X-ray studies are only
possible when several radial data bins exist both above and below
$r_s$.  For \src\ we measure $r_s=12.1\pm 2.2$~kpc so that
approximately 6 annuli lie below $r_s$ and 4 annuli above $r_s$ for
the Cycle 19 data, with $R_{\rm out}\approx 9r_s$ where
$R_{\rm out}=110.7$~kpc is the outer radius of Annulus~10. (The Cycle
16 data contribute 4 annuli below $r_s$, 2 annuli above $r_s$, and one
annulus encloses $r_s$.)  Since $r_s$ is situated well within the data
range with several radial bins above and below it, we believe it is
well constrained, as is reflected in the Bayesian constraints; e.g.,
for the fiducial HE model we obtain $r_s=12.1_{-4.4}^{+7.1}$~kpc (99\%
confidence).

It is instructive to consider the fossil cluster RXJ~1159+5531 for
which the accuracy of some radial extrapolation (i.e.,
from $\rfiveh$ to $\rtwoh$) has been directly tested.  Although it is
$\approx 15$ more massive, RXJ~1159+5531 is similar to \src\ in that
it is a highly relaxed system with an above-average halo
concentration. Using a single \chandra\ observation with data
extending out to $\approx \rfiveh$ with approximately 5 (4) data bins
below (above) $r_s$, \citet{gast07b} obtained $c_{200}=8.3\pm 2.1$ and
$M_{200}=(7.9\pm 5)\times 10^{13}\, M_{\odot}$ with
$r_s\approx 104$~kpc. Later, adding several \suzaku\ observations
covering the entire sky region out to $\rtwoh$, in \citet{buot16a} we
obtained $c_{200}=8.4\pm 1.0$ and
$M_{200}=(7.9\pm 0.6)\times 10^{13}\, M_{\odot}$ with
$r_s\approx 104$~kpc, extremely consistent with the previous
extrapolated measurement. Notice in particular that $r_s$ did not
change between the studies.

Based on a suite of cosmological hydrodynamical cluster simulations,
\citet{rasi13a} reported that significant positive biases (up to 25\%)
in the inferred value of $c_{200}$ are possible if the X-ray HE
analysis only fits data within $\approx \rtwofiveh$.
However, the
largest biases occur for their simulations without AGN feedback. When
AGN feedback is included, the bias is small (i.e., $\la 10\%$, see
Figure~5 of \citealt{rasi13a}). The lowest mass clusters studied by \citet{rasi13a} have
$M_{200}\approx 2\times 10^{14}$; i.e., $\approx 40$ times more
massive than \src. Assuming these results apply to lower-mass halos
like \src, then the $\la 10\%$ bias for the simulations with AGN feedback
is comparable to or less than the statistical error we measure for \src.

In a future paper we will test the extrapolation for \src\ using a new
deep \xmm\ observation that will extend the radial range of the model
fits closer to $\rfiveh$.

\subsection{Spherical Symmetry}
\label{sph}

It has been shown that spherical averaging of an ellipsoidal mass
profile typically introduces only small orientation-averaged biases
for global quantities such as $c_{200}$, $M_{200}$, and $\fgastwoh$ in
X-ray HE studies of galaxy and cluster masses~\citep[e.g.,][and
references therein]{buot12b,buot12c}.  Since we also found such small
biases to be negligible compared to the statistical errors for the
massive elliptical galaxy NGC~6482 (see B17), which has DM properties
very similar to \src, we do not present a specific estimate here.

We have also considered in this paper the mass profile near
$R_e$. Previous studies of the spherical approximation in X-ray studies of
galaxy clusters do not specifically address such a small
radius. However, in \citet[][see also \citealt{chur08a}]{buot12b} we
showed that spherically averaging an ellipsoidal scale-free
logarithmic potential in a HE analysis introduces zero bias in the
inferred mass for any gas temperature profile. We expect this result
to apply to a good approximation near $R_e$ where it is well
established that the total mass profile slope
$\langle\gamma\rangle\approx 2$ for massive ETGs, as we
have found for \src\ (see \S\ref{slope}).

We can test this expectation for the massive elliptical
galaxy NGC~720 for which we previously
reported  a value $M_{\rm tot}/L_B=6.0$~solar at $R_e$ obtained from
spherically averaging ellipsoidal HE model fits to the \chandra\
data~\citep{buot02b}. We compare this measurement to the results
of the fully spherical HE analysis
of NGC~720 we performed using the same \chandra\ data in \citet{hump06b} after accounting for the
slightly different distance used and the fact the latter study quotes
$M_{\rm tot}/L_K$; i.e.,  $M_{\rm tot}/L_B=6.0$~solar becomes
$M_{\rm tot}/L_K=1.1$~solar in excellent agreement with the $M_{\rm tot}/L_K$ profile
displayed in Figure~5 of \citet{hump06b}.  We therefore expect that 
systematic errors associated with spherical
averaging should also be small near $1R_e$ for \src.

\subsection{Stellar Mass Profile}
\label{sys.stars}

To assess the sensitivity of our results to the shape of the stellar
light profile, we also considered a de Vaucouleurs model (i.e., $n=4$
Sersic model) with the half-light radius inferred from the Two Micron
All-Sky Survey (2MASS) Extended Source Catalog~\citep{jarr00a}. We
adopted a median, circularized $R_e=2.1$~kpc and normalized the model
the total $H$-band luminosity (Table~\ref{tab.obs}). We list the
results for this test in the ``Sersic Stars 2MASS'' entries in
Tables~\ref{tab.mass} and \ref{tab.fb}. In all cases using
the Sersic model shifts the paramers by $\le 1\sigma$. 

\section{Discussion}
\label{disc}

\subsection{High Halo Concentration and Formation History}
\label{highc}

\begin{figure}[t]
\begin{center}
\includegraphics[scale=0.42]{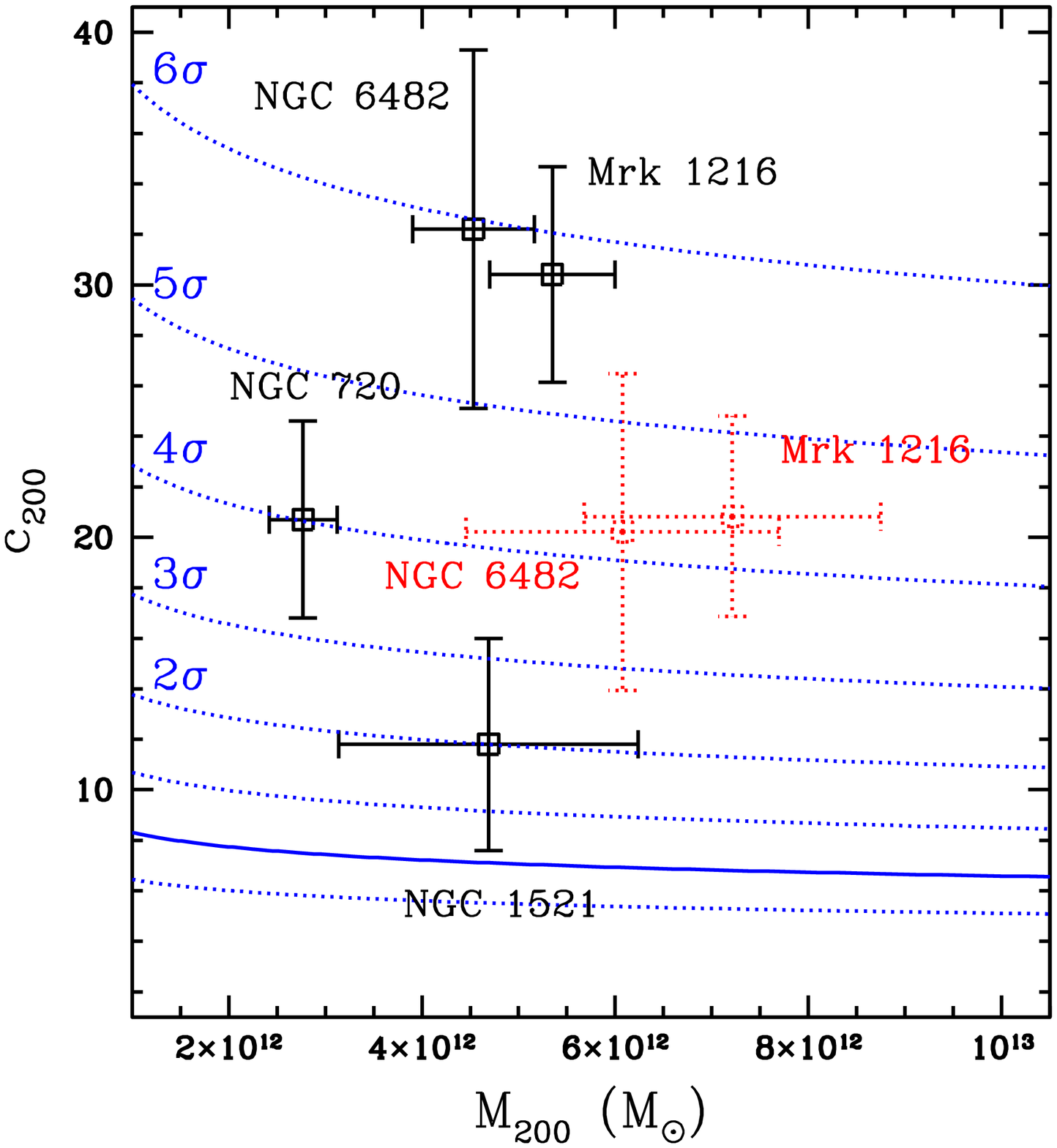}
\end{center}
\caption{\footnotesize Concentration and mass values for \src, the
  fossil group NGC~6482, and two ``nearly fossil''
  massive elliptical galaxies (see \S\ref{highc}). The blue solid
  line is the \lcdm\ relation for relaxed halos from~\citet{dutt14a}
  evaluated at the distance of \src. The blue dotted lines indicate
  the intrinsic scatter in the \lcdm\ relation; i.e., the lines
  closest the mean relation are $\pm 1\sigma$ while the significances
  of the other lines are indicated. The data points indicated by red
  dots are models with Einasto DM halos that have undergone ``weak''
  adiabatic contraction following \citet{dutt15a} (see \S
  \ref{he}). Note that for these Einasto models the deviation from the
mean is less than indicated for the NFW models by roughly $1\sigma$.}
\label{fig.c-m}
\end{figure}

\begin{table*}[t] \footnotesize
\begin{center}
\caption{Comparison to Theoretical \lcdm\ $c_{200}-M_{200}$ Relations}
\label{tab.cm}
\begin{tabular}{ccc|cc|cc|cc}   \hline\hline\\[-7pt]
& \multicolumn{2}{c}{\src} & \multicolumn{2}{c}{Dutton+14} & \multicolumn{2}{c}{Ludlow+16} & \multicolumn{2}{c}{Child+18}\\
DM Profile & $M_{200}$ & $c_{200}$ & $\bar{c}_{200}$ & $\Delta c_{200}$ & $\bar{c}_{200}$ & $\Delta c_{200}$ & $\bar{c}_{200}$ & $\Delta c_{200}$\\
\hline\\[-7pt]
NFW & $(5.3\pm 0.6)\times 10^{12}\, \msun$ & $30.4\pm 4.3$ & 7.0 & $5.8\sigma$ & $\cdots$ & $\cdots$ & 6.1 & $(11.9\sigma,6.3\sigma)$\\
Weak AC NFW & $(6.1\pm 1.0)\times 10^{12}\, \msun$ & $23.9\pm 4.1$ & 6.9 & $4.9\sigma$ & $\cdots$ & $\cdots$ & 6.1 & $(9.0\sigma,5.5\sigma)$\\
Einasto & $(5.8\pm 1.0)\times 10^{12}\, \msun$ & $27.2\pm 4.3$ & 7.9 & $4.1\sigma$ & 7.2 & $4.5\sigma$ & 7.1 & $(8.4\sigma,4.5\sigma)$\\
Weak AC Einasto & $(7.2\pm 1.5)\times 10^{12}\, \msun$ & $20.8\pm 4.0$ & 7.7 & $3.4\sigma$ & 7.1 & $3.7\sigma$ & 7.0 & $(6.1\sigma,3.7\sigma)$\\
\hline \\
\end{tabular}
\tablecomments{Results listed for \src\ pertain to the Best Fit
  Bayesian HE models. For each theoretical $c_{200}-M_{200}$  relation, we list: (1)
  $\bar{c}_{200}$, the median value of $c_{200}$ predicted for the
  Best Fit $M_{200}$ for \src, and (2)
  $\Delta c_{200}=c_{200}-\bar{c}_{200}$ expressed in terms of the
  intrinsic scatter of the theoretical relation. The theoretical
  relations are as follows. (1) Dutton+14~\citep{dutt14a}. \lcdm\ Planck
  cosmological parameters. (2) Ludlow+16~\citep{ludl16a}. \lcdm\ Planck
  cosmological parameters with the same intrinsic scatter as
  Dutton+14. (3) Child+18~\citep{chil18a}. \lcdm\ WMAP-7 cosmological
  parameters. We use the results for the stacked halos quoted in their
  Table~1. We list two values for $\Delta c_{200}$. The first assumes
  a gaussian scatter $\sigma_c=c/3$ preferred by Child+18. The second
  uses the lognormal scatter of Dutton+14.}
\end{center}
\end{table*}

In Table~\ref{tab.cm} we compare the halo concentration $(c_{200})$
and mass $(M_{200})$ we have measured for \src\  using the Bayesian HE
models to some theoretical models of the \lcdm\ $c_{200}-M_{200}$
relation from the literature. Our fiducial model with an NFW DM
halo exceeds the median \lcdm\ relation by $\sim 6\sigma$ in terms of
the intrinsic scatter of the theoretical relations. The size of this
discrepancy implies an outlier so extreme as to be inconsistent with
the \lcdm\ simulations. Indeed, from visual inspection of published
catalogs of $c_{200}$ and $M_{200}$ values we do not see any \lcdm\
halos consistent with the NFW values we have measured for  \src; e.g.,  Figure~16 of
\citet{dutt14a} and Figure~2 of \citet{raga18a}.

When substituting the Einasto DM profile for NFW the value obtained for
$c_{200}$ remains high, though the tension with \lcdm\ is reduced
modestly to $4.1-4.5\,\sigma$ (Table~\ref{tab.cm}). The reduced
tension for the Einasto model is attributed both to the smaller value
of $c_{200}$ measured for \src\ and the larger theoretical intrinsic
scatter compared to the NFW profile~\citep[0.13 dex,][]{dutt14a}.
Nevertheless, we do not see any \lcdm\ halos consistent with the Einasto
values we have measured for \src\ in the simulation results published
by \citet[][ see their Figures~2 and 4]{bens18a}.  Adiabatic
contraction lowers $c_{200}$ more and further lessens the tension with
\lcdm\ to a more modest, though still substantial, $3.4-3.7\, \sigma$.

In Figure~\ref{fig.c-m} we plot the $c_{200}$ and $M_{200}$ values for
\src\ and three fossil and nearly fossil systems with evidence for
above-average concentrations: NGC~720~\citep{hump11a},
NGC~1521~\citep{hump12b}, and NGC~6482 (B17). We refer to NGC~720 and
NGC~1521 as ``nearly'' fossil systems since within their projected
virial radii they obey the fossil classification. NGC~6482 in
particular displays $c_{200}$ and $M_{200}$ behavior extremely similar
to \src\ -- both in terms of its outlier status with respect to \lcdm\
and the way the $c_{200}$ and $M_{200}$ parameters change between NFW,
Einasto, and AC models.

The very high halo concentration we have measured for \src\ has
profound implications for its formation and evolutionary history. The
halo formation time is a major factor determining the scatter in halo
concentrations for a given mass~\citep[e.g.,][]{neto07a,ludl16a}, with
higher concentrations reflecting halos that have formed
earlier. Consequently, the high value of $c_{200}$ indicates that,
compared to the typical halo of its total mass, most of the mass of
\src\ was in place much earlier, which is consistent with largely
passive evolution for most of its existence. {\it Thus the high
concentration provides vital corroborating evidence of \src\ as a
massive ``relic galaxy'' that is a largely untouched descendant of the
red nugget
population at high redshift~\citep[e.g.,][]{remco15a,yild15a,ferr17a,yild17a}.}

While the connection between early formation and high $c_{200}$ is
critical, other factors must also contribute. \citet{ludl16a} estimate
that formation time accounts for a large fraction ($\sim 50\%$) of the
scatter in $c_{200}$, leaving about half the scatter to be likely
explained by the initial conditions or the environment of a given
halo. The recent study by \citet{raga18a} proposes that a measure of
the ``fossilness'' of a halo can account for some of this scatter,
though the proposed effect is too modest to explain the large
$c_{200}$ of  \src. These issues are likely central to understanding
how the DM and total mass profiles of \src\ and NGC~6482 are extremely
similar yet the stellar component of \src\ is more compact obeying the
size-mass relation for $z\sim 2$ galaxies like other CEGs. 

Finally, we note that we have emphasized comparisons with the
theoretical $c_{200}-M_{200}$ relations assuming a lognormal intrinsic
scatter. While a gaussian interpretation of the theoretical scatter is
consistent with cosmological simulations~\citep[e.g.,][]{bhat13a}, the
narrower tails of the gaussian imply \src\ would be an outlier so
extreme (see Child+18 results in Table~\ref{tab.cm}) for all models to be
incompatible with \lcdm. Consequently, the high $c_{200}$ values for
\src\ and NGC~6482 (B17) clearly favor a lognormal distribution for
the intrinsic scatter of the \lcdm\ $c_{200}-M_{200}$ relation.

\subsection{A Key Factor Contributing to Differences in the X-ray and
  Stellar Dynamical Constraints}
\label{keyfactor}

\begin{figure}[t]
\begin{center}
\includegraphics[scale=0.42]{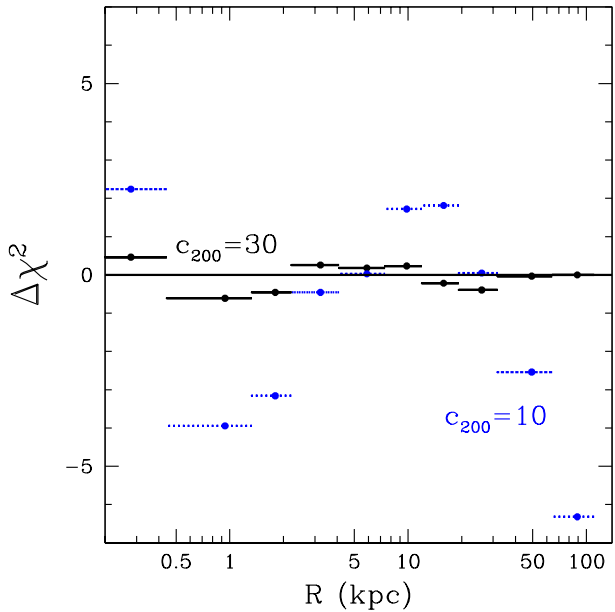}
\end{center}
\caption{\footnotesize $\chi^2$ residuals for frequentist HE model
  fits. We show only the surface brightness residuals for the
  Cycle 19 data. In black we show the fiducial model  ($c_{200}=30$) while
  in blue we show the model with $c_{200}=10$, a value typically
  assumed in stellar dynamical studies that do not constrain the DM halo.}
\label{fig.dchi2}
\end{figure}

We have found significant differences in the mass properties of \src\
inferred from our HE analysis of the hot plasma within a radius $\sim
R_e$ compared to the stellar dynamical (SD) constraints obtained by Y17 and
\citet{wals17a}. There is, however, a major difference in the assumed
DM model between these SD studies and ours. Since the SD studies are
unable to contrain the DM halo, by default they assume an NFW DM halo
with fixed $c_{200}=10$. Here we consider how our HE analysis changes
if we make a similar assumption.

We achieve a similar DM halo in our analysis by fixing the NFW scale
radius to $r_s=60$~kpc in which case we obtain, $c_{200}=9.8\pm 0.2$
and $M_{200} = (2.3\pm 0.1)\times 10^{13}\, \msun$ from our Bayesian
analysis. The fit quality, however, is very poor; i.e., the
frequentist analysis gives $\chi^2=39.5$ for 19 dof compared to
$\chi^2=11.9$ when $c_{200}=$ is allowed to take a value $c_{200}= 30.$ In
Figure~\ref{fig.dchi2} we plot the $\chi^2$ residuals for the
$c_{200}\approx 10,30$ models, showing just the surface brightness data for the
Cycle 19 observation. Compared to the $c_{200}=30$ model, the
$c_{200}=10$ model has much larger residuals over most of the radial
range; i.e., the \chandra\ data clearly exclude the $c_{200}=10$ model
in favor of the  $c_{200}=30$ model. 

While the $c_{200}=10$ model does not provide a good fit to the data,
the derived mass parameters agree very well with those of
Y17\footnote{We note that for NGC~1281 \citet{yild16a} initially
  reported the galaxy to have a ~90\% DM fraction within $R_e$ and a
  total mass of $\sim 10^{14}\,\msun$ indicative of a
  cluster. However, the large amount of DM they inferred, expressed by
  $\log_{10}(M_{\rm DM}/M_{\star})=3.6^{+1.5}_{-0.5}$, was subsequently
  substantially downwardly revised in Y17 to
  $\log_{10}(M_{\rm DM}/M_{\star})=1.49^{+1.57}_{-2.54}$, fully
  consistent with the typical CEG in their sample (and with a typical
  DM fraction within $R_e$).}.  We obtain $\mlh = 1.60\pm 0.05$~solar which agrees
with Y17's value, $\mlhband =1.85^{+0.52}_{-0.40}$, within their
larger $1\sigma$ error bar. While Y17 do not quote values of the total
mass slope and DM fraction within $R_e$ specifically for \src, their
sample-averaged values ($\langle\gamma\rangle = 2.3$,
$f_{\rm DM} = 0.11$) agree very well with what we obtain for \src\
($\langle\gamma\rangle = 2.27\pm 0.01$, $\dmfrac = 0.13\pm 0.01$) when
we assume $c_{200}=10$.  Furthermore, if we also fix the SMBH mass to
the best-fitting ``over-massive'' value ($\mbh=4.9\times 10^9\,\msun$)
obtained by \citet{wals17a}, then we obtain an even worse fit
($\chi^2=50.3$ for 20 dof) but with $\mlh = 1.36\pm 0.05$~solar in
excellent agreement with the value $\mlhband = 1.3\pm 0.4$ solar
obtained by \citet{wals17a}.

We conclude that the assumption of $c_{200}=10$ for the NFW DM halo
fully accounts for the higher $\mlh$ measured with SD by Y17 and
\citet{wals17a} as well as the higher $\langle\gamma\rangle$ and lower
$\dmfrac$ within $R_e$ obtained by Y17. Differences in the form of
the DM halo, however, have negligible impact on our constraints on
$\mbh$ which are manifested only in the central $1\arcsec$ aperture (i.e.,
Annulus~1 of the Cycle 19 data). Below in \S\ref{bigbh} we evaluate the
SMBH constraint further and consider its implications.

\subsection{Evidence for Over-Massive Black Hole}
\label{bigbh}

In \S\ref{smbh} we show that the HE analysis does not favor the
best-fitting over-massive SMBH
$[\mbh=(4.9\pm 1.7)\times 10^9\, \msun]$ found by the SD study of
\citet{wals17a}, but does not clearly exclude the value either when
considering different Bayesian priors and the frequentist fits. In
addition, the SD error estimate on $\mbh$ permits lower values (e.g.,
$1.4\times 10^9\,\msun\, (2\sigma)$ and
$0.65\times 10^9\,\msun\, (2.5\sigma)$) that are very consistent with
our HE models. In sum, the estimated statistical and systematic errors
in the  X-ray HE and SD constraints allow for broad consistency and do
not clearly point to whether the SMBH exceeds the $\msigma$ relation. 

Based on models of the Cycle 16 data of \src\ assuming
$\mbh=4.9\times 10^9\, \msun$, we expected to measure a value of
$\ktemp$ about $20\%$ higher in Annulus~1 than we found with the
Cycle 19 observation (see beginning of \S\ref{smbh}). If we attribute
this underestimate to deviations from HE, it would imply $\sim 40\%$
non-thermal pressure support in Annulus~1 from random turbulent
motions, magnetic fields, rotational support, etc. Future observations
with similar or better spatial resolution to \chandra\ but with much
higher spectral resolution and sensitivity (i.e, {\sl
  Lynx})\footnote{https://www.lynxobservatory.com} 
 will be
necessary to establish the hot gas kinematics within the central
$\approx 1\arcsec$. In the meantime, more
precise constraints on $\mbh$ from SD can better establish the
significance of any tension with the X-ray constraints.

\subsection{Global Baryon Fraction}
\label{disc.baryfrac}

Our Bayesian HE models evaluated at $\rtwoh$ indicate
$\fbtwoh\approx 80\%$ of the cosmic mean value (\S\ref{baryfrac}).
The evidence for near baryonic closure in \src\ is consistent with
results we have obtained previously for other fossil-like massive
elliptical galaxies, NGC~720~\citep{hump11a},
NGC~1521~\citep{hump12b}, and NGC~6482 (B17). These results suggest
that in massive elliptical galaxy / small group halos many of the
``missing'' baryons~\citep{fuku98} can be located in the outer halo as
part of the hot component.

As is clear from Figure~\ref{fig.baryfrac}, most of the baryons
inferred from our HE models within a radius of $\rtwoh$ are in the
form of hot gas located at radii larger than the extent of the
\chandra\ data ($\ga\rtwofiveh$). We have previously noted (\S 8.2 of
B17) the sensitivity of the measured $\fgastwoh$ to the assumed value
of $\zfe$ for $r>\rtwofiveh$. In fact, using a $\beta$-model analysis
of the \chandra\ surface brightness profile of NGC~720,
\citet{ande14a} claim that the gas mass $(M_{\rm gas,200})$ is a
factor 3-4 smaller than we measured in \citet{hump11a}. 

We have reanalyzed the \chandra\ surface brightness profiles using all
presently available observations and performed a similar $\beta$ model
analysis of NGC~720. We find $M_{\rm gas,200}$  values very consistent with those of
\citet{hump11a}. We attribute the lower value of $M_{\rm gas,200}$
obtained by \citet{ande14a} to primarily (1) their quoting a higher
value of $M_{\rm gas,200}$ than was actually reported in Figure~6
of~\citet{hump11a}; i.e., $4\times 10^{11}\, \msun$ instead of the
correct $3\times 10^{11}\, \msun$ at $r=300$~kpc; (2) their using a
different solar abundance standard~\citep{angr} whereas \citet{hump11a}
use~\citet{aspl}; and (3) their assuming $\ktemp=0.5$~keV, $\zfe=0.6$~solar
for the hot plasma appropriate for the center of NGC~720 instead of
quantities appropriate near $\rtwofiveh$ ($\ktemp\approx 0.3$~keV,
$\zfe\approx 0.3$~solar). 

To verify the nearly cosmic baryon fractions in these galaxies it is
necessary to measure both $\ktemp$ and $\zfe$
out to radii beyond $\rtwofiveh$. In future papers we will report the
results of new deep \xmm\ observations of \src\ and NGC~6482 that will
allow the gas to be mapped out closer to $\sim\rfiveh$.

\subsection{Cooling and Feedback}
\label{feedback}

The \chandra\ observations of \src\ suggest the hot plasma properties
within the central $\sim 10$~kpc reflect a balance between radiative
cooling and episodes of gentle AGN feedback. The centrally peaked
radial temperature profile itself (Figure~\ref{fig.best}) does not
implicate AGN shock-heating of the gas, since the continuous injection
of stellar material will produce the same effect~\citep[e.g.,
classical wind models,][]{davi91a,ciot91a} and the observed central
entropy is not peaked (Figure~\ref{fig.he}). Modern models of massive
elliptical galaxies incorporating gentle mechanical AGN feedback
episodes can produce similar centrally peaked temperature
profiles~\citep[e.g.,][]{gasp12a}. The lack of large, significant
disturbances (e.g., cavities) near the center (\S\ref{image}) or
azimuthal spectral variations~(\S\ref{quad}) provides further evidence
that episodic AGN feedback in \src\ has been gentle.

Interior to the sizable radius $\sim 10$~kpc within which $\tc/\tff$
ranges from 14-19 (Figure~\ref{fig.tc}, see \S\ref{tcool}), the
``precipitation-regulated feedback'' scenario~\citep[e.g.,][]{voit17a}
predicts there should be multiphase gas, yet there is little evidence
presently for gas cooler than the ambient hot plasma. However,
\citet{voit17a} emphasizes that condensation will be suppressed by
buoyancy damping if the entropy profile is steeply rising and without
a large isentropic core. We find the entropy profile has a radial
logarithmic slope $\alpha=0.78\pm 0.05$ for $0.66<r<16$~kpc
(Table~\ref{tab.fid}) and no evidence of a core. This slope is larger
than the critical value $\alpha_K=2/3$ for suppressing condensation
derived by \citet{voit17a} using a toy model for isothermal gas.

In fact, this result may be common for massive, X-ray luminous
elliptical galaxies. While the recent study by \citet{baby18a} finds a
universal entropy profile for elliptical galaxies and galaxy clusters
with an entropy slope consistent with $2/3$ within $0.1\rtwofiveh$,
our previous measurements within the central $\sim 10$~kpc of massive
elliptical galaxies have obtained steeper slopes $\alpha\ga 1$
consistent with suppressed condensation: NGC~4649~\citep{hump08a},
NGC~1332, NGC~4261, NGC~4472~\citep{hump09c}, and NGC~6482 (B17).

Although the $\alpha_K=2/3$ criterion for suppressing condensation from
the toy model of \citet{voit17a} is consistent with the single-phase
gas we observe in \src, the three-dimensional hydrodynamical
simulations of \citet{wang18a} predict that $\tc/\tff$ should rise
rapidly with radius for a single-phase gas, in conflict with the
\chandra\ data~(Figure~\ref{fig.tc}) and the lack of evidence for
multiphase gas in the amounts seen in other galaxies with
$\tc/\tff<20$~\citep{wern14a,voit15a}.

In \S\ref{tcool} we remarked that the region where $\tc/\tff$ ranges
from 14-19 (i.e., $r\la 10$~kpc), associated with a steeply
rising entropy profile with $\alpha > 2/3$, also is close to the NFW
DM scale radius $r_s$. For the galaxies listed above with good
constraints on the DM halo, $r_s$ is not far from 10~kpc; i.e.,
$r_s\approx 14$~kpc for NGC~720 and $r_s=11.2\pm 3$~kpc for NGC~6482.

We conclude by noting that the unusually quiescent evolutionary path
indicated by the highly concentrated stellar and DM components of
\src\ imply the presumed episodic AGN feedback regulating the cooling
in the central $\sim 10$~kpc has not been upset by triggering from
galaxy merging but has been solely governed by accretion from the
radiatively cooling hot plasma. We speculate this has also resulted in
unusually quiescent evolution in the hot plasma obeying the HE
condition accurately (see below in \S\ref{disc.he}).

\subsection{Hydrostatic Equilibrium Approximation}
\label{disc.he}

Since the \chandra\ ACIS spectral data do not permit useful
direct measurements of the kinematics of the hot plasma, we instead 
briefly summarize several indirect lines of evidence testifying to the
accuracy of the HE approximation in \src.

\bigskip

\noindent{\it Azimuthal Scatter in Quadrants (\S\ref{quad}) : } When
Annuli~2,3, and 4 of the Cycle 19 observations are divided up into
quadrants, we find the small scatter of the pseudo-pressure and
pseudo-entropy between quadrants in each annulus suggests HE deviations
of $\la 5\%$.  

\medskip 

\noindent{\it Quality of HE Model Fits (Table~\ref{tab.fitqual}): } We
find that reasonable HE models provide excellent fits to the
radial $\ktemp$ and $\xsurf$ profiles of the Cycle 19 and Cycle 16
\chandra\ data.

\medskip 

\noindent{\it Qualified Consistency with Stellar Dynamics Measurements
  (\S\ref{keyfactor}): } If we fix $c_{200}=10$ for the NFW DM halo as
done in the stellar dynamical studies by Y17 and \citet{wals17a}, we
find excellent agreement with the measured $\mlh$ values as well as
the total mass slope ($\langle\gamma\rangle$) and DM fraction
($\dmfrac$) obtained within $1\reff$. The constraints on $\mbh$ are
uncertain and therefore ambiguous with respect to the status of
HE in Annulus~1 of the Cycle~19 data (\S\ref{bigbh}).

\medskip 

\noindent{\it Stellar Mass-to-Light Ratio and SPS Models
  (\S\ref{stars}): }  The value we measure for $\mlh$ agrees very well
with SPS models with a Kroupa IMF. 

\medskip 

\noindent{\it High Halo Concentration (\S\ref{highc}): } The
exceptionally high value of $c_{200}$ we measure for \src\ corroborates  the evidence from
its compact stellar size and old stellar population that it is a
massive relic galaxy; i.e.,  \src\ has evolved passively in isolation since a
redshift of $\sim 2$ and, consequently, is highly evolved and relaxed. 

We therefore expect that \src\ is a highly relaxed system, at least as
relaxed as the massive Perseus cluster with its pronounced cavities
and other surface brightness irregularities for which \hitomi\ found
with gas kinematics measurements that the pressure from turbulence is
only $4\%$ of the thermal gas pressure~\citep{hit16a}. To verify
directly the accuracy of the HE approximation in \src\ awaits the next
generation of X-ray satellites with microcalorimeter
detectors.\footnote{https://heasarc.gsfc.nasa.gov/docs/xrism}\footnote{https://www.the-athena-x-ray-observatory.eu}

\section{Conclusions}
\label{conc}

We present a detailed analysis of the hot plasma X-ray emission and
gravitating mass profile of the CEG \src\ based on a new $\sim 122$~ks
\chandra\ X-ray observation (Cycle 19) and a shallow ($\sim 13$~ks)
archival \chandra\ observation (Cycle 16). The X-ray emission as
revealed by the deep Cycle 19 image exhibits a regular, relaxed
morphology symmetrically distributed about the central emission
peak. We perform a detailed analysis of the image morphology within
the central $\sim 15$~kpc to search for signs of AGN feedback. While
the image does not reveal obvious features of AGN feedback in the form
of cavities or other irregularities, we identify leading candidates
for such feedback signatures as regions with the largest surface
brightness fluctuations with respect to a smooth two-dimensional
model.
 
We search for azimuthal variations in the gas pressure and entropy
within the central $\sim 4$~kpc by dividing up several circular annuli
into four quadrants each. Using proxies for the pressure and entropy
inferred directly from the projected spectra, we find small azimuthal
scatter in these proxies consistent with $\la 5\%$ fluctuations in HE.
Adopting an entropy-based HE method, we then place constraints on the
inner and global mass profile from the radial profiles of temperature
and surface brightness measured from spectra extracted in circular
annuli extending out to $R=100.7$~kpc $(\approx 0.85\rtwofiveh)$. (In
Appendix~\ref{specresults.center} we discuss an anomalous spectral
feature in the central $\approx 1\arcsec$ annulus.)

Our principal conclusion is that the halo concentration of \src\ is a
large, positive outlier in the \lcdm\ $c_{200}-M_{200}$ relation. For
an NFW DM halo we obtain $c_{200}=30.4\pm 4.3$ and
$M_{200}=(5.3\pm 0.6)\times 10^{12}\, \msun$ representing an extreme
outlier in the \lcdm\ $c_{200}-M_{200}$ relation; i.e.,
$\approx 6\sigma$ above the median theoretical value
$c_{200}\approx 7$ (Table~\ref{tab.cm}) considering the intrinsic
scatter and $\approx 5.4\sigma$ considering the measurement error.  If
we replace the NFW DM profile with the Einasto profile modified by
``weak'' adiabatic contraction, the concentration is reduced to a less
extreme (but still significant) $3.4-3.7\,\sigma$ outlier more
compatible with \lcdm. 

The high value of $c_{200}$ we measure implies an unusually early
formation time for \src\ and therefore a more passive and quiescent
evolution compared to the typical halo. It therefore provides new
independent evidence corroborating \src\ as a massive ``relic galaxy''
that is a largely untouched descendant of the red nugget population at
high redshift. Moreover, whereas the stellar-mass-size relation and
old stellar populations previously established the relic nature of the {\it stellar
component} of CEGs, the high $c_{200}$ now firmly establishes the relic
nature of the {\it DM halo} in \src.

The high concentration of DM significantly affects the inferred mass
properties near $R_e$. We measure a DM fraction $(\dmfrac)$ within
$R_e$ about twice as large and a flatter total mass slope
$(\langle\gamma\rangle)$ compared to local massive elliptical
galaxies. Our measured values of $\dmfrac$ and $\langle\gamma\rangle$
also disagree significantly with the mean values of the CEG sample
(which includes \src) by Y17. However, we attribute the discrepancy
with Y17 to their assuming a fixed $c_{200}=10$ in their stellar
dynamical analysis. If we assume $c_{200}=10$ in our HE analysis of
the \chandra\ data, the fits are poor, but the parameters we derive
agree very well with those reported by Y17. We conclude that if the
other CEGs in Y17's sample also have high $c_{200}$ values like \src,
then the sample-average values for $\dmfrac$ and
$\langle\gamma\rangle$ obtained by Y17 will need to be revised
accordingly.  Finally, if \src\ is truly a nearby analog of a
$z\approx 2$ galaxy, the DM fraction we measure is larger than that
produced in recent cosmological simulations of $z=2$ galaxies
(\S\ref{slope}).

If we instead interpret our results using MOND gravity, we find that
MOND requires almost as much DM as in our Newtonian analysis. Owing to
the strong similarity between MOND and the RAR, it would be expected
that \src\ would deviate significantly from the RAR, which we
verify. We attribute the failure of MOND and the RAR to explain the
gravitational field of \src\ to the evidence for highly concentrated
DM occurring well inside the radius where the gravitational
acceleration equals the MOND constant $a_0$; i.e., the evidence for
the DM is safely in the Newtonian regime not addressed by
MOND. Although we have not explored other modified gravity models like
self-interacting DM (SIDM) that predict DM cores, we believe it is
likely that the highly concentrated DM in \src\ will place interesting
constraints on DM models that predict inner halo profiles shallower
than NFW.

Our analysis of the \chandra\ data does not place tight constraints on
the SMBH mass, although our HE models do not favor the over-massive
value $[\mbh=(4.9\pm 1.7)\times 10^9\, \msun]$ obtained with stellar
dynamics by \citet{wals17a}. However, the tension between the X-ray
and SD values of $\mbh$ is modest considering the uncertainties in both
measurements.

Within the extent of the \chandra\ data ($\approx\rtwofiveh$) we
measure a baryon fraction
$\fbtwofiveh=0.071\pm 0.006\approx 0.45f_{\rm b,U}$ with the hot gas
contributing $\approx 40\%$ of $\fbtwofiveh$. When evaluating our
models at $\rtwoh$ we obtain
$\fbtwoh=0.120\pm 0.016\approx 0.80f_{\rm b,U}$ with
$\fgastwoh\approx 0.8\fbtwoh$; i.e., our analysis of the hot plasma
indicates the halo of \src\ contains close to the cosmic fraction of
baryons, where most of the baryons are in the form of hot plasma
between $\rtwofiveh$ and $\rtwoh$.

The radial profile of the ratio of cooling time to free-fall time
varies within a narrow range $(\tc/\tff\approx 14-19)$ over a large
central region ($r\le 10$~kpc) exterior to which it increases
with radius. The observed minimum $\tc/\tff$ suggests
``precipitation-regulated AGN feedback'' for a multiphase
plasma. There is presently little evidence for substantial amounts of
multiphase gas within $r\sim 10$~kpc for \src, which may indicate the
steep radially increasing entropy profile has suppressed condensation
of the hot plasma (see \S\ref{feedback}).  We observe that the
approximately constant region of $\tc/\tff\approx 17$ ends near the
NFW DM halo scale radius $r_s\approx 12$~kpc.

Finally, other than the compact size of the stellar half-light radius
of \src, the stellar, gas, and DM properties of \src\ are remarkably
similar to those of the nearby fossil group NGC~6482 (B17). In
particular, consideration of the nearly identical very high $c_{200}$
values for each, but different compactness of their stellar
components, provides a striking example of factors other than
formation time that must also contribute to the scatter in the
$c_{200}-M_{200}$ relation; e.g., ``fossilness''~\citep{raga18a}.

\acknowledgements 

We thank the anonymous referee for a timely and constructive
report. DAB gratefully acknowledges partial support by NASA through
Chandra Award Numbers GO7-18125X and GO8-19065X issued by the Chandra
X-ray Observatory Center, which is operated by the Smithsonian
Astrophysical Observatory for and on behalf of NASA under contract
NAS8-03060. This research has made use of the NASA/IPAC Extragalactic
Database (NED) which is operated by the Jet Propulsion Laboratory,
California Institute of Technology, under contract with the National
Aeronautics and Space Administration.

\appendix

\section{Joint Spectral Fitting of Individual Cycle 19 Observations}
\label{joint}

\begin{table*}[t] \footnotesize
\begin{center}
\caption{Hot Gas Properties from Joint fit of Cycle 19 Observations}
\label{tab.gas.joint}
\begin{tabular}{lcccccccc}   \hline\hline\\[-7pt]
& & $R_{\rm in}$ & $R_{\rm out}$ & $\Sigma_{\rm x}$ (0.5-7.0~keV) & $k_BT$ & $Z_{\rm Fe}$ & $Z_{\rm Mg}/Z_{\rm Fe}$ & $Z_{\rm Si}/Z_{\rm Fe}$\\
Telescope & Annulus & (kpc) & (kpc) & (ergs cm$^2$ s$^{-1}$ arcmin$^{-2}$) & (keV) & (solar) & (solar) & (solar)\\
\hline \\[-7pt]
&   1 & 0.00 & 0.44 & $\rm   4.65e-11 \pm   1.32e-11$ & $  0.971 \pm   0.027$ & $   0.99 \pm    0.17$ & $   1.54 \pm    0.42$ & $   1.52 \pm    0.31$\\
&   2 & 0.44 & 1.33 & $\rm   1.43e-11 \pm   1.92e-12$ & $  0.904 \pm   0.019$ & $   0.95 \pm    0.09$ & $   1.10 \pm    0.15$ & $   0.75 \pm    0.08$\\
&   3 & 1.33 & 2.21 & $\rm   6.25e-12 \pm   8.67e-13$ & $  0.841 \pm   0.019$ & $   0.95 \pm    0.08$ &  tied &  tied\\
&   4 & 2.21 & 4.10 & $\rm   2.12e-12 \pm   2.26e-13$ & $  0.814 \pm   0.016$ & $   0.89 \pm    0.06$ & $   0.81 \pm    0.11$ &  tied\\
&   5 & 4.10 & 7.42 & $\rm   7.77e-13 \pm   8.68e-14$ & $  0.733 \pm   0.017$ & $   0.76 \pm    0.05$ &  tied &  tied\\
&   6 & 7.42 & 11.96 & $\rm   2.92e-13 \pm   5.03e-14$ & $  0.665 \pm   0.018$ & $   0.78 \pm    0.08$ & $   0.54 \pm    0.10$ & $   1.38 \pm    0.13$\\
&   7 & 11.96 & 19.26 & $\rm   1.19e-13 \pm   2.25e-14$ & $  0.668 \pm   0.020$ & $   0.71 \pm    0.07$ &  tied &  tied\\
&   8 & 19.26 & 31.77 & $\rm   4.27e-14 \pm   1.06e-14$ & $  0.680 \pm   0.024$ & $   0.77 \pm    0.09$ & $   0.81 \pm    0.15$ &  tied\\
&   9 & 31.77 & 64.20 & $\rm   1.14e-14 \pm   2.49e-15$ & $  0.616 \pm   0.026$ &  tied &  tied &  tied\\
&  10 & 64.20 & 110.70 & $\rm   2.66e-15 \pm   6.91e-16$ & $  0.689 \pm   0.090$ & 0.36 &  tied &  tied\\
\hline \\
\end{tabular}
\tablecomments{See notes to Table~\ref{tab.gas}.}
\end{center}
\end{table*}

In Table~\ref{tab.gas.joint} we list the results for the gas
properties obtained by jointly fitting the individual Cycle 19
exposures. The value of the C-statistic for the fit is 6487.8  for
6869 pha bins and 6799 dof. Note for this analysis the spectra of the individual
exposures were rebinned in the same way as the combined spectra (\S\ref{spec}).

\section{Spectral Deprojection Results}
\label{deproj.appendix}

\begin{table*}[t] \footnotesize
\begin{center}
\caption{Hot Gas Properties Obtained from Spectral Deprojection}
\label{tab.gas.deproj}
\begin{tabular}{lcccccccc}   \hline\hline\\[-7pt]
& & $R_{\rm in}$ & $R_{\rm out}$ & $\rho_{\rm gas}$ & $k_BT$ & $Z_{\rm Fe}$ & $Z_{\rm Mg}/Z_{\rm Fe}$ & $Z_{\rm Si}/Z_{\rm Fe}$\\
Telescope & Annulus & (kpc) & (kpc) & (g cm$^{-3}$) & (keV) & (solar) & (solar) & (solar)\\
\hline \\[-7pt]
\\  \chandra\ Cycle~19\\
&   1 & 0.00 & 0.44 &   7.76e-25 $\pm$   1.94e-25 & $  0.988 \pm   0.046$ & $   1.04 \pm    0.30$ & $   1.85 \pm    1.05$ & $   2.03 \pm    0.85$\\
&   2 & 0.44 & 1.33 &   2.58e-25 $\pm$   3.01e-26 & $  0.954 \pm   0.038$ & $   1.04 \pm    0.18$ & $   1.24 \pm    0.32$ & $   0.74 \pm    0.15$\\
&   3 & 1.33 & 2.21 &   1.81e-25 $\pm$   2.28e-26 & $  0.841 \pm   0.034$ & $   0.95 \pm    0.16$ &  tied &  tied\\
&   4 & 2.21 & 4.10 &   6.75e-26 $\pm$   8.04e-27 & $  0.850 \pm   0.029$ & $   1.02 \pm    0.16$ & $   0.91 \pm    0.24$ &  tied\\
&   5 & 4.10 & 7.42 &   3.62e-26 $\pm$   3.94e-27 & $  0.740 \pm   0.038$ & $   0.74 \pm    0.11$ &  tied &  tied\\
&   6 & 7.42 & 11.96 &   1.56e-26 $\pm$   2.11e-27 & $  0.716 \pm   0.036$ & $   0.93 \pm    0.15$ & $   0.55 \pm    0.21$ & $   1.42 \pm    0.20$\\
&   7 & 11.96 & 19.26 &   1.03e-26 $\pm$   1.55e-27 & $  0.617 \pm   0.044$ & $   0.65 \pm    0.13$ &  tied &  tied\\
&   8 & 19.26 & 31.77 &   4.17e-27 $\pm$   4.69e-28 & $  0.717 \pm   0.036$ & $   0.83 \pm    0.13$ & $   0.78 \pm    0.21$ &  tied\\
&   9 & 31.77 & 64.20 &   1.58e-27 $\pm$   1.93e-28 & $  0.601 \pm   0.043$ &  tied &  tied &  tied\\
&  10 & 64.20 & 110.70 &   9.45e-28 $\pm$   1.80e-28 & $  0.677 \pm   0.156$ & 0.36 &  tied &  tied\\
\\  \chandra\ Cycle~16\\
&   1 & 0.00 & 0.78 &   5.21e-25 $\pm$   7.85e-26 & $  1.232 \pm   0.113$ & $   1.16 \pm    0.29$ & $   0.60 \pm    0.29$ & $   0.74 \pm    0.32$\\
&   2 & 0.78 & 1.77 &   1.94e-25 $\pm$   3.27e-26 & $  0.829 \pm   0.060$ &  tied &  tied &  tied\\
&   3 & 1.77 & 3.54 &   7.00e-26 $\pm$   1.30e-26 & $  0.852 \pm   0.066$ &  tied &  tied &  tied\\
&   4 & 3.54 & 6.86 &   3.99e-26 $\pm$   6.48e-27 & $  0.763 \pm   0.062$ & $   0.84 \pm    0.17$ &  tied &  tied\\
&   5 & 6.86 & 14.39 &   1.46e-26 $\pm$   2.34e-27 & $  0.640 \pm   0.056$ &  tied &  tied &  tied\\
&   6 & 14.39 & 28.78 &   5.95e-27 $\pm$   9.38e-28 & $  0.767 \pm   0.068$ & $   0.69 \pm    0.14$ &  tied &  tied\\
&   7 & 28.78 & 73.06 &   1.49e-27 $\pm$   2.80e-28 & $  0.611 \pm   0.086$ &  tied &  tied &  tied\\
\hline \\
\end{tabular}
\tablecomments{These properties of the hot gas have been obtained through spectral deprojection using the {\sc projct} model in \xspec that assumes
that within each annulus the spectrum is spatially constant.  To convert the gas mass density $(\rho_{\rm gas})$  to electron number density ($n_e$) multiply by 
$5.1\times 10^{23}$. See \S\ref{deproj.specresults}, \ref{deproj.appendix} for further details on the deprojection results.}
\end{center}
\end{table*}

In Table~\ref{tab.gas.joint} we list the results for the gas
properties obtained for spectral projection using the {\sc projct}
model in {\sc xspec} (\S\ref{deproj.specresults}). 

\section{Anomalous Line Feature in the Central Spectrum of the Cycle 19 Observation}
\label{specresults.center}

\begin{figure*}
\parbox{0.49\textwidth}{
\centerline{\includegraphics[scale=0.32,angle=-90]{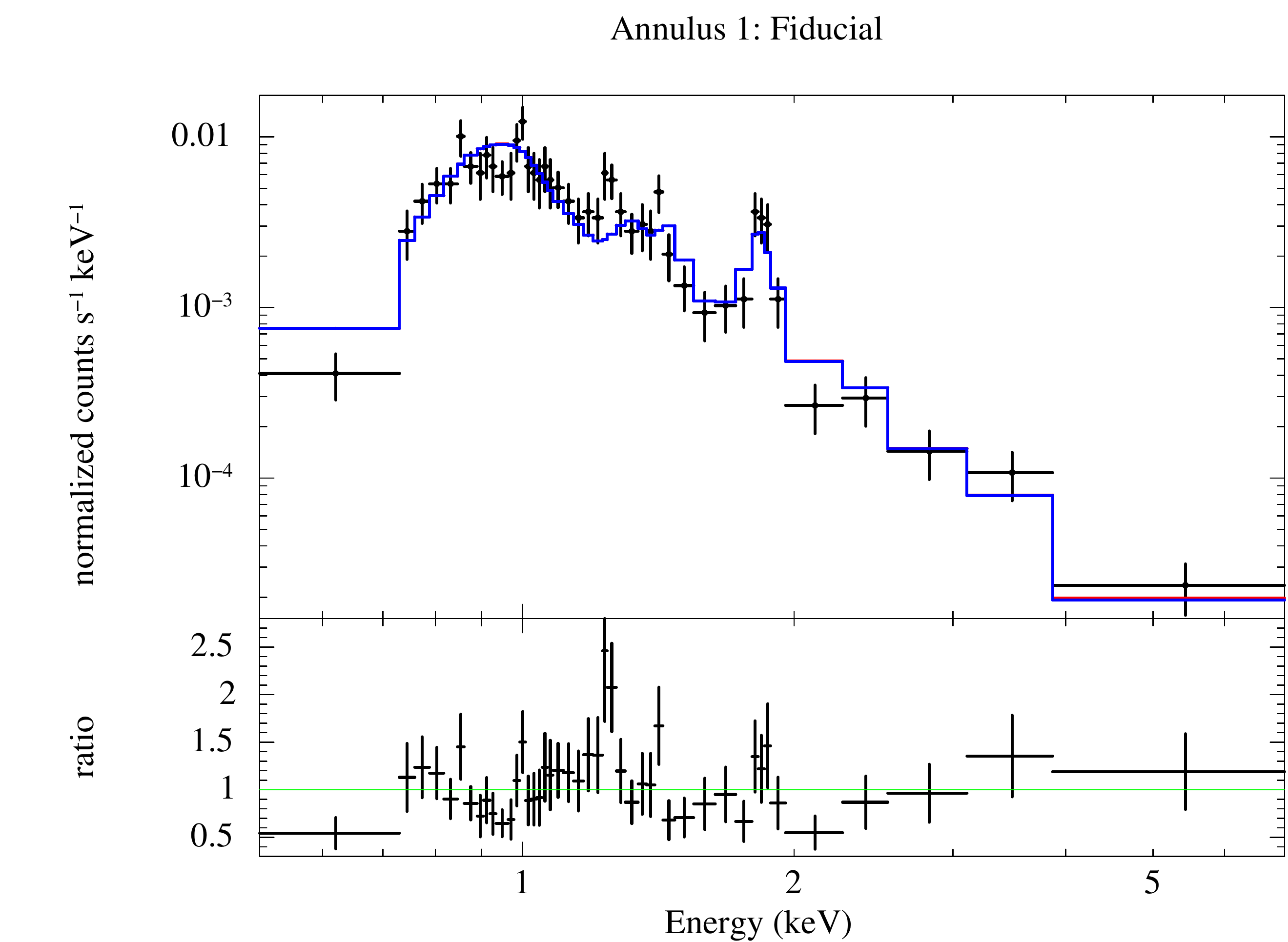}}}
\parbox{0.49\textwidth}{
\centerline{\includegraphics[scale=0.32,angle=-90]{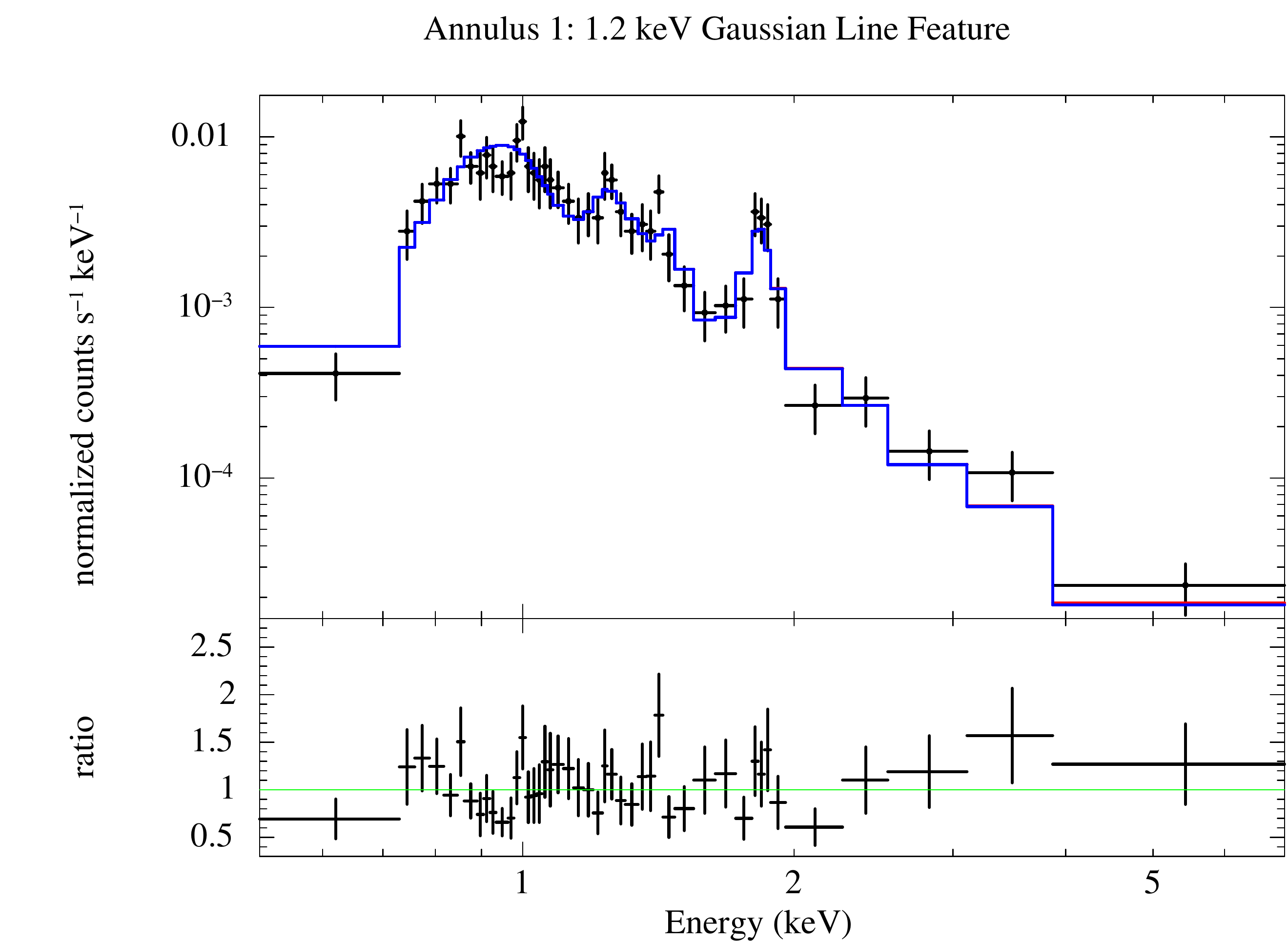}}}
\caption{\label{fig.center} {\sl Left panel:}The Cycle 19 spectrum of
  Annulus~1 plotted as in Figure~\ref{fig.spec} except that the
  residuals are now shown in the bottom panel as data/model
  ratios. {\sl Right panel:} A narrow gaussian emission line with
  $E\approx 1.23$~keV is added to improve the fit.}
\end{figure*}

As noted in \S\ref{specresults}, the Cycle 19 spectrum of Annulus~1
displays some features that deviate from our adopted composite model,
particularly an excess near $E\sim 1.2$~keV. Here we examine the fit
residuals more closely and examine ways to improve the model fit.
In the left panel of Figure~\ref{fig.center} we again
plot the spectrum and best-fitting model of Annulus~1 as in
Figure~\ref{fig.spec} except now we also plot the data / model ratio
in the bottom panel. Despite the presence of these residuals, the fit
is marginally acceptable when judged by the value of the C-statistic
or $\chi^2$; e.g., we obtain a $\chi^2$ null hypothesis probability
of 2.5\% if we consider only the Annulus~1 spectrum and ignore the
negligible background but allow the 7.3~keV bremsstrahlung component
normalization to vary freely (with the spectrum rebinned to at least
20 counts per channel).

To see if this marginal fit can be improved, we begin by considering
well-motivated missing ingredients from our fiducial model. First,
this single-temperature hot plasma component only approximates what is
surely a continuous, radially varying temperature gradient within the
aperture. We therefore expect emission from a range of temperatures
which will produce a slightly broader thermal spectrum than our
single-temperature model.  In fact, the residuals displayed in
Figure~\ref{fig.center} resemble the residual pattern characteristic
of fitting a single-temperature model to a multi-temperature spectrum
with average $\ktemp\sim 1$~keV at \asca/\chandra\ CCD
resolution~\citep[e.g.,][]{buot98c,buot00a,buot03b}. However, we are
unable to improve the fit by adding more discrete temperature
components or a continuous temperature distribution represented by,
e.g., a gaussian differential emission measure. Spectral deprojection
does not improve the fit either (\S\ref{deproj.specresults}). (We
also found no improvement from models allowing for non-equilibrium
ionization and plasma shocks.)

Second, the weak central radio source implies the presence of a
low-luminosity AGN, and we might reasonably expect corresponding X-ray
emission in the \chandra\ bandpass provided the AGN is not too heavily
absorbed. As is readily apparent from Figure~\ref{fig.center}, there
is no significant excess emission at higher energies signaling this
component; i.e., the emission from the unresolved LMXB component is
sufficient to describe the higher energies.

Since neither of these physically well-motivated modifications to the
fiducial model obviously improves the fit, we resort to an empirical
approach. The largest residual excess occurs near an energy 1.2~keV,
and we are unable to adjust our fiducial model (including allowing
other metal abundances to vary) to describe the feature. Consequently,
we tried adding a narrow gaussian emission line and show the
best-fitting result in the right panel of Figure~\ref{fig.center}. The
fit is clearly improved, not only near the line, but for many of the
energy channels below $\sim 1.5$~keV, with a reduction in the
C-statistic of $\approx 19$. 

We obtain good constraints on the fitted parameters of the line:
energy, $E=1.231\pm 0.014$~keV, and flux,
$\log_{10}F_{\rm x}=-14.83^{+0.10}_{-0.13}$ (with the $0.5-10$~keV flux in
ergs~cm$^{-2}$~s$^{-1}$), where the $3\sigma$ lower limit on the flux
is $\log_{10}F_{\rm x}=-15.33$.  Expressing the result in terms of the
line luminosity, we obtain,
$\log_{10}L_{\rm x}=-39.23^{+0.10}_{-0.12}$ (with luminosity in
ergs~s$^{-1}$).

We offer several possible explanations for this line feature and
assess their validity. 

(1) Problem with averaging the response
matrices: Our default procedure combines the spectra from the four
Cycle 19 exposures and averages the RMF and ARF files. However, we
also perform the analysis through joint analysis of the individual
observations and obtain fully consistent results. 

(2) Calibration
problem: We do not believe a calibration error is a viable explanation
since we do not see this line feature in the spectra of the other
annuli, nor are we aware of any reports of anomalous features in the
ACIS-S near 1.2~keV.  

(3) Plasma code: Since the available plasma
codes exhibit some notable differences~\citep[e.g.,][]{mern18a}, we
compared our results using the {\sc vapec} plasma code to those
obtained with the {\sc cie} plasma code from the {\sc spex} v3.0
spectral fitting
package~\citep{spex}\footnote{https://www.sron.nl/astrophysics-spex}.
The {\sc cie} model fit and resulting residual pattern is extremely
similar to what we obtained with {\sc vapec}. In particular, the {\sc
  cie} model cannot explain the 1.2~keV line feature using parameters
consistent with our fiducial {\sc vapec} model. However, the {\sc cie}
model also allows for a variable Na abundance which can reproduce the
1.2~keV line feature reasonably well but only with a large, unphysical abundance
($>100$~solar).  

(4) Decaying Dark Matter: X-ray emission lines may be
signatures of decaying dark matter from a sterile
neutrino~\citep[e.g.,][]{abaz17a,hito16b}. The line flux we
measure for the 1.2~keV feature is too strong and implies a mixing
angle that is too large to be compatible with
the currently allowed parameter space~\citep[e.g.,][]{abaz17a}. 

(5) Charge exchange. The potential importance of charge exchange
emission in clusters has been discussed, though observational evidence
for it remains tentative~\citep[e.g.,][and references
therein]{gu18a,gu18b}. With {\sc spex} we examined whether the charge-exchange
model ({\sc CX})  of \citet{gu16a} could explain the line feature. The
{\sc cx} model can produce a Ne line at the right energy. However,
along with 1.2~keV emission, the {\sc CX} model produces considerably
more Ne Ly$\alpha$ emission near 1~keV that is incompatible with
the observation. In addition, the emission near 1.2~keV  predicted by
the  {\sc CX} model is broader than the observed feature. 
 
In sum, none of the possibilities we have discussed is likely
entirely responsible for the 1.2~keV line, and perhaps the feature is
merely a statistical fluke. We believe, however, that the
charge-exchange model deserves further study. First, the Ne line it
predicts lies at the right energy. Second, for charge exchange to
occur, there must be neutral material. Inspection of the residuals in
the right panel of Figure~\ref{fig.center} reveals that the lowest
energies still show a deficit with respect to the model. When allowing
for absorption in excess of the foreground Galactic absorbing column,
the fit is improved a little more (C-statistic decreases by
$\approx 4$) and is consistent with the presence of cold gas in the
center. (Allowing the column density to be a free parameter in the
other annuli produces no such improvement in the fit; i.e., the
Galactic column is obtained elsewhere.) Although there is presently no
direct evidence for neutral gas at the center of \src, future
observations with {\sl ALMA} could determine whether a substantial
amount of molecular gas surrounds the SMBH.

With our empirical approach here, we mention that adding the narrow
gaussian has some effect on the parameters derived for the hot plasma
component in Annulus~1. While we find $\ktemp$ is unaffected,
$Z_{\rm Fe}$ increases and the gas density decreases. As a systemetics
check on our fiducial results, we have used these modified gas
parameters as input to our HE models and find that the main results
are unchanged within the $\approx 1\sigma$ errors.

Finally, we also mention that in our brief use of the {\sc cie} plasma
model with {\sc spex} in this section, we notice that the derived
temperatures are typically 5-10 percent smaller than those obtained
with the {\sc vapec} model in {\sc xspec}. We would expect this shift
to translate to a similar reduction in magnitude for the masses we
obtained from our HE models, which is comparable to the sizes of the
$1\sigma$ statistical errors (Table~\ref{tab.mass}) and of similar
magnitude to other systematic errors considered. Fully interpreting all
the data with {\sc spex} for comparison to {\sc vapec} is beyond the
scope of our paper.

\bibliographystyle{apj}
\bibliography{dabrefs}

\end{document}